\documentclass{psp-book9x6-modified}

\usepackage{psp-book-van-modified} 
\usepackage{cite,graphicx,color,amsmath,booktabs,microtype,afterpage,subfigure,fixmath,truncate}
\usepackage{mathpazo} 
\usepackage[scaled]{helvet} 
\usepackage{bibunits} 
\usepackage{perpage} 
\usepackage[geometry]{ifsym}
\usepackage[colorlinks,plainpages=false,linkcolor=black,urlcolor=black,citecolor=black,pdfpagemode=UseNone,pdfstartview=FitBH]{hyperref}
\usepackage[dvipsnames]{xcolor}
\usepackage[T1]{fontenc}\usepackage[latin1]{inputenc}

\let\Gamma\varGamma
\let\Delta\varDelta
\let\Theta\varTheta
\let\Lambda\varLambda
\let\Xi\varXi
\let\Pi\varPi
\let\Sigma\varSigma
\let\Upsilon\varUpsilon
\let\Phi\varPhi
\let\Psi\varPsi

\makeatletter
\renewcommand\thefigure{\thechapter.\@arabic\c@figure}
\makeatother

\usepackage{perpage}\MakePerPage{footnote}

\defaultbibliographystyle{inosov-book}

\copyline{Rare-Earth Borides}{Dmytro S. Inosov (ed.)}{2020}
\title{Rare-Earth Borides} 

\makeindex

\graphicspath{{Figures/}} 

\begin{document}









\cleardoublepage
\setcounter{page}{1}
\setcounter{chapter}{8}

%
%
%
%
%
%

\chapter[\bf \mbox{Neutron-scattering studies of spin dynamics in\hspace{-1em}} \hspace{1em}pure and doped CeB$_\text{6}$]{\bf Neutron-scattering studies of spin dynamics in pure and doped CeB$_\text{6}$\label{Chapter:Inosov}}
\addtocontents{toc}{\vspace{-2pt}\hspace{2.7em}\textit{by Pavlo~Portnichenko, Alistair~Cameron, and Dmytro~Inosov}\smallskip}

\chapauth{\mbox{Pavlo~Y. Portnichenko, Alistair~S. Cameron, and Dmytro~S. Inosov$^\ast$}
\chapaff{\noindent
\mbox{Institut f\"ur Festk\"orper- und Materialphysik, TU Dresden, 01069 Dresden, Germany}\\
$^\ast$E-mail address: \href{mailto:dmytro.inosov@tu-dresden.de}{dmytro.inosov@tu-dresden.de}}}

\begin{bibunit}

\vspace{-2pt}\section*{Abstract}\vspace{-1pt}\enlargethispage{2pt}

Magnetism in heavy-fermion metals is governed by the competition between the Kondo screening, which tends to quench the localized magnetic moments and to form a nonmagnetic ground state, and the Ruderman-Kittel-Kasuya-Yosida (RKKY)\index{RKKY interaction} coupling mechanism via the conduction electrons, which supports long-range magnetic order. This competition results in an amazingly rich variety of possible ground states. An important ingredient in the $f$\!-electron system is the strong spin-orbit coupling, which leads to the presence of new eigenstates that may be described in terms of the multipolar moments. Some heavy-fermion materials show long-range ordered multipolar phases, which are invisible to conventional diffraction techniques~\cite{Kusunose08, KuramotoKusunose09}. This so-called ``hidden order'' has been observed in a variety of compounds containing $4f$- and $5f$-elements, like URu$_2$Si$_2$,\index{URu$_2$Si$_2$} NpO$_2$,\index{NpO$_2$} YbRu$_2$Ge$_2$,\index{YbRu$_2$Ge$_2$} and CeB$_6$. Unraveling the underlying structure and a wide range of associated phenomena, ranging from quantum criticality to orbital ordering and unconventional superconductivity, requires a deep understanding of the interplay among these degrees of freedom.

The most well-studied member of this family of compounds is cerium hexaboride, CeB$_6$~\cite{StackelbergNeumann32}, which is considered a textbook example of a system with magnetically hidden order. As a simple-cubic system with only one $f$ electron per cerium ion, CeB$_6$ is of model character to investigate the interplay of orbital phenomena with magnetism. It is difficult to identify the symmetry of hidden-order states in common x-ray or neutron scattering experiments, as there is no signal in zero field, however alternative techniques like neutron diffraction in external field~\cite{ErkelensRegnault87},\index{neutron diffraction} resonant x-ray scattering~\cite{NagaoIgarashi01, NagaoIgarashi06, MatsumuraYonmura09, NagaoIgarashi10},\index{resonant x-ray diffraction} or ultrasonic investigations \cite{NakamuraGoto96, YanagisawaMombetsu18}\index{ultrasound measurements} can be applied.

Another possible method for characterizing hidden order is to look at the magnetic excitation spectrum, which carries the imprint of the multipolar interactions and the hidden order parameter in its dispersion relations \cite{ShenLiu19, PortnichenkoNikitin19}. Using a specific candidate model, the dispersion is calculated and then compared to that measured with inelastic neutron scattering. Until recently, only a limited amount of data which show the presence of dispersing excitations measured along a few high-symmetry directions in an applied magnetic field were available~\cite{Bouvet93}. Early attempts to compare such calculations~\cite{ShiinaShiba03, ThalmeierShiina98, ThalmeierShiina03, ThalmeierShiina04} with experiments showed that only strongest modes at high-symmetry points could be identified. The review of recent neutron-scattering results presented in this chapter is intended to satisfy the need of more accurate inelastic neutron-scattering experiments as a function of field and temperature, explicitly mentioned by theoreticians~\cite{ThalmeierShiina03}, giving us the opportunity to identify existing excitation branches and conclusively compare them with the theoretically predicted multipolar excitations.\index{CeB$_6$!multipolar excitations}\index{multipolar excitations!in CeB$_6$}

\vspace{-2pt}\section{Introduction}\vspace{-1pt}
\label{Ino_SubSec:CeB6Introduction}
\index{CeB$_6$|(}

\begin{figure}[b!]
	\centerline{\psfig{file=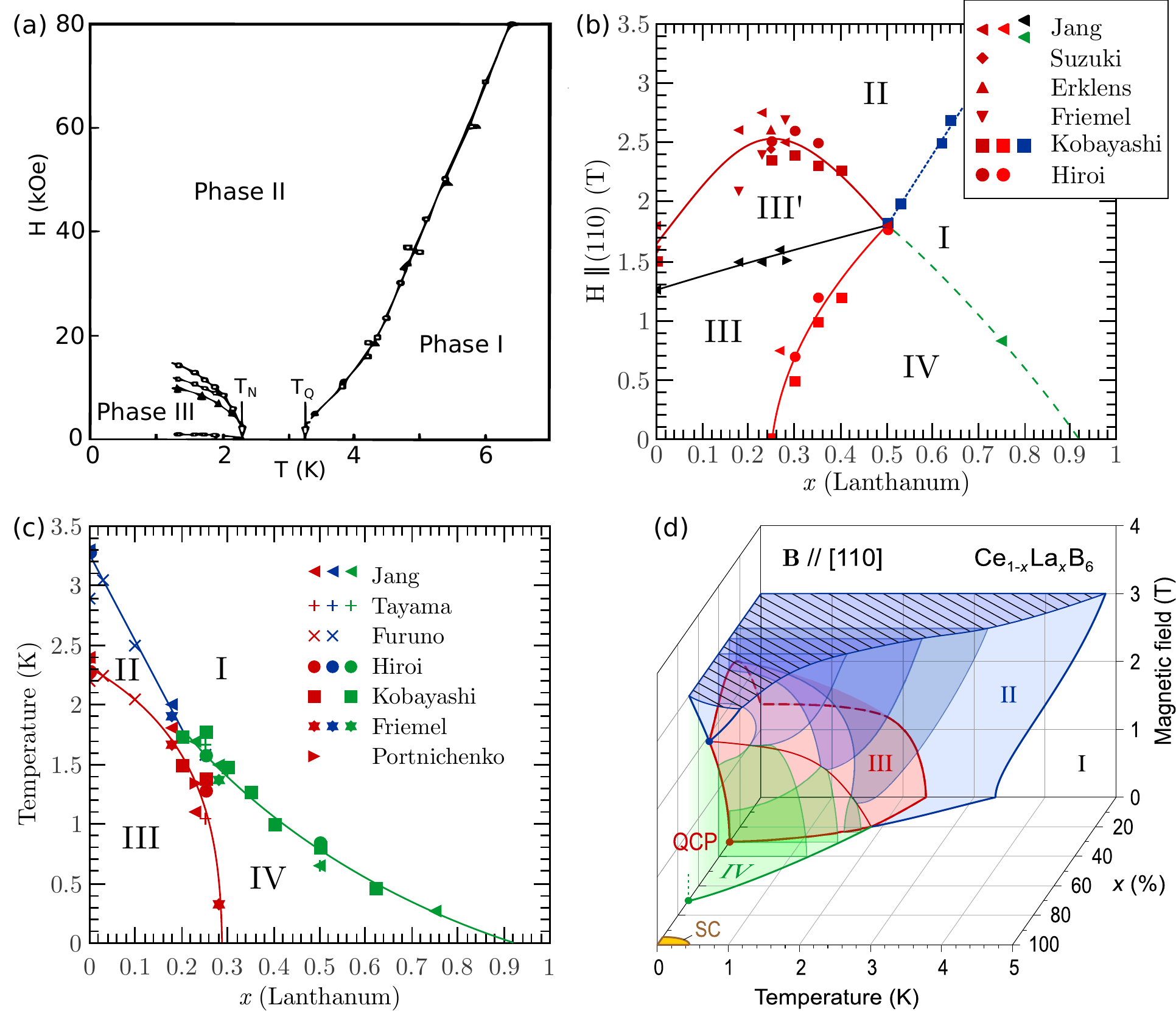, width=1.02\textwidth}}
	\caption{Phase diagrams of cerium hexaboride and its solid solutions. (a)~The temperature\,--\,magnetic-field phase diagram of CeB$_{6}$, reproduced from Ref.~\cite{TakaseKojima80}. Squares are for $\mathbf{B} \parallel [100]$, circles for $\mathbf{B} \parallel [110]$, and triangles for $\mathbf{B} \parallel [111]$. (b)~Doping\,--\,magnetic-field phase diagram for measurements extrapolated to \mbox{$T=0$}, adapted from Refs.~\cite{HiroiSera97, HiroiKobayashi98, KobayashiSera00, ErkelensRegnault87, SuzukiNakamura05, FriemelJang15, JangPortnichenko17}. (c)~Temperature\,--\,doping phase diagram in zero applied field, adapted from Refs.~\cite{HiroiKobayashi98, TayamaSakakibara97, FurunoSato85,KobayashiYoshino03,HiroiSera97,KobayashiSera00, Friemel14, PortnichenkoDemishev16, JangPortnichenko17}. (d)~A schematic temperature\,--\,magnetic-field\,--\,doping phase diagram for the field applied along the $[110]$ direction, reproduced from Ref.~\cite{FriemelJang15}. Phases~I (clear), II~(blue), III~(red) and IV~(green) are shown, alongside the small superconducting dome of LaB$_6$ in yellow~\cite{MatthiasGeballe68}. Quantum critical points are indicated at $x \approx 0.3$ and $x \approx 0.7$ for zero field, with a multi-critical point at $x \approx 0.5$ in an applied field of $\sim$2\,T. Cuts in the field\,--\,temperature plane correspond to reported phase diagrams for lanthanum doping~\cite{HiroiKobayashi98, ErkelensRegnault87, SuzukiNakamura05, KunimoriTanida10}.
\index{CeB$_6$!phase I}\index{CeB$_6$!phase II}\index{CeB$_6$!phase III}\index{Ce$_{1-x}$La$_x$B$_6$!phase IV}}\index{Ce$_{1-x}$La$_x$B$_6$!magnetic phase diagram}\index{magnetic phase diagram!of Ce$_{1-x}$La$_x$B$_6$}
\label{FigInosov:PhaseDiagrams}
\end{figure}

As cerium hexaboride (CeB$_6$) and its doped relatives exhibit magnetic and orbital order at low temperatures, it is natural that neutron scattering would play a key role in the study of these systems. Despite the presence of a magnetically-hidden-order phase, which cannot be directly probed by neutron diffraction in zero magnetic field, this technique has still been key in unlocking the mysteries of the low-temperature behavior in hexaborides. In this chapter we will discuss the neutron scattering research done on CeB$_6$ and its doped derivatives, Ce$_{1-x}$La$_x$B$_6$ and Ce$_{1-x}$Nd$_x$B$_6$.

The parent compound CeB$_6$ has three low-temperature phases, illustrated in the phase diagram of Fig.~\ref{FigInosov:PhaseDiagrams}\,(a). In zero field, the ground state is an antiferromagnetic (AFM) phase, labeled phase~III,\index{CeB$_6$!phase III} which persists up to $T_{\rm N} = 2.4$~K~\cite{ZirngieblHillebrands84}.\index{CeB$_6$!antiferromagnetic order}\index{antiferromagnetic order!in CeB$_6$} Above this is phase~II,\index{CeB$_6$!phase II} which has been identified as an antiferroquadrupolar (AFQ) phase\index{CeB$_6$!antiferroquadrupolar order} with a critical temperature of $T_{\rm Q} = 3.2$~K~\cite{EffantinRossat-Mignod85}. At even higher temperatures, the system is in the paramagnetic phase~I.\index{CeB$_6$!phase I} Under the application of magnetic field $\mathbf{B}$, phase~III\index{CeB$_6$!phase III} undergoes a transition to phase~III$^\prime$,\index{CeB$_6$!phase III$'$} where changes in the AFM structure occur. Interestingly, the phase~II\index{CeB$_6$!phase II} transition temperature $T_{\rm Q}(B)$ possesses a positive slope at low fields, indicating that this state is stabilized with the application of magnetic field. This positive slope continues to around 35\,T, where $T_{\rm Q}$ has a maximum at 10~K~\cite{MurakamiKawada98}. Beyond this, $T_{\rm Q}$ is suppressed to about 8~K at 60\,T, with the field required to fully suppress this phase not yet known. This phase diagram is mildly dependent on field orientation, with the basic features preserved but the critical fields being dependent on field direction as can be seen for the evolution of the boundary between phases II\index{CeB$_6$!phase II} and III\index{CeB$_6$!phase III} in Fig.~\ref{FigInosov:PhaseDiagrams}\,(a).

Figures~\ref{FigInosov:PhaseDiagrams}\,(b,\,c) illustrate the changes in the phase diagram upon lanthanum doping, $x$, as a function of field and temperature respectively. At a doping of \mbox{$x \approx 0.25$}, phase~IV\index{Ce$_{1-x}$La$_x$B$_6$!phase IV} emerges as the ground state of the system, and it is generally believed that this is an antiferrooctupolar (AFO) phase~\cite{MannixTanaka05, KuwaharaIwasa07, LoveseyFernandez-Rodriguez07, KuwaharaIwasa09}.\index{Ce$_{1-x}$La$_x$B$_6$!octupolar order}\index{magnetic octupoles} This is stabilized against the applied field but weakened against temperature with increasing doping to $x = 0.5$, whereupon there is presumably a multicritical point in applied field. Beyond this, the AFO phase is suppressed against field with increasing doping until \mbox{$x > 0.75$}, before the proposed emergence of superconductivity around $x = 1$~\cite{MatthiasGeballe68, SobczakSienko79, SchellWinter82}.\index{LaB$_6$!superconductivity} Figure~\ref{FigInosov:PhaseDiagrams}\,(d) shows a schematic representation of the phase diagram in the full temperature-field-doping space~\cite{FriemelJang15}. Until recently, it was thought that the stabilization of the AFO phase against field continued beyond \mbox{$x = 0.5$}~\cite{TayamaSakakibara97, CameronFriemel16}, but detailed heat-capacity measurements showed more recently that this was not the case, and that phase~IV\index{Ce$_{1-x}$La$_x$B$_6$!phase IV} is continuously suppressed against field with increasing lanthanum doping beyond the multicritical point~\cite{JangPortnichenko17}.

\vspace{-3pt}\section{Electronic properties and ordering phenomena}\vspace{-2pt}
\index{CeB$_6$!electronic properties|(}\index{CeB$_6$!ordering phenomena|(}
\index{neutron scattering|(}

\subsection{Magnetic structure in the antiferromagnetic phases}
\index{CeB$_6$!magnetic structure|(}

\begin{figure}[b!]
\centerline{\psfig{file=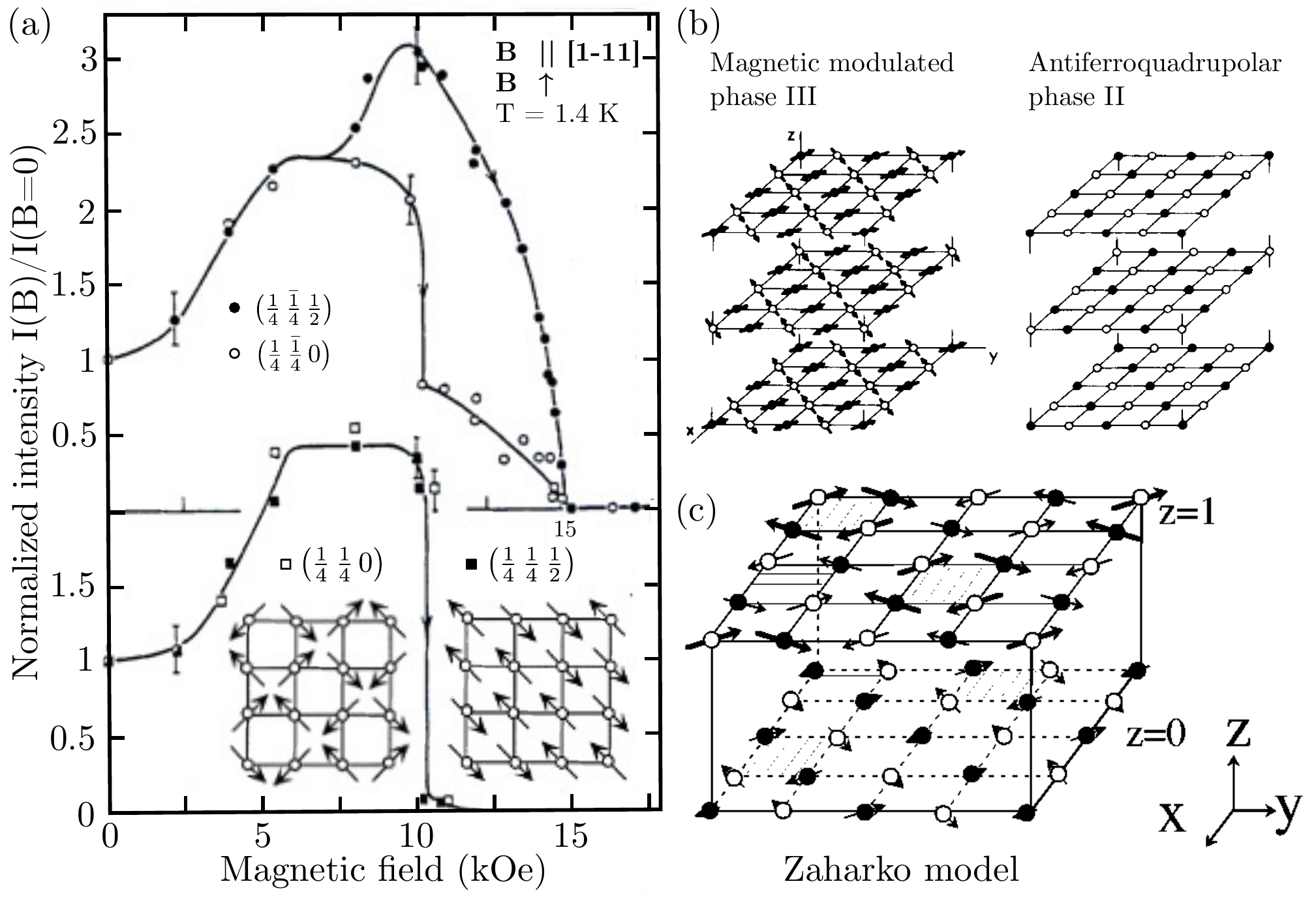, width=1\textwidth}}
\caption{(a)~The field dependence of the magnetic Bragg reflections from phase~III\index{CeB$_6$!phase III} and III$^\prime$,\index{CeB$_6$!phase III$'$} reproduced from Ref.~\cite{EffantinBurlet82}. The proposed in-plane spin structures are shown inset. (b)~Full 3D magnetic structures for phases~III and III$^\prime$, first proposed in Ref.~\cite{Effantin85}. Reproduced from Effantin \textit{et al.}~\cite{EffantinRossat-Mignod85}. (c)~An alternative magnetic structure for phase~III, suggested by Zaharko \textit{et~al.}~\cite{ZaharkoFischer03}.\index{CeB$_6$!antiferromagnetic order}\index{antiferromagnetic order!in CeB$_6$}\index{CeB$_6$!magnetic structure}\index{CeB$_6$!Bragg intensity}}
\label{FigInosov:MagneticStructure}
\end{figure}

The appearance of AFM order in CeB$_6$ \index{CeB$_6$!antiferromagnetic order}\index{antiferromagnetic order!in CeB$_6$} in zero field takes place at \mbox{$T_{\rm N} =  2.4$~K}, which was initially deduced by susceptibility measurements~\cite{PadernoPokrzywnicki67}. Confirmation of the AFM structure came from neutron diffraction~\cite{HornSteglich81, EffantinBurlet82, Effantin85, BurletRossat-Mignod82, ZaharkoFischer03}, which determined the ordering vectors at \mbox{$\mathbf{q}_{1} = \bigl(\frac{1}{4}\frac{1}{4}0\bigr)$}, \mbox{$\mathbf{q}_{2} = \bigl(\frac{1}{4}\frac{\overline{1}}{4}0\bigr)$} and \mbox{$\mathbf{q}_{1}^\prime = \bigl(\frac{1}{4}\frac{1}{4}\frac{1}{2}\bigr)$}, \mbox{$\mathbf{q}_{2}^\prime = \bigl(\frac{1}{4}\frac{\overline{1}}{4}\frac{1}{2}\bigr)$}. This is a \mbox{double-\textbf{q}, or $2\mathbf{q}_{1} - \textbf{q}_{1}^\prime$}, structure which exists in three degenerate domains owing to the cubic symmetry of the crystal. Figure~\ref{FigInosov:MagneticStructure}\,(a) shows the field dependence of peaks in one domain for a magnetic field applied along the $(1\overline{1}1)$ direction, with consecutive data taken in increasing field~\cite{EffantinBurlet82}. First, we note that at low fields the intensity of all peaks rises, which is a result of domain selection. As domain selection should not take place for fields applied exactly along one of the $\langle111\rangle$ directions, when domains exist in the $\langle001\rangle$ planes, this selection process was attributed to a slight misalignment which broke the degeneracy. This domain selection was also observed in further measurements within the same study, with a single domain selected for a field applied along a $(001)$ direction and two domains selected for a field along a $(110)$ direction. In Fig.~\ref{FigInosov:MagneticStructure}\,(a) this process continues until around 6~kOe, at which point domain selection is complete and the intensity flattens as a function of field. Following this, on the approach to 10~kOe, the $\bigl(\frac{1}{4}\frac{\overline{1}}{4}\frac{1}{2}\bigr)$ peak undergoes a second increase in spectral weight, followed by a sharp reduction of the $\bigl(\frac{1}{4}\frac{\overline{1}}{4}0\bigr)$ intensity and the extinction of the $\bigl(\frac{1}{4}\frac{1}{4}0\bigr)$ and $\bigl(\frac{1}{4}\frac{1}{4}\frac{1}{2}\bigr)$ peaks. This indicates a change to a second, single-\textbf{q} AFM structure at the transition to phase~III$^\prime$.\index{CeB$_6$!phase III$'$} The authors of this study proposed a specific ordering of the spins in the $(001)$ plane for these two phases, which are shown in the inset to the figure. The precise transition point is dependent on the field direction, although the structure of phase~III$^\prime$\index{CeB$_6$!phase III$'$} remains the same, and its emergence has also been observed in lanthanum-doped samples~\cite{KunimoriKotani11}.

Despite the abundance of neutron-scattering data on phase~III,\index{CeB$_6$!phase III} determining the precise magnetic structure of this multi-\textbf{q} phase has proven difficult. The original Effantin model, which is shown in the inset of Fig.~\ref{FigInosov:MagneticStructure}\,(a), has spins confined to a $[001]$ plane with moments aligned along the $[110]$ and $[1\overline{1}0]$ directions. This model was expanded in later publications~\cite{BurletRossat-Mignod82, Effantin85}, and a 3D representation is presented in panel (b). However, the diffraction peaks are consistent with multiple structures, and while this model is consistent with the neutron diffraction data, it was not able to explain later $\mu$SR measurements. The model predicts the presence of three muon precession frequencies, whereas eight were found~\cite{FeyerhermAmato94, FeyerhermAmato95}. A following neutron study by Zaharko~\textit{et~al.}, utilizing both powder neutron diffraction and neutron spherical polarimetry to investigate the magnetic structure, proposed several different models to explain the data~\cite{ZaharkoFischer03}. Using polarized neutrons allowed the orientation of magnetic moments to be determined, confirming for instance that the moments were perpendicular to the propagation vector for the $(\pm\!h\,\pm\!h~l)$ reflections. They found the original model to provide a poor fit to their updated data, and their favored model `D', which they find most consistent with both the neutron scattering and $\mu$SR data, is reproduced in Fig.~\ref{FigInosov:MagneticStructure}\,(c). This model is formed of two perpendicular sublattices, with moments oriented along $(110)$ directions and modulated along the $(001)$ direction as well as in the plane. They found moments of $\mu = 0.01\,\mu_{\mathrm{B}}$ and $0.136(7)\,\mu_{\mathrm{B}}$ for the $z = 0$ plane and $\mu = 0.744(16)\,\mu_{\mathrm{B}}$ and $0.543(16)\,\mu_{\mathrm{B}}$ for the $z = 1$ plane, and the complexity of this structure is considered due to competition between the dipolar, quadrupolar and octupolar interactions \cite{ZaharkoFischer03}.\index{multipolar interactions} However, it remains that the available data was not able to completely isolate any single model, and so while the model of Zaharko~\textit{et~al.} is best evidenced by the data, the exact structure still remains uncertain.
\index{CeB$_6$!magnetic structure|)}

\vspace{-2pt}\subsection{Magnetically hidden order in phase~II}\label{Ino_SubSec:CeB6AFQPhase}
\index{CeB$_6$!phase II|(}\index{CeB$_6$!antiferroquadrupolar order|(}\index{CeB$_6$!magnetically hidden order|(}\index{magnetically hidden order!in CeB$_6$|(}

One of the key initial problems presented by phase~II of CeB$_6$ was the absence of magnetic signal from neutron diffraction. Detailed analysis of specific-heat data, combined with information from magnetic susceptibility, had lead to the suggestion that this may be a short-range magnetic order phase~\cite{FujitaSuzuki80, TakaseKojima80}. However, an early neutron study by Horn~\textit{et~al.} found no magnetic Bragg reflections in phase~II in their elastic scattering data~\cite{HornSteglich81}. Interestingly, they also performed inelastic neutron scattering on this material, however the measurements were performed at 10~K as their aim was to search for crystal-field transitions, and so they did not encounter any of the phenomena which relate to the low-field phases which will be discussed in detail later in this chapter. In the following year, Burlet~\textit{et~al.} released a study which presented two seemingly conflicting results, finding the absence of magnetic Bragg peaks in one experiment while a second experiment showed the same Bragg reflections as in phase~III~\cite{BurletRossat-Mignod82}. The conclusion from this was that either the transition from phase~III to phase~II represented a modulated to fully ordered phase transition, or that perhaps phase~II was ferromagnetic. However, following this no further Bragg reflections were found in phase~II in zero field from further experiments, and thus it became known as a magnetically hidden order phase.

Despite the lack of magnetic signal in zero field, it was found that a dipolar order can be induced by the application of field with a wave vector of $\bigl(\frac{1}{2}\frac{1}{2}\frac{1}{2}\bigr)$. This was first reported by Effantin~\textit{et~al.}~\cite{Effantin85}, although the result was first mentioned in an earlier publication~\cite{KomatsubaraSato83}, and the distinct difference in ordering vectors clearly separate it from phase~III. This observation, alongside NMR measurements~\cite{TakigawaYasuoka83}, allowed Komatsubara~\textit{et~al.} to deduce that this phase corresponded to an ordering of the cerium quadrupolar moments~\cite{KomatsubaraSato83}, and the proposed ordering from Ref.~\cite{Effantin85} is reproduced in Fig.~\ref{FigInosov:MagneticStructure}\,(b). Neutrons cannot scatter from quadrupolar moments~\cite{Lovesey15}, which explains why the underlying order to phase~II was invisible to neutron diffraction. However, it is possible for the dipoles to be modulated in accordance with the AFQ ordering vector\index{CeB$_6$!antiferroquadrupolar order} as a result of the Zeeman splitting of the $\Gamma_{8}$ quartet,\index{$\Gamma_8$ quartet} the ground state multiplet of the \textit{f}-levels, in a magnetic field \textbf{B}. The Zeeman term is given by $\mathcal{H}_{\rm Z} = g \mu_{\mathrm{B}}\,\mathbf{J}\cdot\mathbf{B}$, where \textit{g} is the Land\'{e} factor and $\mathbf{J}$ is the total angular momentum. Writing the total angular momentum in terms of the orbital operator $\tau$ and the spin operator $\sigma$, the Zeeman term becomes \cite{Ohkawa85}
\begin{equation}
\hspace{-6pt}\mathcal{H}_{\mathrm{Z}} = 2\mu_{\mathrm{B}} \bigg[ \sigma_{\!x} \bigl(1\kern-1pt+\kern-1pt\frac{8}{7} T^{x}\bigr)B_x + \sigma_{\!y}\bigl(1\kern-1pt+\kern-1pt\frac{8}{7}T^{y}\bigr)B_y + \sigma_{\!z}\bigl(1\kern-1pt+\kern-1pt\frac{8}{7}T^{z}\bigr)B_{z}\bigg],
\end{equation}
where
\begin{equation}
T^{x} = \frac{\sqrt{3}}{2} \tau^{x} - \frac{1}{2}\tau^{z}, \; \; \; T^{y} = -\frac{\sqrt{3}}{2} \tau^{x} - \frac{1}{2}\tau^{z}, \; \; \; T^{z} = \tau^{z}.
\end{equation}
In the case, for instance, of $O_{xz}$-type quadrupolar order, an AFM dipole moment may be induced along \textit{z} for a field applied along the $(110)$ direction~\cite{ShiinaShiba97}. Calculations which supported an AFQ order\index{CeB$_6$!antiferroquadrupolar order} at this wave vector followed shortly after the initial observation of the field-induced dipolar moment~\cite{HanzawaKasuya84, Ohkawa85}, and this remained the general consensus for the order parameter of phase~II until it was directly confirmed by resonant x-ray scattering some years later~\cite{NakaoMagishi01, YakhouPlakhty01}.

\begin{figure}[t]
\centerline{\psfig{file=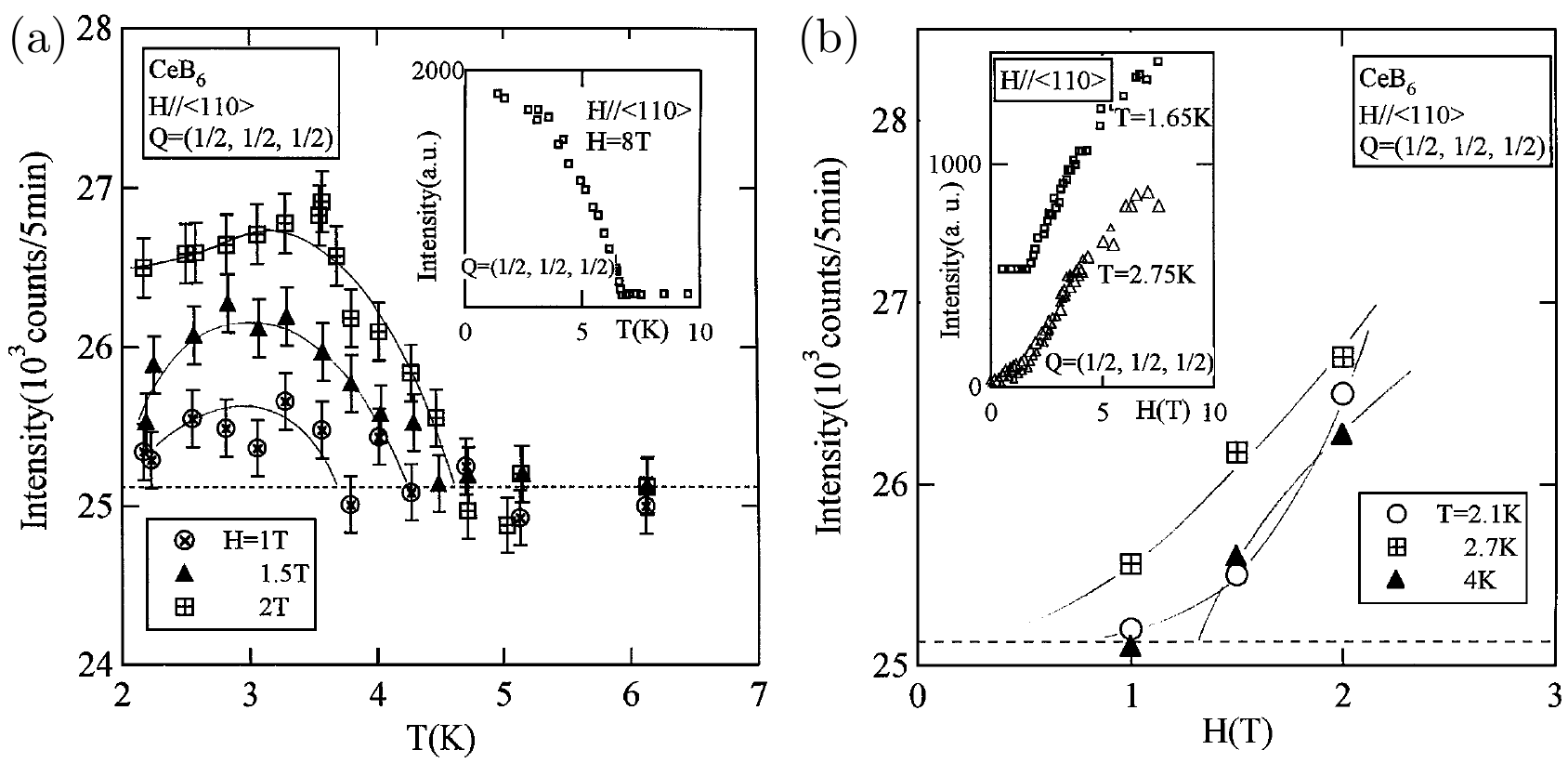, width=\textwidth}}
\caption{(a)~Temperature dependence of the Bragg intensity at the $\mathbf{Q} = \bigl(\frac{1}{2}\frac{1}{2}\frac{1}{2}\bigr)$ for \mbox{$\mathbf{B} \parallel \langle110\rangle$} at $B = 1$, 1.5 and 2~T, reproduced from Sera \textit{et al.}~\cite{SeraIchikawa01}. The inset shows the same temperature dependence at 8\,T, reproduced from an earlier study~\cite{Rossat-Mignod87}. (b)~Field dependence of the same $\mathbf{Q} = \bigl(\frac{1}{2}\frac{1}{2}\frac{1}{2}\bigr)$ peak intensity at $T = 2.1$, 2.7 and 4~K, reproduced from the same study~\cite{SeraIchikawa01}. The inset shows the field dependence of this peak at $T = 1.65$ and 2.75~K~\cite{Rossat-Mignod87}.\index{CeB$_6$!antiferroquadrupolar order}\index{CeB$_6$!Bragg intensity}}
\label{FigInosov:AFQ_neutron_scattering}
\end{figure}

Determination of the ordering vector was not the only contribution of neutron diffraction to the investigation of the magnetically hidden order phase, and studies of the field and temperature dependence of the induced dipole order revealed unusual behavior~\cite{Rossat-Mignod87, SeraIchikawa01}. We reproduce some of the results in Fig.~\ref{FigInosov:AFQ_neutron_scattering}, where in panel (a) we see a nonmonotonic dependence of the $\mathbf{Q} = \bigl(\frac{1}{2}\frac{1}{2}\frac{1}{2}\bigr)$ Bragg intensity as a function of temperature in phase~II, in contrast to the order-parameter-like behavior that one might expect. The intensity rapidly rises after passing below $T_{\rm Q}$, however it possesses a broad maximum around 3~K, and then most unusually appears to decrease with decreasing temperature. However, this behavior is clearly dependent on field, as only the low-field results exhibit the low-temperature decrease in intensity, whereas the data at 8\,T show a much more order-parameter-like dependence. The field dependence of these peaks, shown panel (b), also exhibits unusual behavior and changes between a convex and concave curvature as the temperature is altered. These observations could be explained by considering different models of multipolar order in a mean-field approach~\cite{SeraKobayashi99}. Specifically, the $O_{xy}$-type AFQ ordered state\index{antiferroquadrupolar order} including both AFM exchange and antiferrooctupolar (AFO) interactions reproduced this behavior, which was the first evidence that AFO interactions were playing a role under applied field in \textit{f}-electron compounds.

\begin{figure}[t]
\centerline{\psfig{file=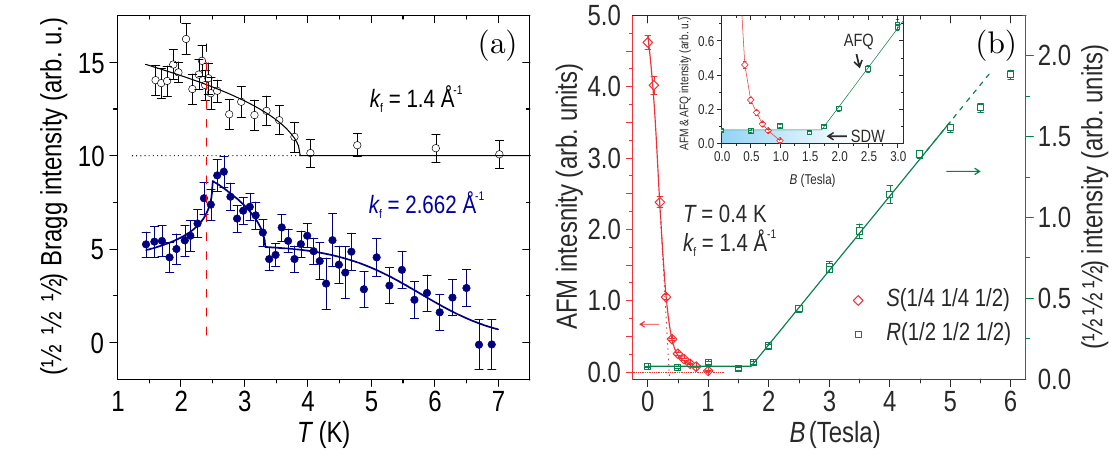, width=\textwidth}}
\caption{(a)~Temperature dependence of the $\bigl(\frac{1}{2}\frac{1}{2}\frac{1}{2}\bigr)$ Bragg peak taken with two different scattered-neutron wave vectors, $k_{\rm f} = 1.4$ and 2.662~\AA$^{-1}$. Reproduced from Ref.~\cite{FriemelLi12}. (b)~Magnetic-field dependence of the AFM and AFQ Bragg intensities at $T = 0.4$~K. The low-field region is enlarged in the inset. Reproduced from Ref.~\cite{Friemel14}.\index{CeB$_6$!antiferroquadrupolar order}}
\label{FigInosov:SDW_neutron_scattering}
\end{figure}

While initial measurements indicated no elastic magnetic scattering in the AFQ phase in zero field,\index{CeB$_6$!antiferroquadrupolar order} later neutron spin-flip scattering measurements found a signal at the $R\bigl(\frac{1}{2}\frac{1}{2}\frac{1}{2}\bigr)$ point~\cite{PlakhtyRegnault05}. This appeared already at 7~K, within the paramagnetic phase, and increased with decreasing temperature until it peaked at $T_{\rm AFM}$ before being suppressed within the AFM phase. This was surprising, as Bragg scattering would not originate from either the paramagnetic state nor from ordered quadrupoles. More recent elastic-scattering measurements, presented in Fig.~\ref{FigInosov:SDW_neutron_scattering}\,(a), shed light on the origin of this signal~\cite{FriemelLi12, Friemel14}. The two datasets were taken using two different triple-axis spectrometers,\index{triple-axis spectrometer} one from the cold-neutron spectrometer IN14 with the neutron wave vector of $k_{\rm f} = 1.4$~\AA$^{-1}$, and another from the thermal-neutron spectrometer IN3 with $k_{\rm f} = 2.662$~\AA$^{-1}$. The cold-neutron spectrometer has a higher resolution, and therefore shows a signal appearing just below 4~K and rising continuously with decreasing temperature. The thermal-neutron spectrometer, which has a lower resolution, records a signal appearing at 7~K, which rises smoothly until entering the AFQ phase,\index{CeB$_6$!antiferroquadrupolar order} whereupon it rises sharply to a maximum at $T_{\rm N}$ before being moderately suppressed in the AFM phase. Such differing behaviors under identical conditions from instruments with differing resolutions indicates that there is more than one contribution to the signal. This would suggest that the instrument with the higher resolution is probing a true Bragg peak, indicating that there is some sort of measurable long-range order at the \textit{R} point which appears below $T_{\rm Q}$ in zero field, while the lower-resolution instrument is additionally detecting a quasielastic signal,\index{diffuse neutron scattering}\index{quasielastic magnetic scattering}\index{neutron scattering!quasielastic} which will be discussed in a later section. The appearance of Bragg scattering at this wave vector and temperature is, however, intriguing. While this may coincide with the quadrupolar ordering vector, we know that neutrons cannot probe this order parameter, and its temperature dependence indicates that it is unrelated to the magnetic order of phase~III. However, its width in $\mathbf{q}$-space is similar to that of the AFM Bragg peaks~\cite{FriemelLi12}, so this does suggest that it represents some sort of magnetic long-range order. Figure~\ref{FigInosov:SDW_neutron_scattering}\,(b) presents the field dependence of both this signal and the AFM Bragg peak.

\index{CeB$_6$!$R$-point resonant mode}\index{CeB$_6$!magnetic resonant mode}\index{neutron resonant mode}
We see that the $R$-point signal is both weak and flat within the AFM phase, before rising continuously with field within the AFQ state,\index{CeB$_6$!antiferroquadrupolar order} as the induced dipole order also possesses this wave vector. The localized models of multipolar order cannot explain the flat intensity with respect to field within phase~III, and this suggests a different origin. It has been suggested that this signal is from a spin density wave (SDW)~\cite{FriemelLi12}, which correlates to the suggestions of Sluchanko \textit{et~al.} from transport and magnetization measurements that a SDW may exist below $T_{\rm Q}$~\cite{SluchankoBogach07}, and its suppression at $T_{\rm Q}$ does suggest that it is somehow related to the AFQ state.\index{CeB$_6$!antiferroquadrupolar order} On the other hand, from the field independence of the intensity below 1.7\,T we know that this order parameter cannot be associated with the $O_{xz}$-type quadrupolar order itself, as it is known to result in field-induced dipolar and octupolar ordered moments\index{antiferrooctupolar order!field-induced}\index{magnetic octupoles} that would lead to a linear increase of intensity starting from zero field. Judging from this result, the AFQ order parameter does not coexist with the AFM order at base temperature and sets in as a second-order phase transition only after the AFM phase is suppressed in magnetic field. However, at elevated temperatures the transition is broadened, as can be seen in the inset to Fig.~\ref{FigInosov:AFQ_neutron_scattering}\,(b), and a gradual increase of the $R$-point Bragg intensity can be seen starting already from zero magnetic field. This can be explained by the slow fluctuations of both phases coexisting in the vicinity of the critical point.
\index{CeB$_6$!phase II|)}\index{CeB$_6$!antiferroquadrupolar order|)}\index{CeB$_6$!magnetically hidden order|)}\index{magnetically hidden order!in CeB$_6$|)}

\subsection{Mean-field description of the ordering phenomena in CeB$_\text{6}$}\label{Ino_SubSec:MeanFieldDescription}
\index{multipole interaction model}

In order to understand the relevance and significance of more recent inelastic neutron scattering results, it will be necessary to briefly review some of the earlier results of theoretical descriptions of cerium hexaboride. The available orbital degrees of freedom to compounds with 3\textit{d} or 4\textit{f} ions allows for the expression of multipolar order, and constructing a model of the interacting multipoles in CeB$_6$ has been far from simple. The unusual behavior of the AFQ phase,\index{CeB$_6$!antiferroquadrupolar order} namely its stabilization upon the application of magnetic field, did not possess any obvious explanation, and several scenarios were proposed. Initially, Effantin \textit{et~al.} suggested that the magnetic behavior of CeB$_6$ was governed by the competition between single-site Kondo fluctuations and the interaction between the dipoles and quadrupoles~\cite{EffantinRossat-Mignod85}. In this scenario, a dense Kondo state suppresses the interactions which create the AFQ order,\index{CeB$_6$!antiferroquadrupolar order} and the application of field suppresses the Kondo state leading to an enhancement in $T_{\rm Q}$. One of the earlier models constructed to describe CeB$_6$ was a study by Hanzawa and Kasuya \cite{HanzawaKasuya84}, which assumed a ground-state $\Gamma_7$ doublet\index{$\Gamma_7$ doublet} and an excited $\Gamma_8$ quartet,\index{$\Gamma_8$ quartet} where the dominant interaction was a quadrupole--quadrupole coupling between neighboring cerium ions. This description was able to reproduce the behavior of $T_{\rm Q}$, as well as explain other experimental observations. However it could not reproduce, for instance, the $^{11}$B line spitting seen in NMR measurements, and they ascribed this to an over-simplification of their level scheme. However, it was later determined that the level scheme was incorrect, and the $\Gamma_8$ quartet\index{$\Gamma_8$ quartet} was in fact the ground state~\cite{SatoKunii84, ZirngieblHillebrands84, LoewenhauptCarpenter85}, with a 545~K splitting separating it from the excited states.

\begin{figure}[b]
\centerline{\psfig{file=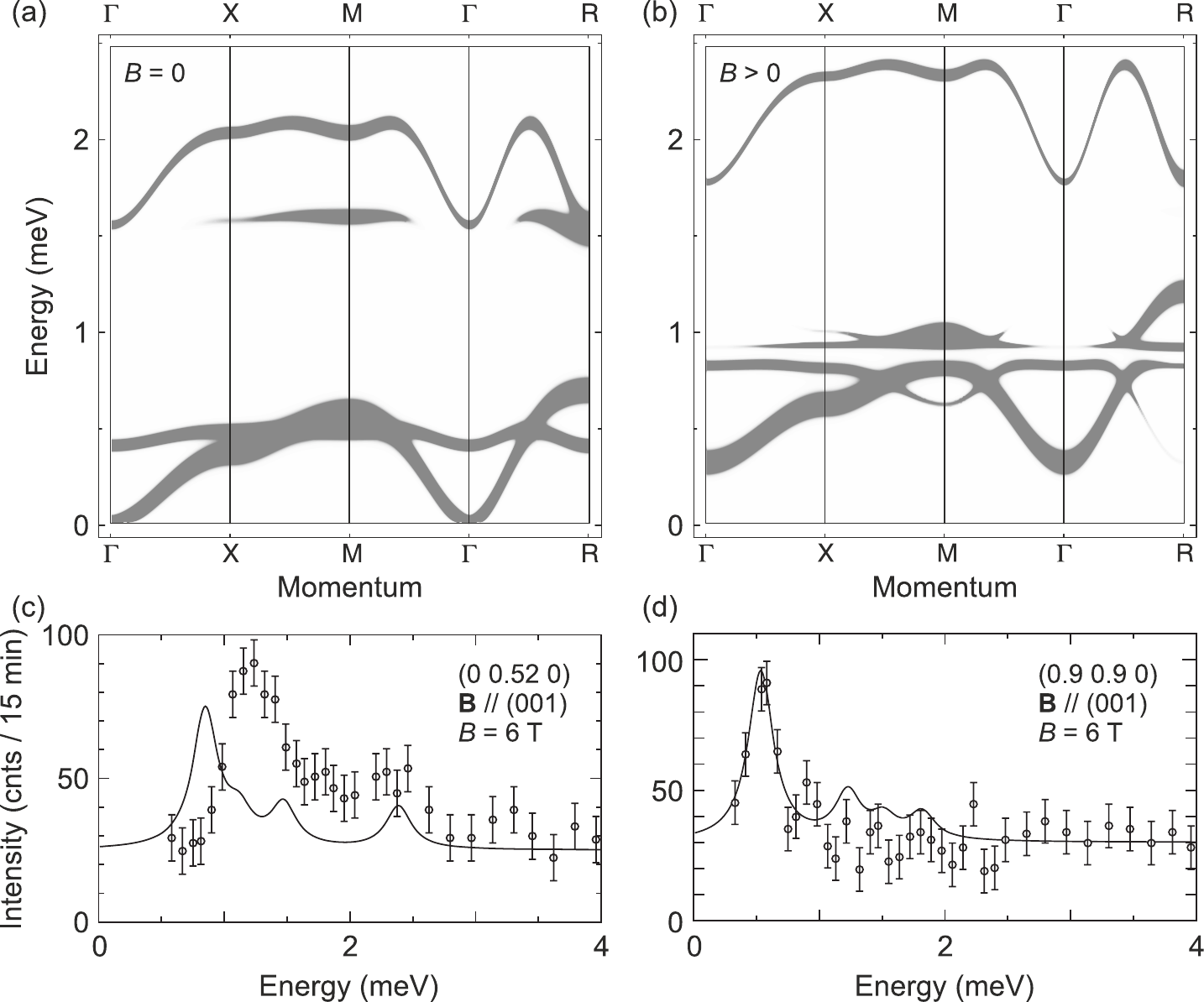, width=\textwidth}}
\caption{(a)~Dipolar excitation spectrum, $S(\mathbf q, \omega)$, calculated for the AFQ phase in zero field by Thalmeier~\textit{et~al.} (b)~Same calculation as in (a), but for an applied field along the $\langle 001 \rangle$ direction. Reproduced from Ref.~\cite{ThalmeierShiina03}. (c)~Comparison of the calculation (solid line) and measured (data points) spectra for the $(0~0.52~0)$ wave vector, which is near the \textit{X} point, in an applied field of 6\,T along the $[001]$ direction. (d)~Same calculation and measurements as for (b), but taken at the $(0.9~0.9~0)$ wave vector, which is near the $\Gamma$ point. Reproduced from Ref.~\cite{ShiinaShiba03}. \copyright\ Physical Society of Japan.}
\label{FigInosov:Mean_field_calculations}
\end{figure}

Alternative explanations to the quadrupole-quadrupole interaction were also proposed. Ohkawa \textit{et~al.} suggested a model consisting of multipoles in the $\Gamma_8$ ground state\index{$\Gamma_8$ quartet} interacting through the RKKY interaction~\cite{Ohkawa83, Ohkawa85}.\index{RKKY interaction} Here, higher-order terms resulting from spin and orbital degeneracies caused the enhancement of $T_{\rm Q}$ at low fields. Alternatively, a study by Uimin \textit{et~al.} showed that large fluctuations of the quadrupole moment at low fields would also lead to a suppression of the AFQ phase~\cite{UiminKuramoto96}.\index{CeB$_6$!antiferroquadrupolar order} Following this, Shiina \textit{et~al.} published a model restricting themselves to a $\Gamma_8$ ground state,\index{$\Gamma_8$ quartet} in which there are 15 allowed multipolar moments, five of which are quadrupolar~\cite{ShiinaShiba97, ShiinaSakai98}. From this set, three are of the $\Gamma_5$ type and two are of the $\Gamma_3$-type, and it was predicted that in zero field that only the $\Gamma_5$ ordering would be realized, resulting in $O_{xy}$, $O_{yz}$ and $O_{zx}$-type order. This model is elaborated in detail in Chapter~8 [\href{https://arxiv.org/abs/1907.10967}{arXiv:1907.10967}]. Further developments of this approach by Saki \textit{et~al.} were able to resolve the long-standing dispute between neutron scattering and NMR results~\cite{SakaiShiina97}. Including the octupolar interaction\index{multipolar interactions} in the scenario where the $O_{xy}$ quadrupoles are selected by a field applied along the $\langle001\rangle$ direction, they were able to describe the results of both techniques without the triple-\textbf{q} structure which was proposed from NMR~\cite{KawakamiKunii81, TakigawaYasuoka83}.

These mean-field models, which have been able to describe much of the behavior of CeB$_6$, are also able to predict the formation of collective multipolar modes associated with the quadrupolar ordering. Figure~\ref{FigInosov:Mean_field_calculations} shows the results from two separate studies of the excitation spectrum. Panels (a) and (b) present predictions of the excitation spectrum along a path through several high-symmetry points in the Brillouin zone, from a study by Thalmeier \textit{et~al.}~\cite{ThalmeierShiina03}, who calculated the dipolar scattering function using both the random-phase approximation and the Holstein Primakoff approach. They find a Goldstone mode emanating from the $\Gamma$ point\index{CeB$_6$!ferromagnetic resonance}\index{CeB$_6$!zone-center excitations} which forms a dispersion branch that extends across the Brillouin zone~\cite{ThalmeierShiina98, ThalmeierShiina03, ThalmeierShiina04}. All the low-energy modes increase linearly in energy under the application of magnetic field, which can be seen by comparison of the zero-field results of panel (a) with the in-field calculation of panel (b). A comparison of calculations by Shiina \textit{et~al.}~\cite{ShiinaShiba03} to INS measurements in an applied field of 6\,T applied along the $[001]$ direction are shown in Figs.~\ref{FigInosov:Mean_field_calculations}\,(c,\,d). The calculation struggles to reproduce the data near the \textit{X} point in panel (c), but gives a reasonable approximation of the data near the zone center in panel (d). It was determined that the calculations were able to describe the leading mode of the spectrum in field, however zero-field experiments found a featureless quasielastic response,\index{diffuse neutron scattering}\index{quasielastic magnetic scattering}\index{neutron scattering!quasielastic} which disagreed with the theory. This situation remained until more recent neutron scattering measurements, which will be detailed in the next section, improved the coverage of the neutron data in terms of both exploration of momentum space and the phase diagram.
\index{CeB$_6$!electronic properties|)}\index{CeB$_6$!ordering phenomena|)}

\section{Spin excitations in the absence of magnetic field}
\index{CeB$_6$!magnetic excitations|(}\index{magnetic excitations!in CeB$_6$|(}\index{inelastic neutron scattering|(}

\subsection{Collective excitations in the antiferromagnetic phase}\label{Ino_SubSec:CollectiveExcitations}
\index{CeB$_6$!antiferromagnetic order}\index{antiferromagnetic order!in CeB$_6$}

\begin{figure}[t]
\centerline{\psfig{file=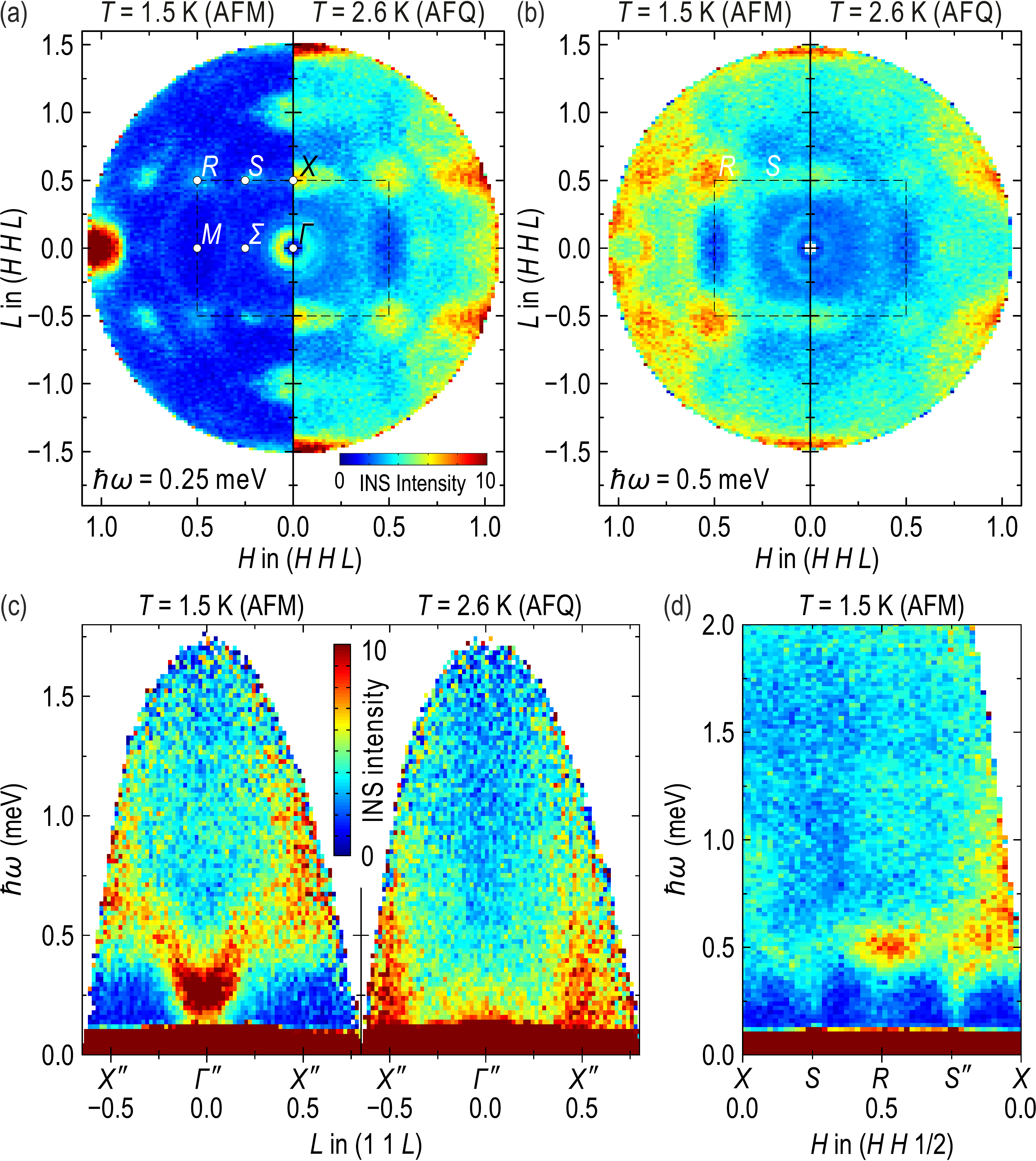, width=\textwidth}}
\caption{A summary of INS data from TOF measurements on CeB$_6$. (a,\,b)~Constant-energy cuts within the $(HHL)$ plane of momentum space, at energy transfers of $\hbar\omega = 0.25$ and 0.5~meV, respectively, with an integration range of $\pm 0.1$~meV. The left and right sides of each figure were measured in the AFM (1.5~K) and AFQ (2.6~K) phases, respectively. (c)~Energy-momentum cuts along the $(11L)$ direction in $\mathbf{Q}$ space in the AFM (left) and AFQ (right) phases. (d)~Energy-momentum cut along the $(HH\frac{1}{2})$ direction in $\mathbf{Q}$ space in the AFM phase. Adapted from Jang \textit{et al.}~\cite{JangFriemel14}.\index{multipolar excitations!dispersion}}
\label{FigInosov:CeB6INS}
\end{figure}

Investigation of the spin dynamics of CeB$_6$ has provided a wealth of information, unearthing multiple unexpected phenomena and challenging the prevailing theories. However, it was not until recent developments in inelastic neutron scattering (INS) that the full picture became clear, as many of the features of interest appeared in unexpected locations in momentum space. Figure~\ref{FigInosov:CeB6INS} presents a summary of the INS data from Jang~\textit{et~al.}~\cite{JangFriemel14}, who probed the full Brillouin zone of CeB$_6$ using time-of-flight (TOF) neutron scattering.\index{time-of-flight spectrometer} Panels (a) and (b) present constant-energy cuts in the $(HHL)$ plane, for an energy transfer of $\hbar\omega = 0.25$~meV and 0.5~meV respectively. Each panel shows data from the AFM ($T = 1.5$~K) and AFQ ($T = 2.6$~K) phases.\index{CeB$_6$!antiferroquadrupolar order} Panels (c) and (d) show energy-momentum cuts along the $(11L)$ and $(HH\frac{1}{2})$ paths in the reciprocal space, respectively, with the data taken in the AFM and AFQ phases.

\begin{figure}[t]
\centerline{\psfig{file=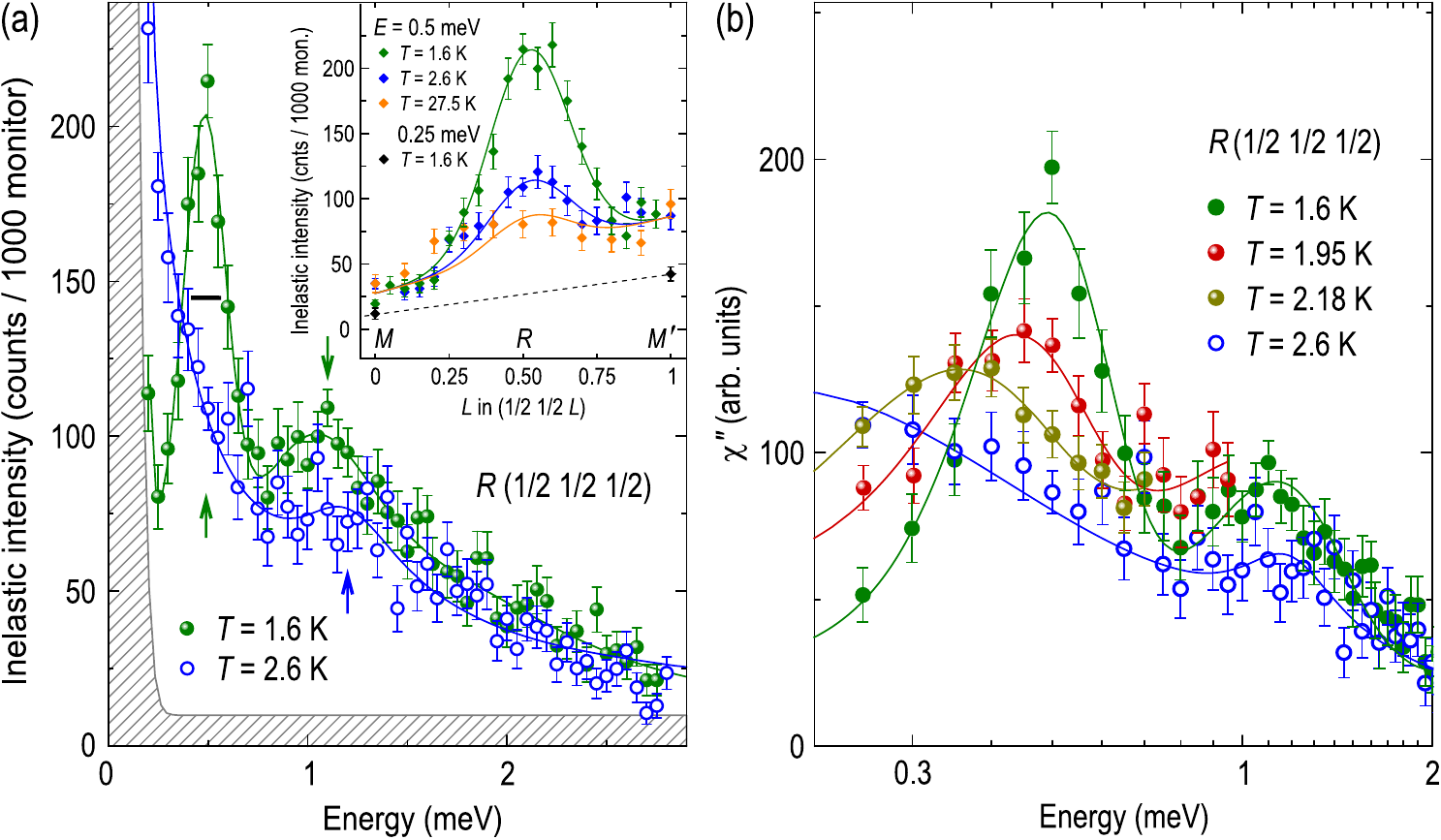, width=\textwidth}}
\caption{(a)~Inelastic neutron scattering spectrum of CeB$_6$ at $R\bigl(\frac{1}{2}\frac{1}{2}\frac{1}{2}\bigr)$ in the AFM state ($T = 1.5$~K) and AFQ state ($T = 2.5$~K). The horizontal black bar inside the peak shows the experimental energy resolution, the hatched region represents background intensity, and the lines are guides to the eyes. The inset shows constant-energy scans across the peak maximum ($\hbar\omega=0.5$~meV) along the $MRM'$ line at different temperatures; the dashed line is the background level suggested by the intensity minima at $M\bigl(\frac{1}{2}\frac{1}{2}0\bigr)$ and $M'\bigl(\frac{1}{2}\frac{1}{2}1\bigr)$ within the spin gap ($\hbar\omega = 0.25$~meV).\index{CeB$_6$!spin gap}\index{spin gap!in CeB$_6$} (b)~Temperature dependence of the imaginary part of the dynamic spin susceptibility,\index{dynamic spin susceptibility} $\chi''(E)$, at the $R$ point from within the AFM state to the AFQ state, illustrating the gradual development of the spin-exciton\index{exciton modes!in CeB$_6$}\index{CeB$_6$!exciton modes} peak below the AFM transition. Reproduced from Friemel \textit{et al.}~\cite{FriemelLi12}.\index{CeB$_6$!$R$-point resonant mode}\index{CeB$_6$!magnetic resonant mode}}
\label{FigInosov:CeB6RpointResonance}
\end{figure}

Within the AFM phase, we can immediately see that the inelastic signal is not dominated by magnon modes emanating from the magnetic ordering vectors, which is indicative of the complex set of competing interactions at low temperature which leads to the correspondingly complex low-temperature phase diagram. One of the strongest features is the signal at the $R$ point [Figs.~\ref{FigInosov:CeB6INS}\,(b,\,d)],\index{CeB$_6$!$R$-point resonant mode}\index{CeB$_6$!magnetic resonant mode} which is especially pronounced in panel (b) at an energy transfer of $\hbar\omega = 0.5$~meV, and comes from a resonant exciton mode\index{exciton modes!in CeB$_6$}\index{CeB$_6$!exciton modes} which was initially identified by Friemel~\textit{et~al.}~\cite{FriemelLi12}, as shown in Fig.~\ref{FigInosov:CeB6RpointResonance}. This sharp peak appears with an order-parameter-like behavior below $T_{\rm N}$, alongside the appearance of a spin gap below 0.35~meV,\index{CeB$_6$!spin gap}\index{spin gap!in CeB$_6$} and it is similar in nature to the resonant modes\index{neutron resonant mode} observed in some high-$T_{\rm c}$ and heavy-fermion superconductors~\cite{FongKeimer95, InosovPark10, StockBroholm08, StockertArndt11}. These are considered to be spin excitons\index{spin exciton}\index{exciton modes!in CeB$_6$}\index{CeB$_6$!exciton modes} that form below the onset of a particle-hole continuum,\index{particle-hole continuum} owing to a divergence in the dynamical spin susceptibility~\cite{LiuZha95, AbanovChubukov99, EreminMorr05, ChubukovGorkov08}.\index{dynamic spin susceptibility} Point-contact spectroscopy measurements found a charge gap in CeB$_6$\index{charge gap!in CeB$_6$} within the AFM phase~\cite{PaulusVoss85}, with a size of \mbox{$2\Delta_{\rm AFM} \approx 1.2$}~meV, which suggests that the \textit{R}-point mode lies within this gap. However, while the exciton modes\index{exciton modes!in CeB$_6$}\index{CeB$_6$!exciton modes} observed in the unconventional superconductors tend to be localized to the magnetic propagation vector of their parent compounds, the one observed in CeB$_6$ appears at multiple wave vectors within the Brillouin zone, and the strong feature at the \textit{R}-point is just a local maximum in intensity. Furthermore, this is not the only hexaboride to possess a resonant exciton,\index{SmB$_6$!exciton modes}\index{exciton modes!in SmB$_6$} as the topological Kondo insulator SmB$_6$,\index{SmB$_6$} discussed in Chapter~6, has been demonstrated to exhibit similar excitons at the $X$\index{SmB$_6$!$X$-point resonance} and $R$\index{SmB$_6$!$R$-point resonance} points at approximately 14~meV~\cite{AlekseevMignot95, FuhrmanLeiner15}.

\index{CeB$_6$!$R$-point resonant mode}\index{CeB$_6$!magnetic resonant mode}
Akbari and Thalmeier~\cite{AkbariThalmeier12} proposed an elegant theoretical model for the formation of the $R$-point exciton in CeB$_6$\index{exciton modes!in CeB$_6$}\index{CeB$_6$!exciton modes} in the itinerant heavy-quasiparticle picture, where AFQ and AFM order parameters are treated as particle-hole condensates,\index{particle-hole condensate} and the resonant mode from the INS measurements~\cite{FriemelLi12} is interpreted as a feedback spin exciton,\index{feedback spin exciton}\index{spin exciton} where the feedback effect results in a change in magnetic spectral properties across a respective transition due to the appearance of an order parameter. This model is described in detail in Chapter~8 [\href{https://arxiv.org/abs/1907.10967}{arXiv:1907.10967}]. In the formalism of the random phase approximation (RPA), the dynamic spin susceptibility\index{dynamic spin susceptibility} for the interacting quasiparticles can be written as
\begin{equation}
\chi_\text{RPA}(\mathbf{q},\omega)=\chi_0(\mathbf{q},\omega)\big/\bigl[1-J_\mathbf{q}\,\chi_0(\mathbf{q},\omega)\bigr],
\end{equation}
where $\chi_0(\mathbf{q},\omega)$ is the Lindhard function\index{Lindhard function|(} of the heavy conduction electrons, and the interaction parameter $J_\mathbf{q}$ is assumed to have a Lorentzian form centred at the exciton wavevector $\mathbf{q}_0$. The charge gaps\index{charge gap!in CeB$_6$} that appear in the AFQ and AFM phases have the effect of pushing the quasiparticle response to higher energies, and if the condition $J_{\kern-.5pt\mathbf{q}_0\kern.5pt}\chi_0(\mathbf{q}_0,\omega)=1$ is met, the RPA susceptibility manifests a pole at $\mathbf{q}=\mathbf{q}_0$. It is suggested that this condition is satisfied in the AFM state of CeB$_6$, which explains the sharp excitation seen at the $R$ point in neutron spectroscopy (Fig.~\ref{FigInosov:CeB6RpointResonance}).

The exciton mode\index{exciton modes!in CeB$_6$}\index{CeB$_6$!exciton modes} was not the only unpredicted feature of the INS spectrum in the AFM phase. We see from the data in Fig.~\ref{FigInosov:CeB6INS}\,(c) that there are intense dispersive modes at low energy which emanate from the $(110)$ and $(001)$ zone-center points. The temperature dependence of this mode observed by Jang~\textit{et~al.}~\cite{JangFriemel14} indicated that it was magnetic in origin, and the dispersion followed the parabola expected for ferromagnons. Surprisingly, these ferromagnetic fluctuations\index{ferromagnetic fluctuations}\index{CeB$_6$!ferromagnetic resonance} are much more intense than the spin waves\index{CeB$_6$!spin waves}\index{spin waves!in CeB$_6$} associated with the AFM order, which suggests that CeB$_6$ is very close to a ferromagnetic instability. Interestingly, it has long been considered that ferromagnetic fluctuations\index{ferromagnetic fluctuations} are necessary for the observation of an electron spin resonance (ESR)\index{electron spin resonance} in Kondo lattice systems~\cite{SchaufusKataev09},\index{Kondo lattice} and CeB$_6$ has been known to produce a sharp ESR signal~\cite{DemishevSemeno05, DemishevSemeno09}.\index{CeB$_6$!electron spin resonance}\index{electron spin resonance!in CeB$_6$} However, the possibility of a ferromagnetic instability and the corresponding fluctuations had largely been ignored in the case of CeB$_6$ until the recent INS results, because due to the ordered AFM and AFQ phases, all the relevant interactions in the system were assumed to be antiferromagnetic. This assumption represents a dramatic oversimplification of the real situation, as more explicit calculations of the RKKY interaction parameters\index{RKKY interaction} between dipoles and various multipoles in CeB$_6$ based on the band-structure theory~\cite{YamadaHanzawa19, HanzawaYamada19} indicate that the coupling constants have an oscillatory character in direct space, and up to 8 effective interactions of different sign would have to be considered up to the 3rd-nearest neighbor in a realistic description of the magnetic Hamiltonian. Some of these interactions are frustrated,\index{frustration} and therefore a competition among different ground states is entirely plausible.

The conventional magnon (spin-wave)\index{CeB$_6$!spin waves}\index{spin waves!in CeB$_6$} excitations from the AFM order are among the weaker features of the INS spectrum~\cite{JangFriemel14}. In Fig.~\ref{FigInosov:CeB6INS}\,(d), they form a cone-shaped dispersion emanating from the ordering vectors $S(\frac{1}{4}\frac{1}{4}\frac{1}{2})$ and $S''(\frac{3}{4}\frac{3}{4}\frac{1}{2})$ of phase~III,\index{CeB$_6$!phase III} with a small spin gap of 0.3--0.4~meV,\index{CeB$_6$!spin gap}\index{spin gap!in CeB$_6$} as one would expect for an antiferromagnet with a small spin anisotropy. Peaked at a fairly low energy of around 0.7~meV at the zone boundary, this dispersion results in a narrow magnon bandwidth only twice larger than the spin gap.\index{CeB$_6$!spin gap}\index{spin gap!in CeB$_6$} The spin waves\index{CeB$_6$!spin waves}\index{spin waves!in CeB$_6$} hybridize with the excitations from the $R$-point\index{CeB$_6$!$R$-point resonant mode}\index{CeB$_6$!magnetic resonant mode} and ferromagnetic modes to form a continuous magnon band in the energy range from 0.2 to 0.7~meV.

The emergence of so many features in the inelastic scattering data of the AFM phase is an indication of the complexity of CeB$_6$, the nontrivial nature of the interactions which govern its low-temperature behavior, and the competing order parameters which arise as a result. Indeed, the emergence of the $R$-point exciton\index{exciton modes!in CeB$_6$}\index{CeB$_6$!exciton modes}\index{CeB$_6$!magnetic resonant mode} and the ferromagnetic mode at the $\Gamma$ point\index{CeB$_6$!ferromagnetic resonance}\index{CeB$_6$!zone-center excitations} in the absence of magnetic field within in phase~III\index{CeB$_6$!phase III} could not be explained by the prevailing theories of interacting multipoles which had been developed to describe systems such as CeB$_6$, leading to a development of these models beyond the localized approach.

\subsection{Quasielastic magnetic scattering}\label{Ino_SubSec:QuasielasticMagneticScattering}
\index{quasielastic magnetic scattering|(}

The early theoretical descriptions in the framework of mean-field models were used to predict the formation of collective multipolar excitations\index{CeB$_6$!multipolar excitations}\index{multipolar excitations!in CeB$_6$} associated with the quadrupolar ordering and their corresponding dipolar dynamic structure function. Specifically, calculations by Thalmeier~\textit{et~al.} predicted a Goldstone mode at the $\Gamma$ point within phase~II,\index{CeB$_6$!phase II}\index{CeB$_6$!ferromagnetic resonance}\index{CeB$_6$!zone-center excitations} which would form a dispersion branch that stretches across the entire Brillouin zone within the AFQ phase~\cite{ThalmeierShiina98, ThalmeierShiina03, ShiinaShiba03, ThalmeierShiina04}. While the predicted excitations under an applied magnetic field found reasonable agreement with the early INS data~\cite{Bouvet93, RegnaultErkelens88}, the predicted zero-field behavior turned out to be in disagreement with the corresponding measurements, which revealed a featureless quasielastic response~\cite{Bouvet93}.\index{diffuse neutron scattering}\index{quasielastic magnetic scattering}\index{neutron scattering!quasielastic} However, these data were limited in their coverage of momentum space, while the more recent measurements \cite{FriemelLi12, JangFriemel14}, partially represented in Figs.~\ref{FigInosov:CeB6INS} and \ref{FigInosov:CeB6RpointResonance}, encompass the entire Brillouin zone. We can see here that while dispersing modes are indeed absent from the AFQ phase, the quasielastic response both below and above $T_{\rm Q}$ is far from featureless, with peaks in intensity at the $\Gamma$, $X$, and $R$ points,\index{CeB$_6$!$R$-point resonant mode}\index{CeB$_6$!magnetic resonant mode} which also correspond to minima in the spin relaxation rate, determined from the quasielastic line width~\cite{JangFriemel14}. The fact that these features correspond with the dominant, intense excitations of the AFM phase indicates that the strongly overdamped quasielastic signal of the AFQ state already contains important information about the ordering vectors and excitations which form at lower temperature. These features in the quasielastic scattering also appear not just at the AFM ordering vectors, but also at several other high-symmetry points of the Brillouin zone ($\Gamma$, $R$, $X$). This suggests that the intense modes seen in the AFM phase are indirectly associated with itinerant quasiparticles. Within the self-consistent renormalization theory for heavy-fermion systems~\cite{MoriyaTakimoto95}, a modulated relaxation rate and dynamic susceptibility\index{dynamic spin susceptibility} are expected from the emergence of the local dynamic susceptibility and intersite interactions in the terms which describe these quantities. The hot spots of intensity in the neutron-scattering data would therefore correspond to maxima in the intersite interaction, which is governed by the band structure according to the RKKY mechanism.\index{RKKY interaction|(} This suggests that the scattering function would be dominated by nesting vectors of the Fermi surface,\index{nesting}\index{Fermi surface!nesting properties} in a similar manner to CeCu$_2$Si$_2$~\cite{StockertFaulhaber04} and potentially URu$_2$Si$_2$~\cite{WiebeJanik07}. This conclusion is supported by calculations of the intersite interaction in rare-earth hexaborides, which found maxima at the $R$ point, near the $X$ point, and around the AFM zone center $\mathbf{q}^\prime_1 = \bigl(\frac{1}{4}\frac{1}{4}\frac{1}{2}\bigr)$.

\begin{figure}[b!]
\centerline{\psfig{file=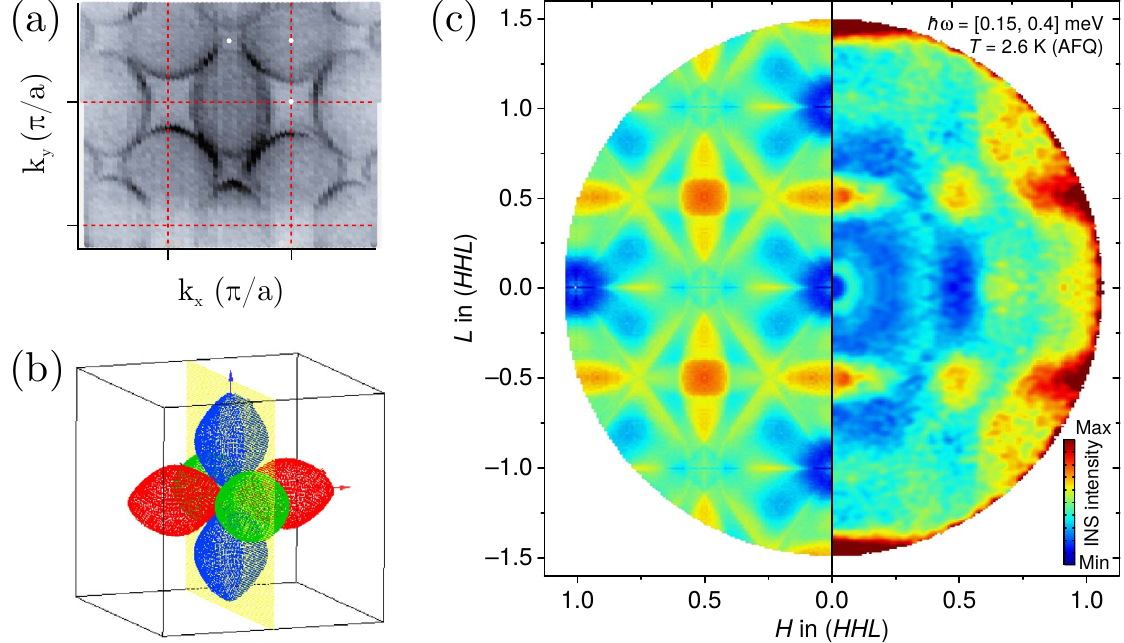, width=\textwidth}}
\caption{(a)~Fermi surface map in the $(100)$ plane,\index{Fermi surface!of CeB$_6$}\index{CeB$_6$!Fermi surface} taken with a photon energy of $h\nu = 700$~eV. (b) A 3D representation of the measured Fermi surface, with the $(100)$ plane indicated. (c)~A two-dimensional representation of the Lindhard function\index{Lindhard function}\index{Lindhard function!of CeB$_6$} for the $(HHL)$ plane (left), compared with the quasielastic magnetic scattering\index{diffuse neutron scattering}\index{quasielastic magnetic scattering}\index{neutron scattering!quasielastic} in the same plane, as measured by INS in the AFQ phase. Adapted from Koitzsch \textit{et al.}~\cite{KoitzschHeming16}.\vspace{-2pt}}
\label{FigInosov:CeB6PhotoTOM}
\end{figure}

To investigate this further, Koitzsch~\textit{et~al.} performed photoemission tomography, i.e. soft x-ray angle-resolved photoemission (ARPES)\index{photoelectron spectroscopy|(}\index{angle-resolved photoemission|(} measurements on several low-index cleavage planes of the crystal~\cite{KoitzschHeming16}, to determine the full 3D Fermi surface of CeB$_6$ with high accuracy. The aim of this study was to test the hypothesis that the mediation of interactions between magnetic moments through the conduction electrons could lead to magnetic order through the propensity of the electronic structure to nesting instabilities of the Fermi surface.\index{nesting}\index{Fermi surface!nesting properties}\index{Fermi surface!instabilities} The electronic structure of Ce$_{1-x}$La$_x$B$_6$ had been previously studied by both de Haas\,--\,van Alphen (dHvA)\index{de Haas\,--\,van Alphen} measurements~\cite{OnukiUmezawa89, JossRuitenbeek87, DeursenPols85, EndoNakamura06, GoodrichHarrison99, IshizawaTanaka77, HarrisonMeeson93, HarrisonHall98, TekluGoodrich00}, and ARPES~\cite{SoumaIida04, NeupaneAlidoust15}. However, these were not detailed enough to provide the required accuracy for the determination of the nesting vectors through Lindhard-function calculations. The improved bulk sensitivity from the use of soft x-ray ARPES, together with the innovative approach of Koitzsch~\textit{et~al.}~\cite{KoitzschHeming16} that consisted in analyzing the ARPES data measured on samples cleaved along all high-symmetry crystallographic planes $(100)$, $(110)$, and $(111)$, enabled them to reconstruct the complete 3D bulk electronic structure of CeB$_6$ with much better accuracy than typically achievable in theoretical band-structure calculations or conventional ARPES measurements. CeB$_6$ presents some natural complications to this technique, which explains the sparsity of previous studies. Firstly, the electronic structure is fully three-dimensional, yet is it known that in a conventional surface-sensitive ARPES experiment, the momentum resolution in one direction orthogonal to the crystal surface is always inferior to the lateral resolution. Further, hexaborides are prone to surface reconstruction and surface states~\cite{PatilAdhikary10, Trenary12, ZhangButch13, HemingTreske14}, such that the bulk structure may be masked to ARPES. These problems were to a large extent overcome by performing tomographic measurements in the soft x-ray regime, which spans a large $k_z$ interval. Figure~\ref{FigInosov:CeB6PhotoTOM} presents an example of the photoemission tomography results, alongside a calculation of the Lindhard function compared to the INS data in phase~II.\index{CeB$_6$!phase II} Figure~\ref{FigInosov:CeB6PhotoTOM}\,(a) shows the Fermi surface map taken for the $(100)$ plane. By combining the results from all high-symmetry directions in the material, the authors were able to build up a full 3D picture of the Fermi surface, which is shown in Fig.~\ref{FigInosov:CeB6PhotoTOM}\,(b), with the corresponding $(100)$ plane of panel (a) also illustrated. These results were also consistent with de-Haas\,--\,van Alphen measurements, which had indicated ellipsoidal Fermi-surface sheets centered around the $X$ points. From this model of the Fermi surface, the Lindhard function was calculated, with its cross-section in the $(HHL)$ plane shown next to the INS data (quasielastic intensity)\index{diffuse neutron scattering}\index{quasielastic magnetic scattering}\index{neutron scattering!quasielastic} in Fig.~\ref{FigInosov:CeB6PhotoTOM}\,(c). The two maps are exceptionally similar, reproducing not only the peaks in intensity but also the qualitative shape of the dominant features. This demonstrates the itinerant character of the magnetic excitations in CeB$_6$, and suggests that the propagation vector of the AFQ order is determined by the geometry of the Fermi surface.\index{angle-resolved photoemission|)}\index{photoelectron spectroscopy|)} This does not disagree with the conclusion that phase~II\index{CeB$_6$!phase II} is AFQ in nature, as the Ce~4\textit{f} states may still be considered local, but rather indicates the role of itinerant electrons in determining the RKKY interactions between the local dipolar and multipolar moments. This interpretation is additionally supported by the new results on La- and Nd-doped CeB$_{6}$ \cite{FriemelJang15, NikitinPortnichenko18}, to be discussed in Sec.~\ref{Ino_Sec:DopedCeB6} later in this chapter.
\index{Lindhard function|)}
\index{quasielastic magnetic scattering|)}

\vspace{-2pt}\section{Magnetic-field dependence of the collective excitations}\vspace{-3pt}
\index{CeB$_6$!multipolar excitations|(}\index{multipolar excitations!in CeB$_6$|(}
\label{Ino_Sec:BdepCeB6}

\subsection{General remarks}

The first INS measurements on CeB$_{6}$, discussed in Sec.~\ref{Ino_SubSec:CeB6AFQPhase}, revealed an appearance of intense dispersive magnon branches induced by magnetic field. Unfortunately, these data were limited to only several high-symmetry directions of the reciprocal space and a few field values, however these measurements motivated the development of theoretical models which could describe the magnetic excitation spectrum in phase~II\index{CeB$_6$!phase II} both with and without the application of magnetic field under the assumption of quadrupolar symmetry breaking. To simplify the calculations, RKKY-type interactions\index{RKKY interaction|)} between the multipoles were restricted to the nearest neighbors only, and the competing dipolar AFM phase that replaces the AFQ ground state in weak magnetic fields was completely neglected (see Chapter~8 [\href{https://arxiv.org/abs/1907.10967}{arXiv:1907.10967}] for more details). Comparison of the calculations at certain points of reciprocal space with experiments indicated relatively good agreement only close to the zone center, and the limited number of measured data available at that time did not allow for a convincing identification of the assumed quadrupolar and octupolar order parameters from such a comparison.

Due to the dipole-dipole interaction between the magnetic moments of the unpaired electron spins in the sample and the neutron magnetic moment, neutron scattering became a highly effective probe of condensed matter physics. However, in contrast to the magnetic order formed by electrons' dipolar moments, ordering phenomena associated with higher-order multipoles are more difficult to characterize. The theory of neutron scattering beyond the dipole approximation, which considers the contribution of the neutron's interactions with the multipoles to the double-differential cross-section~\cite{BalcarLovesey89, JensenMackintosh91}, states that neutron scattering is sensitive to all odd-rank \textit{magnetic} multipoles (such as dipole, octupole, dotriakontapole etc.)~\cite{Lovesey15}.\index{magnetic octupoles} At short scattering vectors, \mbox{$|\mathbf{Q}|\rightarrow0$}, the form factor\index{magnetic form factor}\index{form factor} is expected to suppress all higher-order multipolar contributions, which justifies the so-called \textit{dipolar approximation} that has been employed for calculating the dynamical structure factor even for multipolar-ordered systems. The nonmonotonic form factor\index{magnetic form factor!nonmonotonic}\index{nonmonotonic form factor} of higher-order multipoles vanishes at \mbox{$\mathbf{Q}=0$} and then starts to increase until reaching a maximum at some finite momentum transfer. This has been established from both theory and experiment for elastic neutron scattering \cite{ShiinaSakai07, KuwaharaIwasa07, KuramotoKusunose09, SantiniCarretta09, Shiina12}, but to the best of our knowledge, the highly involved theory of INS beyond the dipolar approximation~\cite{BalcarLovesey89, JensenMackintosh91} was never successfully applied to calculate the dynamical response functions for any compound with a multipolar-ordered ground state. On the other hand, experiments clearly demonstrate that the intensity of the $\Gamma$-point excitation in CeB$_6$\index{CeB$_6$!ferromagnetic resonance}\index{CeB$_6$!zone-center excitations} increases with wave vector when going from the shortest $\Gamma'(001)$ to the second-shortest $\Gamma''(110)$ reciprocal-lattice vector~\cite{JangFriemel14}, suggesting a nonmonotonic form factor characteristic of multipolar moments that cannot be described in the dipolar approximation. Access to the same excitation at even longer wave vectors is restricted by the worsening of energy resolution imposed by the kinematic constraints, and therefore the actual share of nondipolar spectral weight in the experimental INS spectrum remains unknown. We expect that it should change with the applied magnetic field, as it induces secondary dipolar and octupolar order parameters on top of the primary AFQ order, thereby activating additional magnetic degrees of freedom that are subject to collective fluctuations~\cite{PortnichenkoAkbari20}. The multipolar corrections should increasingly influence the measured intensities of magnetic excitations towards higher $|\mathbf{Q}|$, leading to deviations from the established spin-dynamical models.

In order to characterize hidden-order phases, one usually looks at the magnetic excitation spectrum which carries the imprint of the multipolar interactions and the hidden order parameter\index{CeB$_6$!magnetically hidden order}\index{magnetically hidden order!in CeB$_6$} in its dispersion relations. Using a specific candidate model for the hidden order, the dispersion is usually calculated and then compared to the dynamical structure factor measured by single-crystal INS. Such attempts to describe the hidden-order phase of CeB$_{6}$ have been made, yet an overall determination of the dispersion in the whole BZ still remains elusive. The absence of systematic investigations of the spin dynamics does not allow us to identify some of the experimentally observed excitation modes, and the need for more accurate INS experiments as a function of field and temperature, explicitly mentioned by theoreticians, motivated us to extend the data within phase~II\index{CeB$_6$!phase II} with high-quality measurements that would cover the complete reciprocal space under various magnetic fields and different field directions, thereby allowing for a detailed quantitative comparison with theoretical models. In contrast to the earlier INS measurements, which yielded only a very fragmented dataset on pure CeB$_6$ and lacked the energy resolution to resolve low-lying magnetic excitations in this system, modern instrumentation allowed us for the first time to obtain systematic and conclusive high-resolution data over the multidimensional parameter space, which will be the topic of the following sections.

\index{CeB$_6$!ferromagnetic resonance|(}\index{CeB$_6$!zone-center excitations|(}

\begin{figure}[t!]
\centerline{\includegraphics[width=0.8\textwidth]{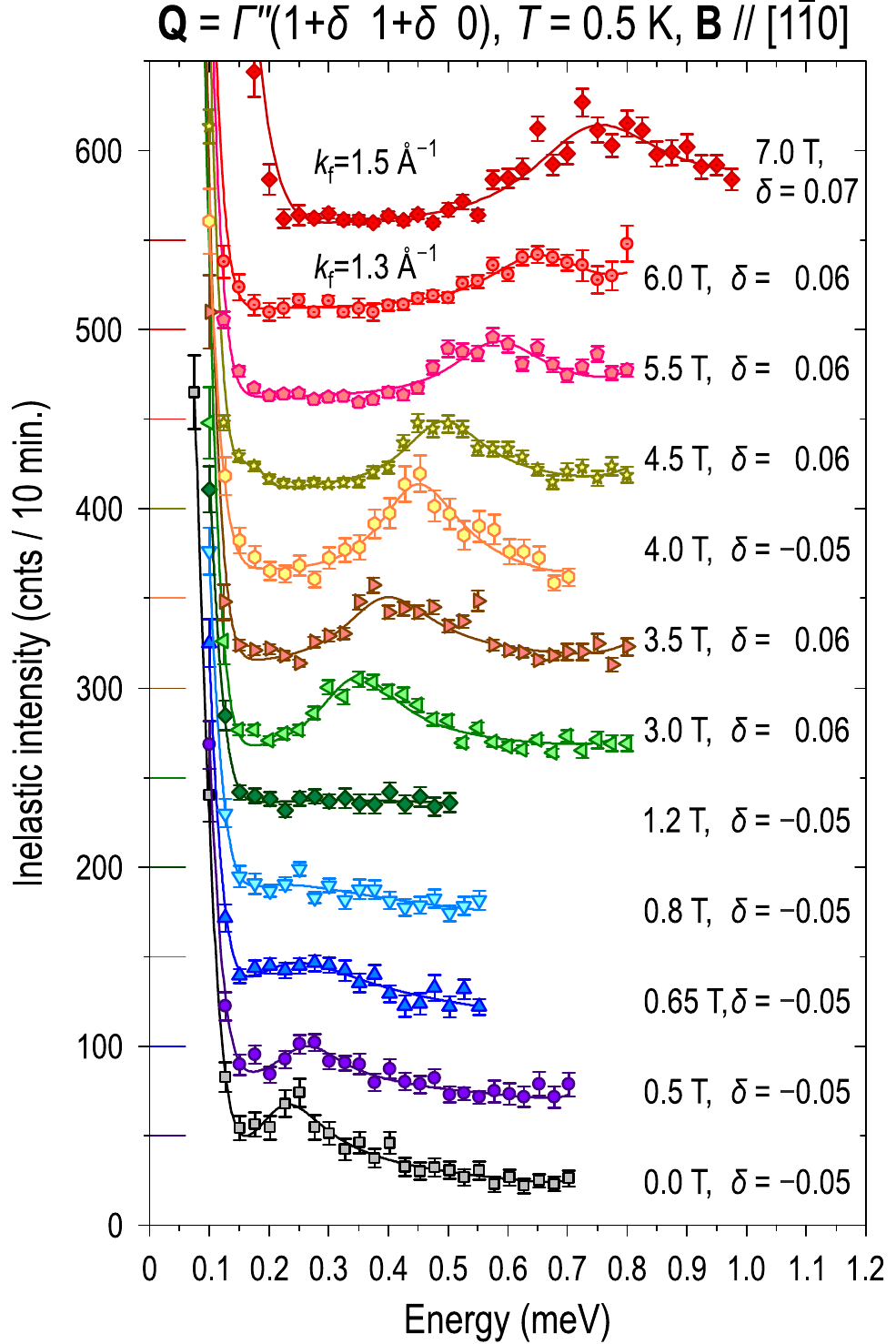}}
\caption{INS spectra measured near the zone center $\Gamma''(110)$ at a slightly incommensurate wave vector as indicated in the legend, to avoid contamination from the Bragg tail. The spectra are shifted vertically for clarity with horizontal lines at the left indicating the background baseline for each spectrum. Solid lines represent Lorentzian fits on top of a nonmagnetic background. Reproduced from Ref.~\cite{PortnichenkoDemishev16}.\index{CeB$_6$!ferromagnetic resonance!field dependence}\index{CeB$_6$!zone-center excitations!field dependence}}
\label{FigInosov:CeB6HdepGamma_PANDA}
\end{figure}

\vspace{-2pt}\subsection{Zone-center excitations}\label{Ino_SubSec:CeB6GammaPoint}

\begin{figure}[t!]\vspace{-1pt}
\centerline{\includegraphics[width=0.8\textwidth]{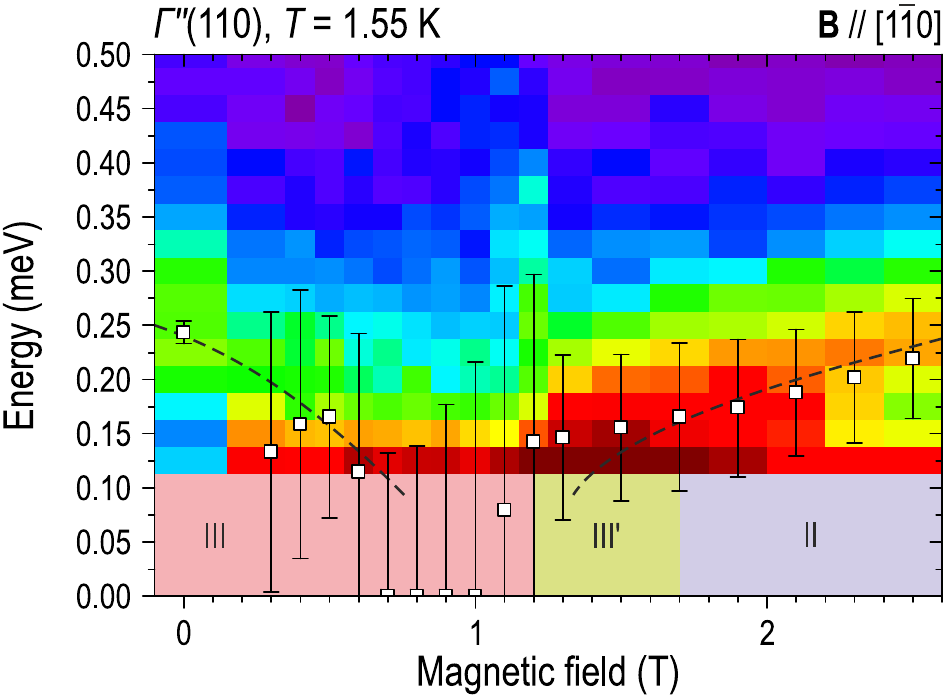}\vspace{-1pt}}
\caption{Magnetic-field dependence of the resonance peaks at $\Gamma$ point. Each one-dimensional energy profile obtained by integration of the four-dimensional TOF data \cite{PortnichenkoDemishev16} within $\pm 0.15$~r.l.u. around the $\Gamma$ point. Markers show fitted positions of peak maxima. Dashed lines are guides to the eyes, and the shaded areas at the bottom of the figure mark the resolution cutoff and indicate the field regions corresponding to the AFM (III, III$^\prime$) and AFQ (II) phases.\index{CeB$_6$!phase II}\index{CeB$_6$!phase III}\index{CeB$_6$!phase III$'$} Reproduced from Ref.~\cite{PortnichenkoDemishev16}.\index{CeB$_6$!ferromagnetic resonance!field dependence}\index{CeB$_6$!zone-center excitations!field dependence}
\vspace{-2pt}}
\label{FigInosov:CeB6BdepGamma_IN5}
\end{figure}

We start by presenting the evolution of the ferromagnetic mode in magnetic fields up to 7\,T at $T=0.5$\,K, obtained with the magnetic field \mbox{$\mathbf{B}\parallel[1\overline{1}0]$}, because this field direction has been most extensively studied in the past. Figure~\ref{FigInosov:CeB6HdepGamma_PANDA} shows energy scans near the zone center $\Gamma''(1\!+\!\delta \;1\!+\!\delta\; 0)$. The sharp resonance reported in Sec.~\ref{Ino_SubSec:CollectiveExcitations} initially gets suppressed and broadens with the application of an external magnetic field as long as the system remains in the AFM state. The observed signal can be described by a Lorentzian line shape \cite{GoremychkinOsborn00},
\begin{multline}\label{EqInosov:Quasielastic}
S(\mathbf{Q},\omega)\propto\,\frac{\omega}{1-\exp(-\hslash\omega/k_\text{B}T)}\\
\times\biggl(\frac{\Gamma}{\hslash^2(\omega-\omega_0)^2+\Gamma^2}+\frac{\Gamma}{\hslash^2(\omega+\omega_0)^2+\Gamma^2}\biggr),
\end{multline}
where $\Gamma$ is the half width at half maximum of the Lorentzians centered at $\pm\hslash\omega_0$, whereas $\hslash$ and $k_{\rm B}$ are the Planck and Boltzmann constants. An even more detailed evolution of the ferromagnetic excitation is shown in Fig.\,\ref{FigInosov:CeB6BdepGamma_IN5}, which illustrates the nonmonotonic behavior of the zone-center excitation as it initially softens to zero upon entering phase~III$^\prime$~\cite{KunimoriKotani11}\index{CeB$_6$!phase III$'$} and then reappears within phase~II\index{CeB$_6$!phase II} at an energy that continuously increases with the applied field.

\begin{figure}[t]\vspace{-1pt}
\centerline{\includegraphics[width=\textwidth]{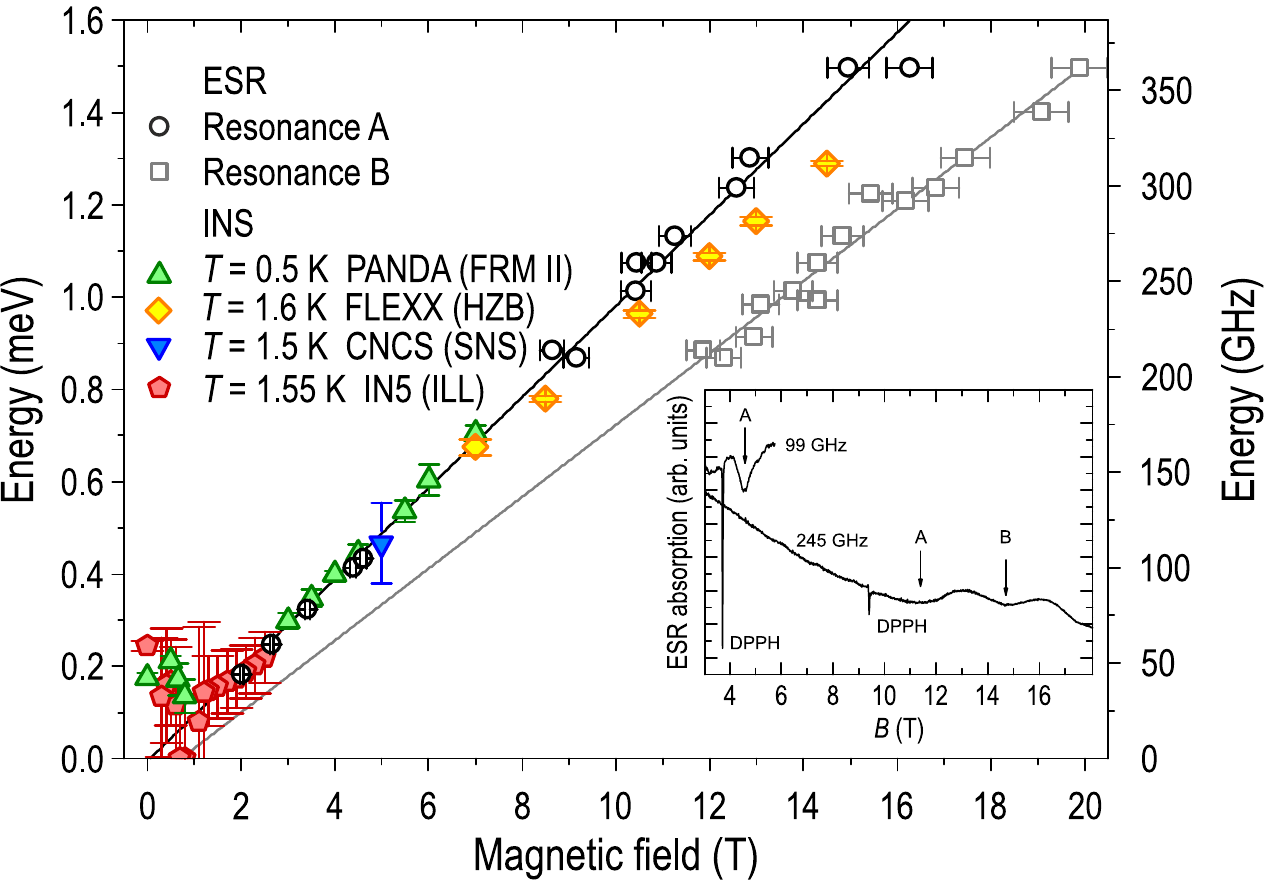}\vspace{-1pt}}
\caption{Summary of the magnetic-field dependence of zone-center excitations obtained from both INS and ESR spectra.\index{CeB$_6$!electron spin resonance} Solid lines are linear fits of resonances A and B. The inset shows a field dependence of the cavity transmission at 99\,GHz and ESR spectrum obtained at 245\,GHz using a quasioptical technique\index{electron spin resonance!quasioptical technique} as typical examples of unprocessed datasets from which the points in the main plot were obtained. Sharp lines marked as DPPH originate from a small 2,2-diphenyl-1-picrylhydrazyl reference sample. Adapted from Ref.~\cite{PortnichenkoDemishev16}, with higher-field INS data from Ref.~\cite{PortnichenkoAkbari20} added (diamond symbols).
\index{CeB$_6$!electron spin resonance}\index{electron spin resonance!in CeB$_6$}\index{CeB$_6$!ferromagnetic resonance!field dependence}\index{CeB$_6$!zone-center excitations!field dependence}}
\label{FigInosov:CeB6ESRandFieldDep}
\end{figure}

According to the discussion in Sec.~\ref{Ino_SubSec:CollectiveExcitations}, ESR spectroscopy\index{electron spin resonance} can also probe the zone-center excitation, therefore experimental results at the $\Gamma$ point measured with neutron scattering can be directly compared with the ferromagnetic resonance seen earlier using ESR \cite{DemishevSemeno06, DemishevSemeno08, DemishevSemeno09}.\index{CeB$_6$!electron spin resonance|(}\index{electron spin resonance!in CeB$_6$|(} These measurements have shown that the frequencies of the two observed resonances A and B change linearly with field within phase~II,\index{CeB$_6$!phase II} as shown in Fig.~\ref{FigInosov:CeB6ESRandFieldDep}. We find perfect agreement between the INS data (closed symbols) and the resonance A (open symbols) in the intermediate field range within phase~II,\index{CeB$_6$!phase II} suggesting that the same ferromagnetic excitation is probed in both experiments. This comparison nicely demonstrates the complementarity of the ESR and INS methods.

\begin{figure}[t!]
\centerline{\includegraphics[width=0.95\textwidth]{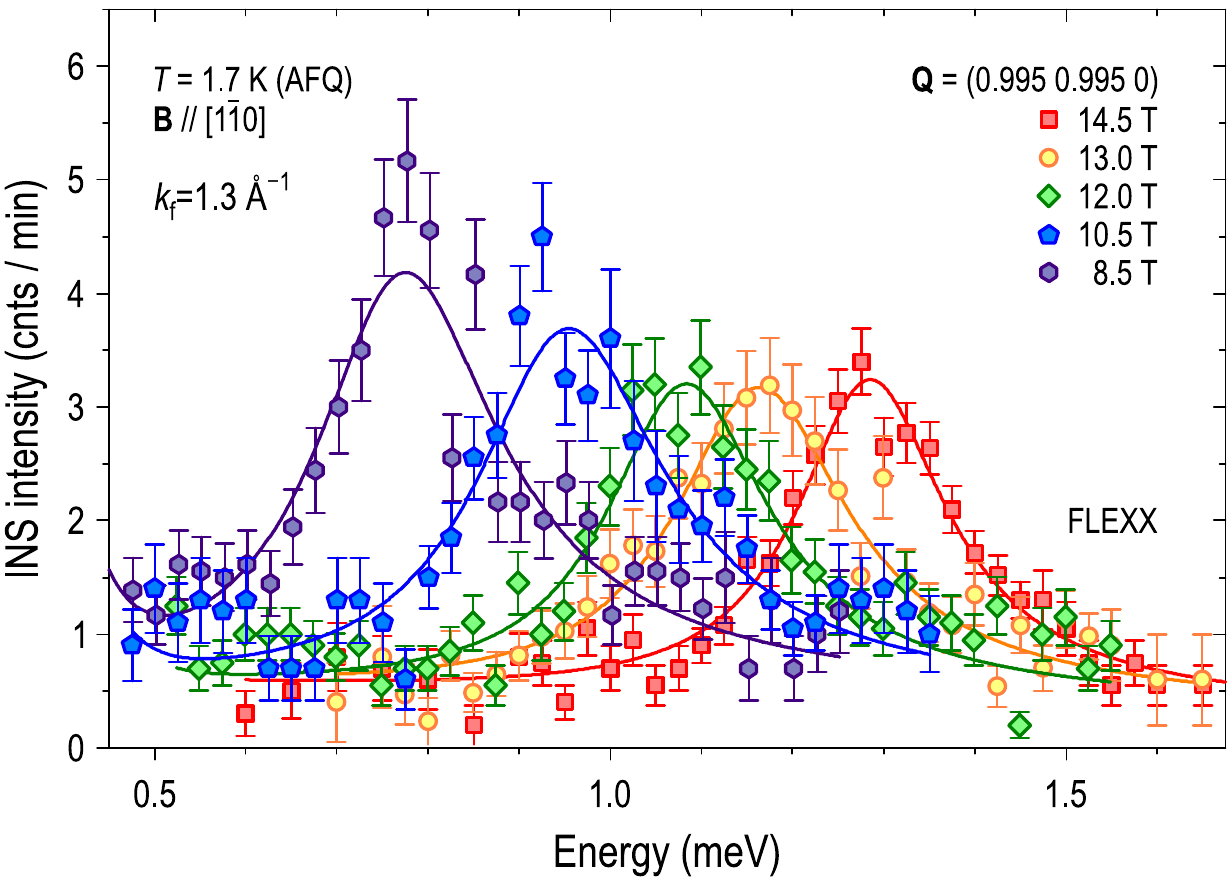}}
\caption{Unprocessed INS spectra measured near the zone center $\Gamma''(110)$ for $\mathbf{B} \parallel [1\overline{1}0]$. Reproduced from Ref.~\cite{PortnichenkoDemishev16}.
\index{CeB$_6$!ferromagnetic resonance!field dependence}\index{CeB$_6$!zone-center excitations!field dependence}\vspace{-2pt}}
\label{FigInosov:CeB6Hdep110FLEXX}
\end{figure}

At fields above 12\,T, which we did not at first cover in our neutron-scattering experiment, the appearance of a second ESR resonance (resonance~B) was observed. The presence of only one resonance, rather than four, within the fourfold degenerate $\Gamma_8$ ground multiplet at relatively low fields, as well as the appearance of the second resonance in the ESR spectrum at high fields above $\sim$12\,T was explained by Schlottmann~\cite{Schlottmann12, Schlottmann13, Schlottmann18}. The AFQ ordering in CeB$_{6}$ introduces two interpenetrating sublattices and simultaneously quenches three out of four resonances for each site. The coherence between sublattices prevents two distinct resonances for each site being occupied simultaneously, thus hybridizing them into a single resonance. This scenario successfully explained the presence of only one resonance at 60~GHz. The emergence of the second ESR line at frequencies above 200~GHz was interpreted as the result of a crossover of the excited state to the free-ion limit, as the field at which it appears is comparable with the condensation energy of the AFQ phase, $\sim$$1.75{\kern.5pt}k_\text{B}T_\text{Q}$~\cite{Schlottmann12, Schlottmann13}.

Since simultaneous observation of the resonance by INS and ESR was previously confirmed, we extended the available neutron-scattering data in order to uncover the second resonance~B that escaped direct observation in our previous experiments due to the limited magnetic-field range. We expected to observe the emergence of the resonance B at lower energies relative to the main resonance, as suggested by the ESR data, however a systematic study of the $\Gamma$ point, shown in Fig.~\ref{FigInosov:CeB6Hdep110FLEXX}, demonstrates that the energy of resonance~A increases continuously with the applied field, experiencing a slight broadening. The corresponding energy scans show no signatures of any additional peak in the expected energy range of 0.9--1.1~meV up to the maximal field of 14.5\,T. The background in our range of interest is clean and essentially constant, and the energy resolution is sufficiently narrow,\footnote{The measurements were done at the triple-axis spectrometer\index{triple-axis spectrometer} FLEXX (Helmholtz-Zentrum Berlin, Germany) with the final wave vector of the neutrons fixed to $k_{\rm f}=1.3$\,\AA$^{-1}$, and therefore the instrument resolution was less than 0.1~meV~\cite{LeQuintero-Castro13}.} so that the tail of the peak corresponding to resonance~A does not overlap with the expected position of resonance~B at such high magnetic fields. For instance, at 14.5\,T the expected peak splitting is 0.35~meV. Therefore, we can conclude that resonance~B is not visible to neutrons, possibly due to certain selection rules that are different from those of ESR.

\begin{figure}[t!]
\centerline{\includegraphics[width=\textwidth]{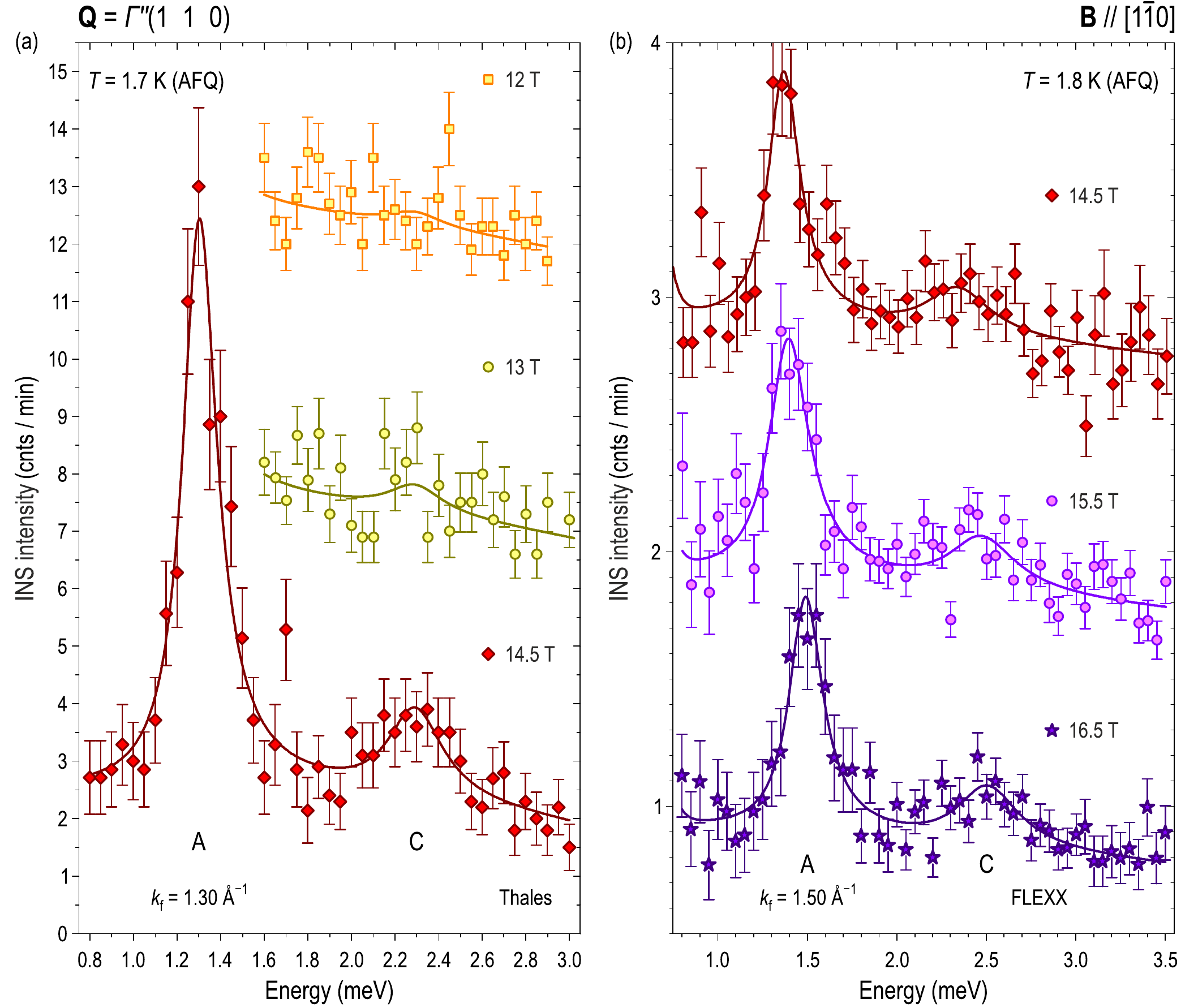}}
\caption{Unprocessed INS spectra measured near the zone center $\Gamma''(110)$ for $\mathbf{B}\parallel[1\overline{1}0]$. The data are offset vertically for clarity. They illustrate the appearance of resonance~C (smaller peak) at a higher energy with respect to resonance~A (stronger peak) at magnetic fields above 14\,T. Both peak positions shift upwards in energy with increasing field. Solid lines are Lorentzian fits. Reproduced from Ref.~\cite{PortnichenkoAkbari20}.\index{CeB$_6$!ferromagnetic resonance!field dependence}\index{CeB$_6$!zone-center excitations!field dependence}}
\label{FigInosov:CeB6Hdep110ThalesFLEXXBooster}
\end{figure}

It is important to pay attention to the fact that at higher fields, the excitation that followed the energy dependence of the A resonance starts to deviate towards lower energies, as shown in Fig.\,\ref{FigInosov:CeB6ESRandFieldDep}. To confirm this unexpected result and exclude the possibility that the observed deviation is a consequence of incorrect magnetic-field calibration, while the second resonance located at a lower energy simply cannot be distinguished from the background due to insufficient statistics, we conducted an additional experiment, where we repeated the same measurements up to 14.5\,T at the higher-flux spectrometer ThALES (Institute Laue-Langevin, Grenoble), using the same single crystal and sample environment~\cite{PortnichenkoAkbari20}. The main result of this measurements is shown in Fig.\,\ref{FigInosov:CeB6Hdep110ThalesFLEXXBooster}\,(a). The absence of the peak associated with resonance~B was reproduced, but extending the spectrum to higher energies revealed a new peak at a higher energy of $\sim$2.25~meV in the 14.5\,T dataset that rapidly vanished when the field was decreased below its maximal value. This result suggests that an additional resonance (resonance~C) forms above $\sim$14\,T, at approximately the same field as resonance~B, but on the opposite side of the intense peak corresponding to resonance~A.
\index{CeB$_6$!electron spin resonance|)}\index{electron spin resonance!in CeB$_6$|)}

In order to avoid any doubt about the magnetic origin of resonance~C, it was necessary to extend the measurements to even higher fields. Significant efforts have been made in order to extend the maximal field up to 16.5\,T~\cite{PortnichenkoAkbari20}. We used an additional dysprosium ``booster'', which concentrates magnetic field lines in a smaller sample volume, so that a higher magnetic field can be achieved at the expense of the smaller sample size. The resulting spectra, shown in Fig.\,\ref{FigInosov:CeB6Hdep110ThalesFLEXXBooster}\,(b), show that the energy of both resonances A and C shifts upwards with increasing field with approximately the same slope, which unequivocally confirms the magnetic nature of resonance~C.
\index{CeB$_6$!ferromagnetic resonance|)}\index{CeB$_6$!zone-center excitations|)}

\vspace{-2pt}\subsection{Dispersion of the field-induced collective modes}\label{Ino_SubSec:CeB6Dispersion}

At present we have successfully studied the evolution of the resonant mode at the $\Gamma$ point upon application of the magnetic field along the $[1\overline{1}0]$ crystallographic direction over a broad range of magnetic fields. A strong ferromagnetic mode at the zone center $\Gamma$ within the AFM phase, which is hybridized with the maximum of intensity at the $R$ point\index{CeB$_6$!$R$-point resonant mode}\index{CeB$_6$!magnetic resonant mode} and the spin-wave modes\index{CeB$_6$!spin waves}\index{spin waves!in CeB$_6$} emanating from the AFM wave vectors, initially softens to zero upon entering the phase~III$^\prime$\index{CeB$_6$!phase III$'$} and then reappears at higher fields. Together they form a continuous dispersive magnon with a bandwidth that continuously increases with the application of magnetic field~\cite{JangFriemel14}. At 14\,T and above, we clearly observe the onset of the new resonance at the $\Gamma$ point, which still requires further clarification.

\begin{figure}[t!]
\centerline{\includegraphics[width=0.8\textwidth]{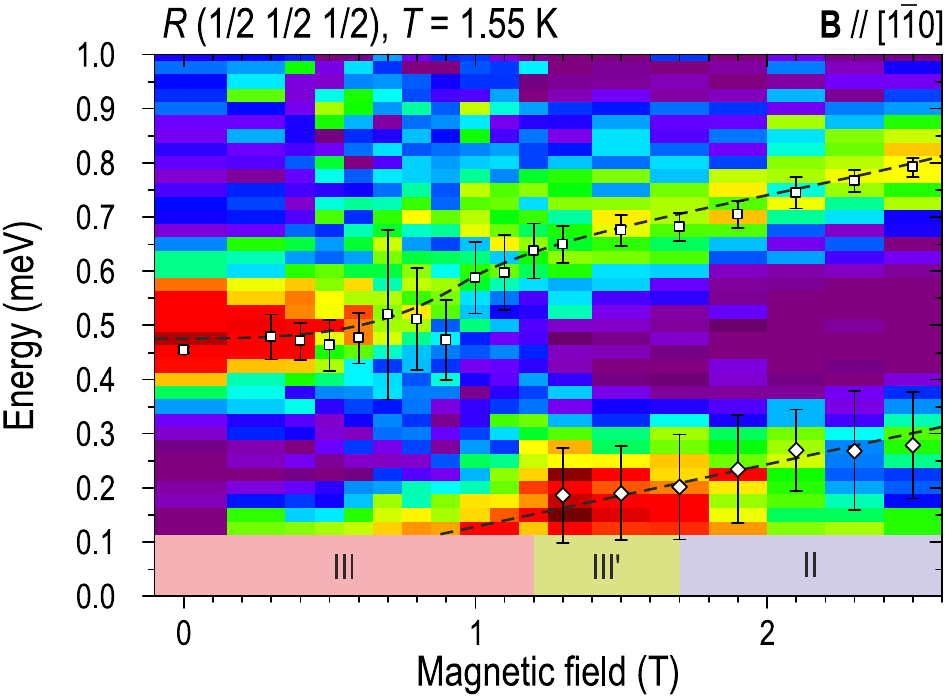}}
\caption{Magnetic field dependence of the resonance peaks at the $R$ point. Every energy profile was obtained by integration of the TOF data within $\pm 0.15$~r.l.u. around the $R$ point. Markers show fitted positions of peak maxima. Dashed lines are guides to the eyes, and the shaded areas at the bottom of the figure mark the resolution cutoff and indicate the field regions corresponding to the AFM (III, III$^\prime$)\index{CeB$_6$!phase III}\index{CeB$_6$!phase III$'$} and AFQ (II)\index{CeB$_6$!phase II} phases. Reproduced from Ref.~\cite{PortnichenkoDemishev16}.\index{CeB$_6$!$R$-point resonant mode!field dependence}\index{CeB$_6$!magnetic resonant mode!field dependence}}
\label{FigInosov:CeB6BdepR_IN5}
\end{figure}

A possible way to shed light on the nature of resonance~C is to study its dispersion at a constant magnetic field and to understand its relationship to the features in INS spectra that were observed at other high-symmetry points in the Brillouin zone, as mentioned in Sec.~\ref{Ino_SubSec:MeanFieldDescription}. One piece of information can be found in previous studies~\cite{FriemelJang15, PortnichenkoDemishev16}, which show a qualitatively different behavior of the resonance peak at the $R$ point, illustrated in Fig.~\ref{FigInosov:CeB6BdepR_IN5}. Increasing the field within phase~III\index{CeB$_6$!phase III} keeps the resonance energy constant while it decreases in amplitude and broadens, transferring a significant part of its spectral weight to the second low-energy mode whose tail can be seen above the elastic line already above $\sim$0.5\,T. Upon crossing through the phase~III--III$^\prime$\index{CeB$_6$!phase III}\index{CeB$_6$!phase III$'$} transition, the amplitude of the low-energy mode is maximized, whereas the higher-energy mode shifts up in energy. Both excitations then follow a linear trend with the same slope and approximately equal amplitudes in phase~II.

\begin{figure}[t!]
\centerline{\includegraphics[width=\textwidth]{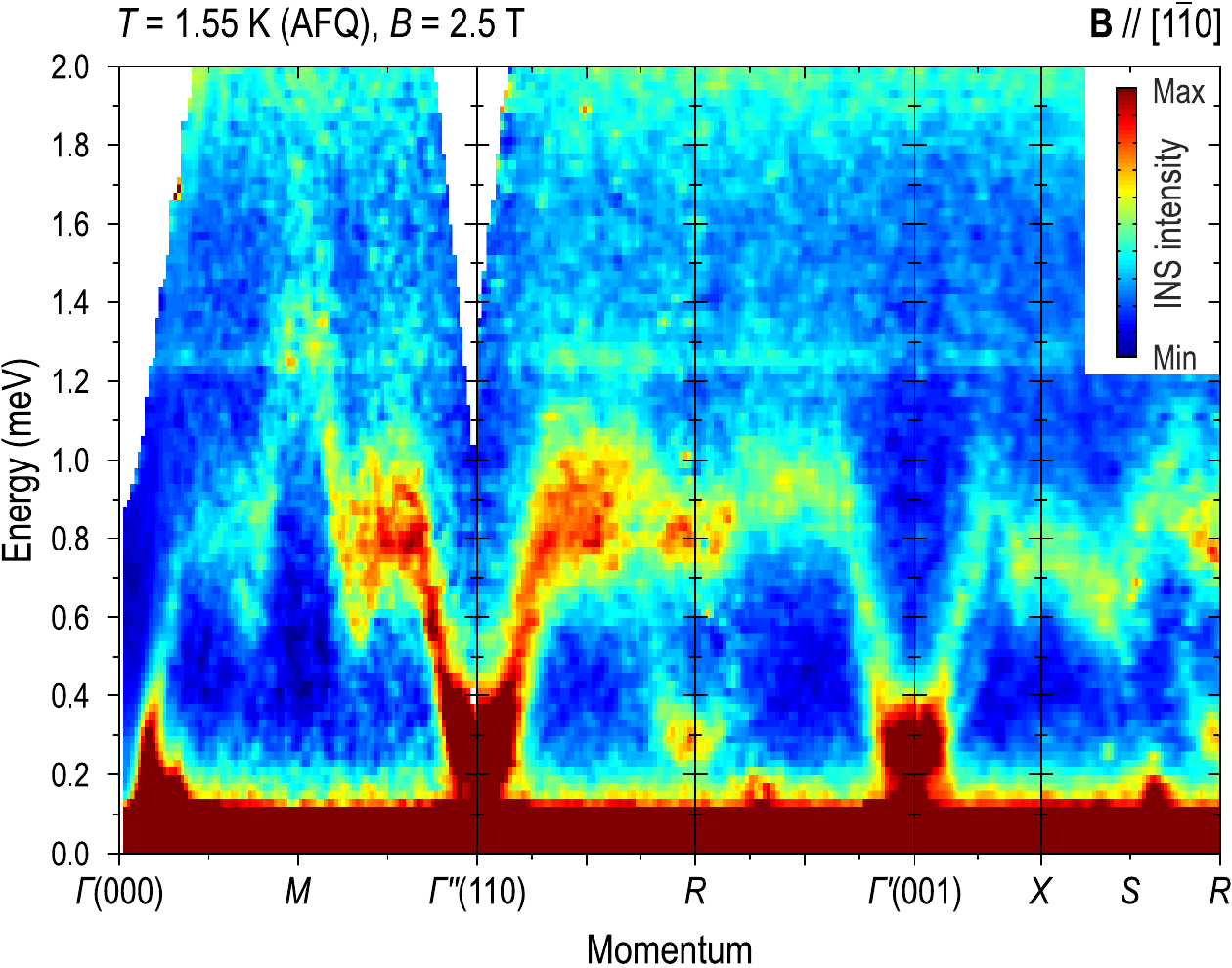}}
\caption{Distribution of INS intensity in energy-momentum space along high-symmetry directions in the AFQ state in a magnetic field of $B = 2.5$\,T. Reproduced from Ref.~\cite{PortnichenkoDemishev16}.
\index{multipolar excitations!dispersion}\index{multipolar excitations!field dependence}\index{CeB$_6$!zone-center excitations!field dependence}\index{CeB$_6$!$R$-point resonant mode!field dependence}\vspace{-2pt}}
\label{FigInosov:CeB6Map2p5T_IN5}
\end{figure}

Since the presence of a continuous dispersive magnon band within the AFM phase was previously confirmed, it is logical to assume the presence of similar bands also within the AFQ phase, as it was predicted earlier~\cite{ShiinaShiba03, ThalmeierShiina98, ThalmeierShiina03, ThalmeierShiina04} in the calculations shown in Fig.~\ref{FigInosov:Mean_field_calculations}. The presence of two resonances at the $R$ point\index{CeB$_6$!$R$-point resonant mode}\index{CeB$_6$!magnetic resonant mode} at low fields leaves an open question about the detailed evolution of these branches along the $\Gamma$--$R$ line. Another question is the momentum dependence of the resonance~C, which emerges above 14\,T.

\begin{figure}[t]\vspace{-2pt}
\centerline{\includegraphics[width=0.74\textwidth]{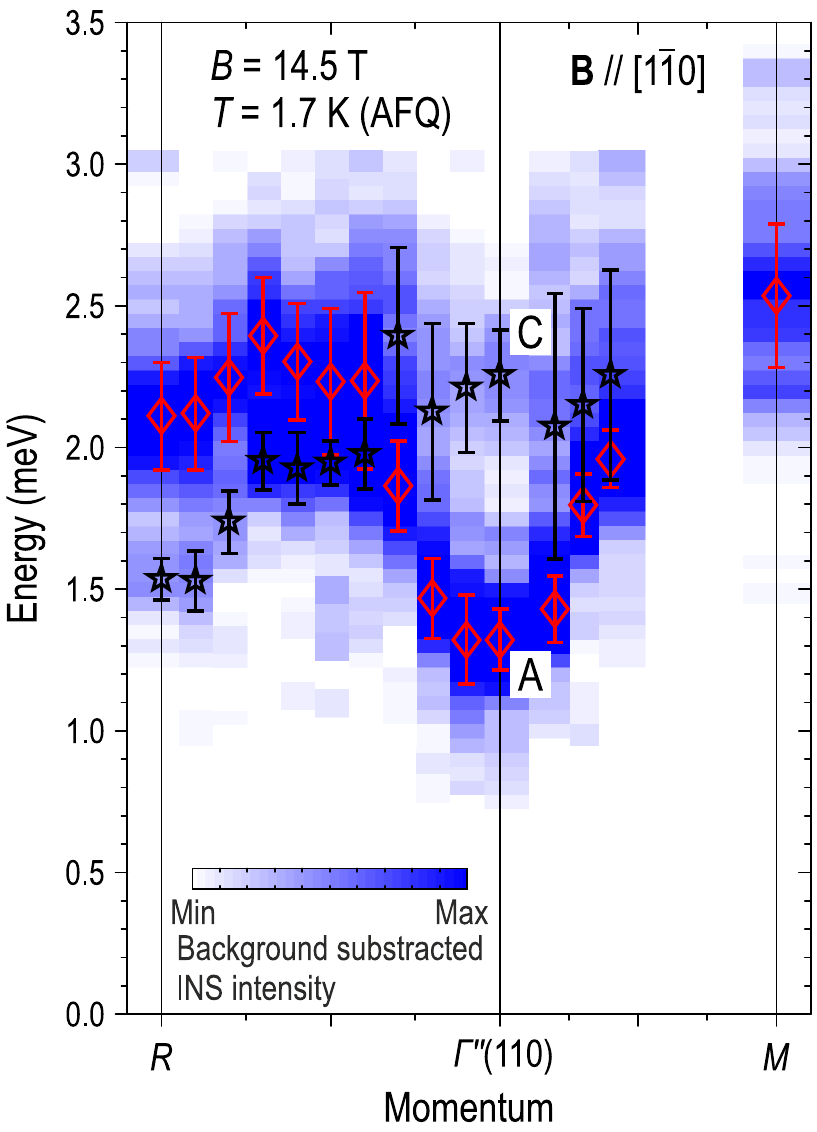}}
\caption{INS intensity vs.~energy and momentum along straight segments connecting the $R(\frac{1}{2}\frac{1}{2}\frac{1}{2})$, $\Gamma''(110)$, and $M(\frac{1}{2}\frac{1}{2}0)$ points in a magnetic field of 14.5\,T, applied along the $[1\overline{1}0]$ direction. The color map shows raw data after background subtraction and subsequent smoothing with a one-dimensional Gaussian filter characterized by a FWHM of 0.1\,meV (to reduce statistical noise). Darker blue color corresponds to higher INS intensity. Markers show fitted peak positions. Reproduced from Ref.~\cite{PortnichenkoAkbari20}.\index{multipolar excitations!dispersion}\index{multipolar excitations!field dependence}}
\label{FigInosov:CeB6cmap110ThalesFLEXX14p5T}
\end{figure}

To get a more complete picture about the momentum dependence of magnetic excitations, we performed a cold-neutron TOF experiment, which is particularly suitable for mapping out the dispersion relationships in the energy-momentum space. Figure\,\ref{FigInosov:CeB6Map2p5T_IN5} shows a continuous dispersive magnon band measured at 2.5\,T. Its intensity distribution along the main high-symmetry directions of reciprocal space suggests an anomalous nonmonotonic behavior of the dynamic form factor,\index{magnetic form factor!nonmonotonic}\index{nonmonotonic form factor} shown in Fig.\,14 in Chapter~8 [\href{https://arxiv.org/abs/1907.10967}{arXiv:1907.10967}], that is characteristic of multipolar moments~\cite{KuwaharaIwasa07,KuramotoKusunose09, Shiina12}, because the signal is more intense around the $\Gamma''(110)$ point rather than at the equivalent $\Gamma'(001)$ or $\Gamma(000)$ positions. A magnetic field of 2.5\,T does not change the excitation energy at the zone center with respect to 0\,T measurements significantly, but increases the magnon bandwidth twofold, as the dispersion now reaches $\sim$1.4~meV at the $M$ point in contrast to 0.7\,meV in zero field~\cite{JangFriemel14}. A second field-induced low-energy magnetic excitation which appears at the AFQ propagation vector, $R(\frac{1}{2}\frac{1}{2}\frac{1}{2})$, merges with the more intense branch emanating from the zone center.

\begin{figure}[t!]\vspace{-2pt}
\centerline{\includegraphics[width=\textwidth]{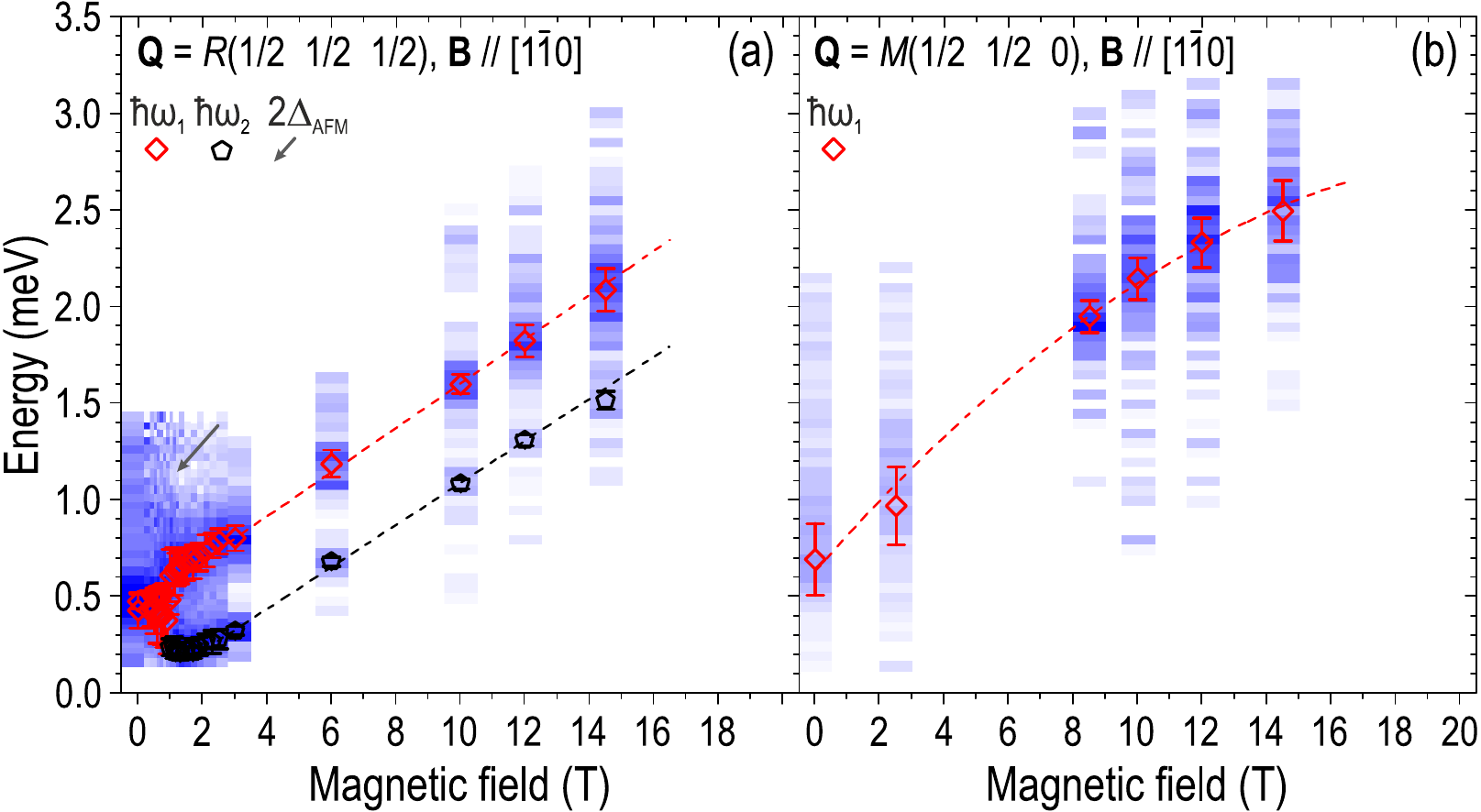}}
\caption{Summary of the neutron spectra, plotted as a function of magnetic field at the (a)~$R(\frac{1}{2}\frac{1}{2}\frac{1}{2})$ and (b)~$M(\frac{1}{2}\frac{1}{2}0)$ points for \mbox{$\mathbf{B} \parallel [1\overline{1}0]$}. Data points mark the fitted peak positions. The dashed line in panel (a) shows the expected position of resonance~B seen in ESR data (see text). Dotted lines are guides to the eyes. Reproduced from Ref.~\cite{PortnichenkoAkbari20}.\index{multipolar excitations!field dependence}\index{CeB$_6$!$R$-point resonant mode!field dependence}\index{CeB$_6$!magnetic resonant mode!field dependence}\vspace{-2pt}}
\label{FigInosov:SummaryFieldRandM}
\end{figure}

In order to follow the momentum dependence of the resonance~C at the $\Gamma$ point, which appears above 14\,T, we had to use conventional triple-axis spectrometers,\index{triple-axis spectrometer} because to the best of our knowledge all magnets that are compatible with TOF spectrometers cannot provide sufficiently high magnetic fields. Therefore we mapped out the dispersion by measuring the $\mathbf{Q}$ dependence of the INS spectrum along the $R(\frac{1}{2}\frac{1}{2}\frac{1}{2})$\,--\,$\Gamma''(110)$\,--\,$M(\frac{1}{2}\frac{1}{2}0)$ polygonal path in momentum space. These data were measured with the maximal field of 14.5\,T applied along $[1\overline{1}0]$. Figure\,\ref{FigInosov:CeB6cmap110ThalesFLEXX14p5T} shows that the more intense higher-energy excitation at the $R$ point\index{CeB$_6$!$R$-point resonant mode}\index{CeB$_6$!magnetic resonant mode} appears to be continuously connected to resonance~A at the zone center, whereas the weaker low-energy resonance at $R$ seems to cross this branch and re-emerge as resonance~C at the $\Gamma$ point. At the same time, both resonances approach each other along the $\Gamma M$ line, so that only a single peak is observed at the $M$ point.

From the earlier results, we know that upon transition into the AFQ phase the magnon bandwidth increases significantly. However, within phase~II\index{CeB$_6$!phase II} up to 14.5\,T the total magnon bandwidth does not change and stays at approximately 1.2~meV in the whole field range. The bottom of the magnon band is located at the $\Gamma$ point, and the maximum of the dispersion is reached at the $M$ point. Figure \ref{FigInosov:SummaryFieldRandM}\,(a) shows that the two excitations that were observed earlier at the $R(\frac{1}{2}\frac{1}{2}\frac{1}{2})$ point in phase~II\index{CeB$_6$!phase II} (Fig.~\ref{FigInosov:CeB6BdepR_IN5}) persist also in higher fields, following the same linear trends with an approximately equal slope. The data at $M(\frac{1}{2}\frac{1}{2}0)$ in Fig.\,\ref{FigInosov:SummaryFieldRandM}\,(b) indicate the presence of at least one excitation, whose energy increases monotonically with the magnetic field. These results clearly demonstrate that the whole spectrum shifts upward in energy with an increasing field.

\vspace{-2pt}\subsection{Anisotropy with respect to the field direction}
\index{CeB$_6$!magnetic excitations!field-angular anisotropy|(}

So far we have shown that both the energy and intensity as well as the number of experimentally observable multipolar excitations changes as a function of field. However, the full mapping of the reciprocal space requires significant efforts. At the moment, most of the TOF spectrometers that are naturally suited for these tasks, have significant limitations on the maximum possible value of the magnetic field. The only option left is to use conventional triple-axis spectrometers,\index{triple-axis spectrometer} which usually allow us to reach high enough fields. The magnon dispersion measured in this way, as shown in Fig.\,\ref{FigInosov:CeB6cmap110ThalesFLEXX14p5T}, provides a lot of information for characterizing field-induced collective excitations in the hidden-order phase, as it carries the imprint of the multipolar interactions\index{multipolar interactions}\index{multipolar excitations} and the hidden order parameter in their dispersion relations.\index{CeB$_6$!magnetically hidden order}\index{magnetically hidden order!in CeB$_6$} However, the extreme complexity of such measurements, and in particular the large amount of required neutron beam time, would require alternative approaches to be sought.

\begin{figure}[b!]
\centerline{\includegraphics[width=1\textwidth]{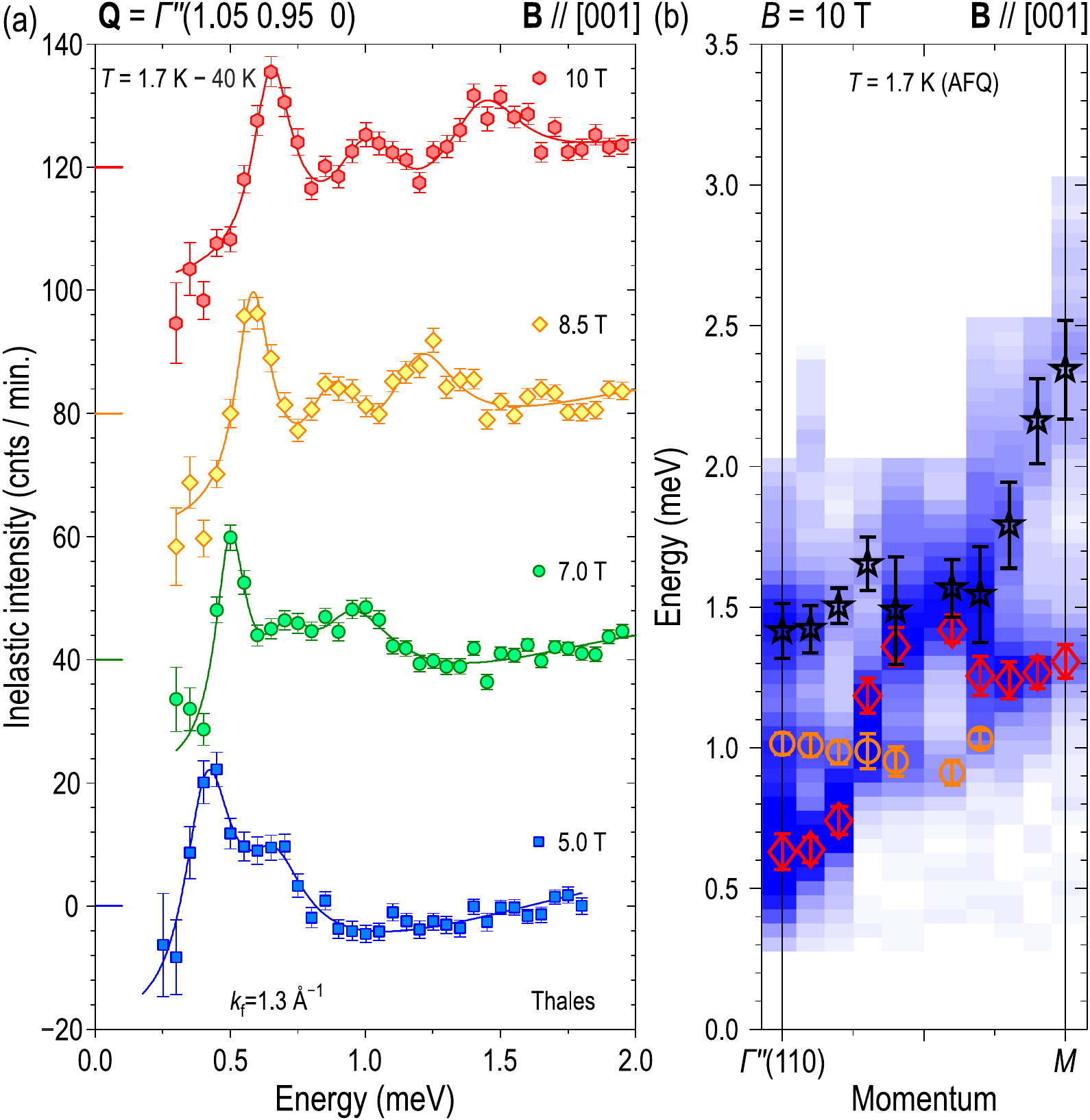}}
\caption{(a)~INS spectra measured near the zone center $\Gamma''(110)$ for $\mathbf{B}\parallel[001]$. A 40~K background scan was subtracted from each spectrum. (b)~INS intensity vs.~energy and momentum along straight segments connecting the $\Gamma''(110)$, and $M(\frac{1}{2}\frac{1}{2}0)$ points in a magnetic field of 10\,T, applied along the $[001]$ direction, presented in the same way as the data in Fig.\,\ref{FigInosov:CeB6cmap110ThalesFLEXX14p5T}. Reproduced from Ref.~\cite{PortnichenkoAkbari20}.\index{CeB$_6$!ferromagnetic resonance}\index{CeB$_6$!zone-center excitations}\index{multipolar excitations!field dependence}\index{multipolar excitations!dispersion}}
\label{FigInosov:CeB6Hdep001ThalesMap}
\end{figure}

We have recently proposed another way of analyzing the symmetry of hidden order\index{CeB$_6$!magnetically hidden order}\index{magnetically hidden order!in CeB$_6$} parameters and multipolar interactions in the magnetic excitation spectrum, which appears to be more promising in providing quantitative information or at least in yielding complementary data for comparison with future theoretical models~\cite{PortnichenkoAkbari20}. Rather than changing continuously the momentum transfer and following the dispersion of multipolar excitations in momentum space, one may fix the wave vector at some high-symmetry point with large INS intensity (e.g. $R$ or $\Gamma$) and instead change the direction and strength of the applied magnetic field. Ideally, the field could be continuously rotated, so that changing the field alone would give access to a three-dimensional parameter space for every single value of the momentum-transfer vector $\mathbf{Q}$. However, due to the limited measurement time, we were practically restricted only to several high-symmetry directions of the magnetic field in the available measurements.

From the calculations presented in Chapter~8 [\href{https://arxiv.org/abs/1907.10967}{arXiv:1907.10967}] by Thalmeier~\textit{et~al.}, a significant field-angular dependence\index{multipolar excitations!field-angular dependence} of the multipolar excitations is expected. Moreover, it has been argued that by keeping the momentum transfer fixed close to the zone center and by varying the field strength and direction, one can achieve better comparison between the theory and experiment. Thermodynamically, the anisotropy of the critical fields of phase~II\index{CeB$_6$!phase II} in CeB$_6$ was found to be determined by the underlying AFQ/AFO hidden order.\index{CeB$_6$!magnetically hidden order}\index{magnetically hidden order!in CeB$_6$} Therefore, a similar anisotropy in the pattern of collective modes under the field rotation is not at all surprising and can be calculated within the framework of available models. In the following, we demonstrate the usefulness of this new approach for understanding multipolar ordering phenomena by studying the field-angular anisotropy of magnetic excitations in CeB$_6$, which is a canonical example of a hidden-order compound in which the multipolar order parameter is already known from other methods.

\begin{figure}[b!]\vspace{-1pt}
\centerline{\includegraphics[width=\textwidth]{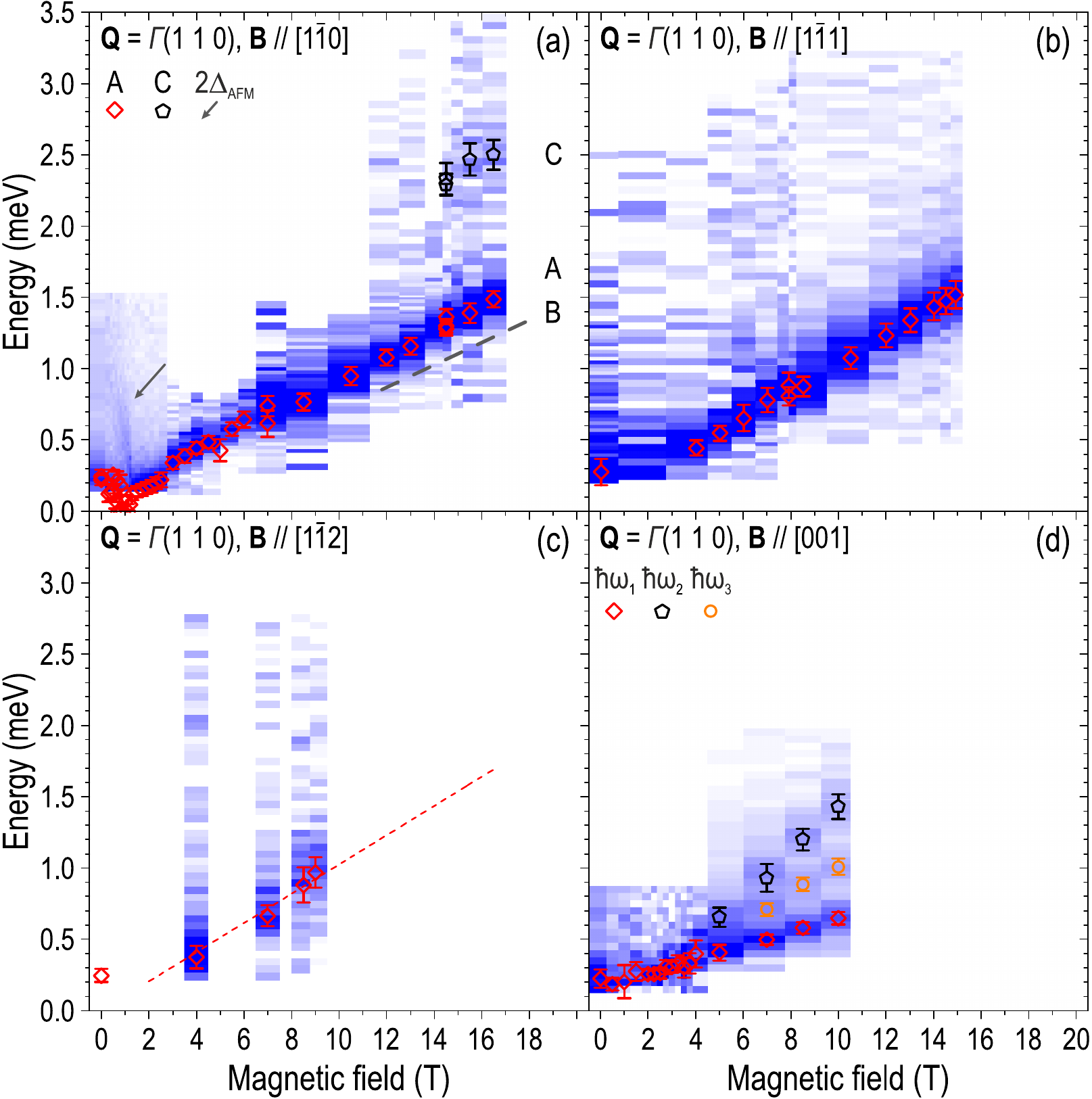}}
\caption{Summary of the neutron spectra, plotted as a function of magnetic field at the $\Gamma''(110)$ point for (a) $\mathbf{B}\parallel[1\overline{1}0]$; (b) $\mathbf{B}\parallel[1\overline{1}1]$; (c) $\mathbf{B}\parallel[1\overline{1}2]$; (d) $\mathbf{B}\parallel[001]$ field directions. Data presented in the same way as the data in Fig.\,\ref{FigInosov:SummaryFieldRandM}. Reproduced from Ref.~\cite{PortnichenkoAkbari20}.\index{multipolar excitations!field-angular dependence}\index{multipolar excitations!field dependence}\vspace{-2pt}}
\label{FigInosov:CeB6SummaryCMap}
\end{figure}

In order to study the anisotropy effects in CeB$_6$, we have measured the same single crystal after its crystallographic $[001]$ axis was reoriented parallel to the magnetic field. We observe an appearance of additional spectral lines, as shown in Fig.\,\ref{FigInosov:CeB6Hdep001ThalesMap}\,(a). For this field orientation, one can see three peaks that shift as a function of field with a different slope. Qualitative differences are also seen at other points in the BZ. For instance, the dispersion along the $\Gamma M$ direction is depicted in Fig.~\ref{FigInosov:CeB6Hdep001ThalesMap}\,(b). An additional peak is observed at the $M$ point, which appears to correspond to two excitations that split closer to the zone center.

Figure\,\ref{FigInosov:CeB6SummaryCMap}\,(a) summarizes all the data discussed in Sec.~\ref{Ino_SubSec:CeB6GammaPoint} for the field direction $\mathbf{B}\parallel[1\overline{1}0]$ in the form of a color map. The fitted positions of resonances~A and C are shown with red diamonds and black pentagon symbols, respectively. One can see that the energy of both resonances shifts upwards with increasing field with approximately the same slope. The expected position of resonance~B, according to ESR results, is shown with a dashed gray line. In the low-field region within phase~III,\index{CeB$_6$!phase III} an arrow marks the previously reported feature at twice the AFM charge gap~\cite{FriemelJang15}.\index{charge gap!in CeB$_6$} The behavior of the resonance for the field direction $\mathbf{B}\parallel[001]$ is shown for comparison in Fig.\,\ref{FigInosov:CeB6SummaryCMap}\,(d), exhibiting significantly different behavior. At first, in the low-field region, the magnetic field does not soften the energy to zero with the suppression of the AFM order parameter, and its position does not change until the transition to the AFQ phase happens. Then, at higher magnetic fields, the differences become even more apparent. We observe three spectral lines for $\mathbf{B}\parallel[001]$, which shift as a function of field with different slopes.

\begin{figure}[b!]
\centerline{\includegraphics[width=0.8\textwidth]{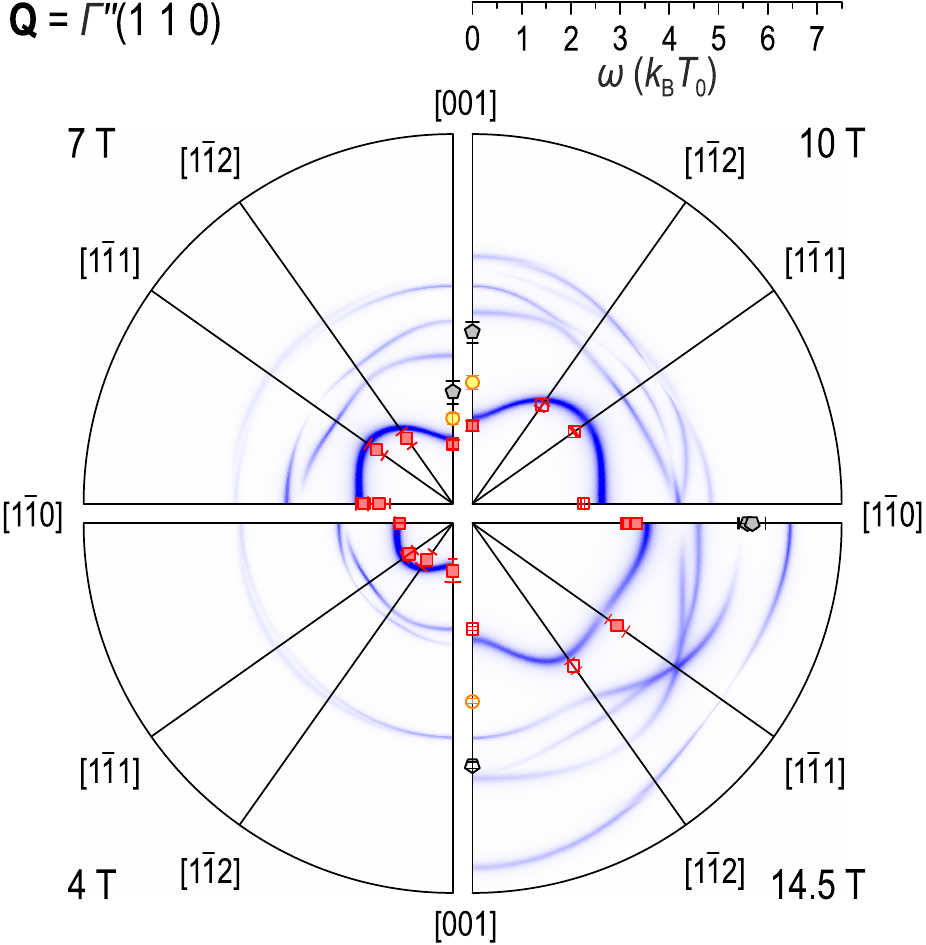}}
\caption{Multipolar excitation branches for the $\Gamma''(110)$ point in polar representation, where the radial coordinate corresponds to the excitation energy, plotted from 0 to $7.5 T_0$, and the angular coordinate to the magnetic field direction, continuously rotating from $[001]$ to $[1\overline{1}0]$ in the plane orthogonal to $[110]$. The four segments of the figure show results for 4, 7, 10, and 14.5\,T. The color map shows the results of the calculation, with the experimental data points along high-symmetry field directions overlayed for comparison. The symbol shapes for different modes are consistent with those in Figs.~\ref{FigInosov:CeB6cmap110ThalesFLEXX14p5T},~\ref{FigInosov:CeB6Hdep001ThalesMap}\,(b)~and~\ref{FigInosov:CeB6SummaryCMap}. Filled symbols correspond to direct measurements of peak position, whereas open symbols are the results of interpolation from the field dependencies plotted in Fig.~\ref{FigInosov:CeB6SummaryCMap}. Reproduced from Ref.~\cite{PortnichenkoAkbari20}.\index{multipolar excitations!field-angular dependence}\index{$g$-factor anisotropy}
\vspace{-2pt}}
\label{FigInosov:CeB6EnergyFieldAngleExperiment}
\end{figure}

The observed excitations have a very clear anisotropy with respect to the field direction. In order to understand this behavior and to determine how the transition from two to three excitations occurs in between the limiting cases when the magnetic field is applied parallel to $[1\overline{1}0]$ or $[001]$ crystallographic axis, we also did systematic measurements of the excitation spectra as a function of magnetic field for several intermediate field directions. Measurements were done with the magnetic field applied parallel to $[1\overline{1}1]$ and $[1\overline{1}2]$ directions, as shown in Fig.\ref{FigInosov:CeB6SummaryCMap}\,(b) and (c), respectively. They demonstrate that the deviation of the magnetic-field direction by approximately 35$^{\circ}$ with respect to $[1\overline{1}0]$ or $[001]$ is sufficient to change the number of observed excitations. In both cases, only one resonance can be clearly identified, and to exclude the possibility of missing additional resonances, all scans were done over a broad energy range.

This example clearly demonstrates that INS measurements in a rotating field offer additional information about the collective modes. Polar field-angle anisotropy plots of multipolar excitations in CeB$_6$ at high-symmetry points~\cite{PortnichenkoAkbari20} represent a large set of information on mode positions and intensities that may be very useful for comparison with our results. If the model so far accepted for CeB$_6$ is reasonable, some features of the field-anisotropy plots described by Thalmeier~\textit{et~al.} should be identified in the INS experiment. Such a comparison is presented in Fig.~\ref{FigInosov:CeB6EnergyFieldAngleExperiment}.

A reliable comparison with experiment is so far only possible at the $\Gamma$ point,\index{CeB$_6$!ferromagnetic resonance}\index{CeB$_6$!zone-center excitations} where detailed field-angular dependences\index{multipolar excitations!field-angular dependence}\index{$g$-factor anisotropy} of the INS spectra were obtained. In Fig.~\ref{FigInosov:CeB6EnergyFieldAngleExperiment}, the corresponding experimental peak positions for the fields of 4, 7, 10, and 14.5\,T are plotted on top of the calculated field-angular polar maps for the same magnetic fields. Every segment of the plot covers the same irreducible 90$^\circ$ range of field directions between $[001]$ and $[1\overline{1}0]$ in the plane orthogonal to $[110]$. The experimental data points were obtained either by directly fitting the peak positions from measurements at the corresponding field values (closed symbols) or via linear interpolation of the field dependences (open symbols).

We see a remarkably good agreement of the calculation with the field-directional anisotropy of the most intense low-energy mode (resonance~A) with a quasi-linear field dependence, which has been followed experimentally over a large field range for all four high-symmetry directions of the magnetic field: $\mathbf{B}\parallel[001]$, $[1\overline{1}2]$, $[1\overline{1}1]$, and $[1\overline{1}0]$. At high magnetic fields of the order of 10\,T, we observe a very considerable anisotropy of $\sim$60\% in the effective $g$-factor between its minimal and maximal values reached for the $[001]$ and $[1\overline{1}1]$ field directions, respectively. For lower fields, however, the anisotropy is reduced, which is a direct consequence of the initially nonlinear field dependence of the low-energy mode. Indeed, the $g$-factor anisotropy reported from previous ESR measurements \cite{SemenoGilmanov16, SemenoGilmanov17, Gilmanov19}, which were performed at magnetic fields of $\sim$\,3\,T, is several times smaller and constitutes a relative change of less than 10\% between the [100] and $[111]$ field directions. However, the overall shape of the anisotropic $g$-factor dependence turns out to be the same in ESR and INS measurements and agrees well with the results of model calculations by Schlottmann~\cite{Schlottmann12, Schlottmann13, Schlottmann18}.

Figure~\ref{FigInosov:CeB6EnergyFieldAngleExperiment} also shows the positions of the new resonance peaks that have been so far measured only for $\mathbf{B}\parallel[1\overline{1}0]$ and $\mathbf{B}\parallel[001]$ in high magnetic fields. Their energies fall in the range where the theoretical model predicts multiple modes resulting from the hybridization of high- and low-energy excitations. With the available data, it is not possible to assign one of these modes uniquely to the experimentally observed resonances. The model also predicts no additional resonances at the $\Gamma$ point\index{CeB$_6$!ferromagnetic resonance}\index{CeB$_6$!zone-center excitations} below the energy of resonance~A, which means that it is unable to explain the appearance of the resonance~B in ESR data at fields above 12\,T~\cite{DemishevSemeno08}. Theoretical considerations about the possible origin of this high-field resonance were proposed earlier~\cite{Schlottmann12, Schlottmann13, Schlottmann18}, but the reason why it cannot be seen in INS measurements remains unclear.\enlargethispage{2pt}

The limited agreement of theory and experiment is not surprising due to several reasons. First of all, to minimize the number of adjustable parameters in the theoretical models, they have so far been restricted to only nearest-neighbor interactions among the multipoles, which is not really justified. Generalized RKKY interactions\index{RKKY interaction|(} that are mediated by the conduction electrons are expected to be long-range with an oscillatory character in direct space. Recent calculations of the effective RKKY-type exchange terms between different types of multipoles in CeB$_6$~\cite{YamadaHanzawa19, HanzawaYamada19} demonstrate that second-nearest-neighbor contributions are always stronger than the nearest-neighbor ones, and even third-nearest-neighbor terms are not negligible. Another difficulty in comparing the results of neutron scattering experiments with the theoretical models is that the calculated inelastic magnetic response is limited to the dipolar response function, whereas neutron scattering is sensitive to all odd-rank magnetic multipoles.
\index{CeB$_6$!magnetic excitations!field-angular anisotropy|)}
\index{CeB$_6$!multipolar excitations|)}\index{multipolar excitations!in CeB$_6$|)}

\vspace{-3pt}\section{Spin dynamics in Ce$_{\text{1}-x}$La$_x$B$_\text{6}$ and Ce$_{\text{1}-x}$Nd$_x$B$_\text{6}$}
\index{diffuse neutron scattering|(}\index{quasielastic magnetic scattering|(}\index{neutron scattering!quasielastic|(}\index{paramagnon fluctuations|(}
\index{Ce$_{1-x}$La$_x$B$_6$!spin dynamics|(}\index{Ce$_{1-x}$Nd$_x$B$_6$!spin dynamics|(}
\index{nesting|(}\index{Fermi surface!nesting properties|(}\index{CeB$_6$!La doped|(}
\index{Ce$_{1-x}$La$_x$B$_6$!Fermi-surface nesting|(}\index{Ce$_{1-x}$Nd$_x$B$_6$!Fermi-surface nesting|(}
\index{Ce$_{1-x}$La$_x$B$_6$!magnetic excitations|(}\index{magnetic excitations!in Ce$_{1-x}$La$_x$B$_6$|(}
\label{Ino_Sec:DopedCeB6}

\subsection{The influence of La and Nd substitution on the electronic structure}

In the previous sections, we mainly focused our attention on the collective multipolar excitations and their field dependencies.\index{CeB$_6$!multipolar excitations}\index{multipolar excitations!in CeB$_6$} The resonant mode at the $R$ point\index{CeB$_6$!$R$-point resonant mode}\index{CeB$_6$!magnetic resonant mode} with a spin-gap of $\sim$0.5~meV,\index{CeB$_6$!spin gap}\index{spin gap!in CeB$_6$} discussed in Sec.~\ref{Ino_SubSec:CollectiveExcitations}, seriously questioned the validity of the established localized approach~\cite{Ohkawa85, ShiinaShiba97, ShiinaSakai98} to the description of spin excitations in CeB$_6$, at least in low magnetic fields within phase~III.\index{CeB$_6$!phase III} An alternative theory which considers the spin dynamics of itinerant heavy quasiparticles, developed by Akbari and Thalmeier~\cite{AkbariThalmeier12}, was able to reproduce the above-mentioned resonance as well as its momentum dependence on the qualitative level. However, this theory is not based on the real band structure of CeB$_6$ and overlooks the intense collective mode at the zone center (ferromagnetic resonance\index{CeB$_6$!ferromagnetic resonance} at the $\Gamma$ point)\index{CeB$_6$!ferromagnetic resonance}\index{CeB$_6$!zone-center excitations} that dominates the magnetic excitation spectrum of CeB$_6$.

The itinerant character of magnetic excitations is further supported by a good agreement between the Lindhard susceptibility,\index{Lindhard function} calculated from the measured electronic structure, and the distribution of quasielastic magnetic intensity measured by INS~\cite{KoitzschHeming16}. According to discussion in Sec.~\ref{Ino_SubSec:QuasielasticMagneticScattering}, itinerant electrons determine the RKKY interactions between Ce~$4f$ multipolar moments, which can be still considered local, and the propagation vector of the AFQ order is determined by the geometry of the Fermi surface and its nesting vectors.

Despite the fact that the angle-resolved photoelectron spectroscopy is the most direct experimental probe of the Fermi surface (unlike macroscopic measurements of quantum oscillations which can be used for an indirect reconstruction of the Fermi-surface geometry), in order to extract the nesting vectors of the Fermi surface from ARPES data,\index{angle-resolved photoemission}\index{photoelectron spectroscopy} a technically demanding procedure has to be performed, as the whole low-energy band structure has to be first fitted to a tight-binding model. This method is, therefore, rather indirect and can be to some extent model-dependent~\cite{NikitinPortnichenko18}.

\begin{figure}[b!]
\centerline{\psfig{file=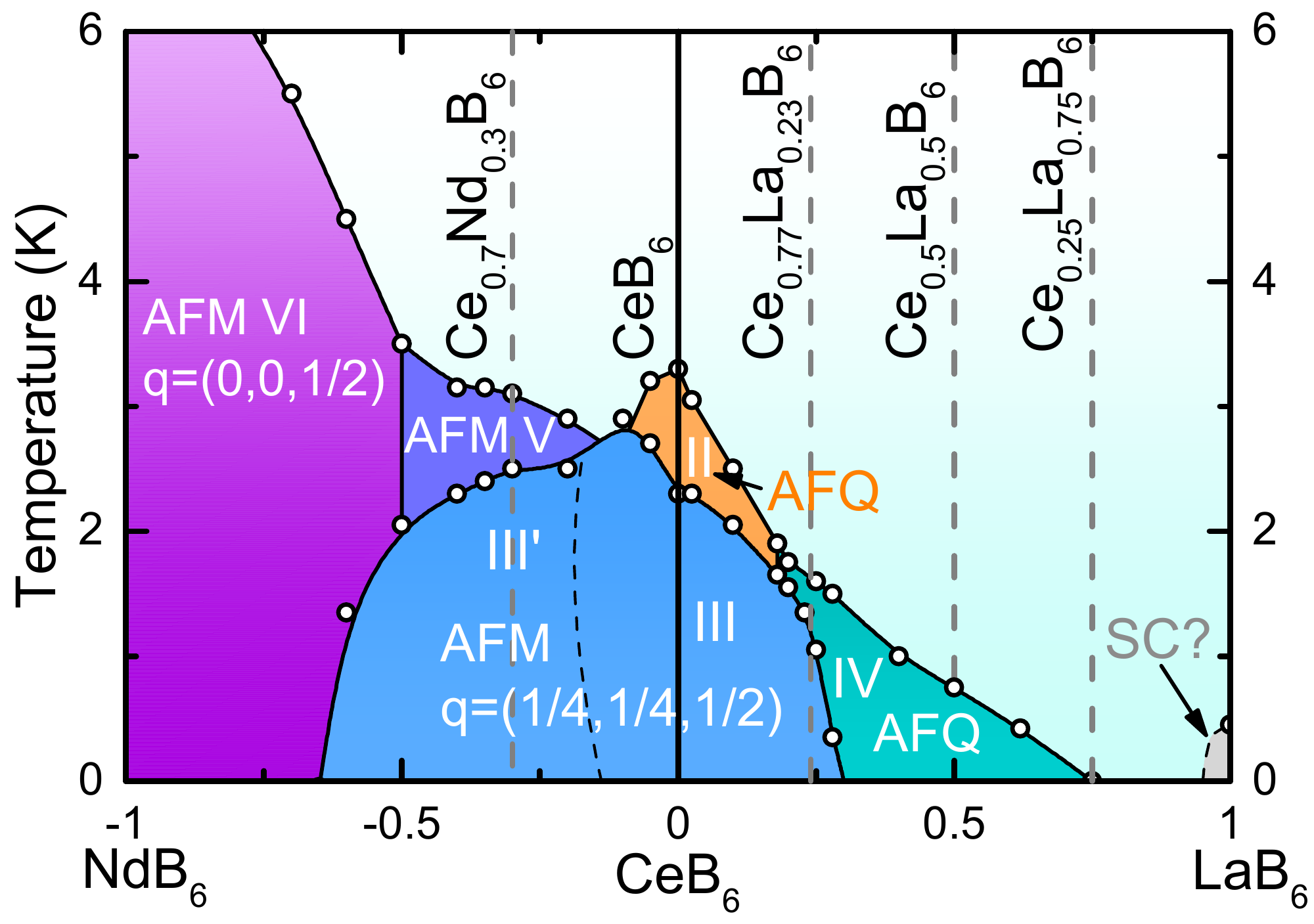, width=0.85\textwidth}}
\caption{Schematic magnetic phase diagram of the solid solutions Ce$_{1-x}R_x$B$_6$ ($R$~=~La, Nd) after Refs.~\cite{SeraSato87, TayamaSakakibara97, KobayashiSera00, KobayashiYoshino03, YoshinoKobayashi04, SchenckGygax07, MignotRobert09, KunimoriTanida10, FriemelJang15, CameronFriemel16} at zero field. Phases II and IV are associated with the two multipolar phases with AFQ and octupolar ordering.\index{magnetic octupoles} The ``SC'' dome at the bottom-right corner schematically indicates the dubious superconducting phase of the pure LaB$_6$~\cite{MatthiasGeballe68, ArkoCrabtree75, BatkoBatkova95, CameronFriemel16}. Phases III, V and VI are three different types of AFM ordering. Vertical gray lines indicate the available sample compositions. Reproduced from Nikitin \textit{et al.}~\cite{NikitinPortnichenko18}.\vspace{-2pt}
\index{Ce$_{1-x}$La$_x$B$_6$!magnetic phase diagram}\index{Ce$_{1-x}$Nd$_x$B$_6$!magnetic phase diagram}\index{magnetic phase diagram!of Ce$_{1-x}$La$_x$B$_6$}\index{magnetic phase diagram!of Ce$_{1-x}$Nd$_x$B$_6$}
\index{CeB$_6$!phase I}\index{CeB$_6$!phase II}\index{CeB$_6$!phase III}\index{Ce$_{1-x}$La$_x$B$_6$!phase II}\index{Ce$_{1-x}$La$_x$B$_6$!phase III}\index{Ce$_{1-x}$Nd$_x$B$_6$!phase III}
\index{Ce$_{1-x}$Nd$_x$B$_6$!antiferromagnetic order}\index{antiferromagnetic order!in Ce$_{1-x}$Nd$_x$B$_6$}
\index{Ce$_{1-x}$La$_x$B$_6$!phase IV}\index{Ce$_{1-x}$Nd$_x$B$_6$!phase V}\index{Ce$_{1-x}$Nd$_x$B$_6$!phase VI}}
\label{FigInosov:CeLaNdB6PhaseDiagram}
\end{figure}

Earlier works~\cite{StockertFaulhaber04, InosovEvtushinsky09, ButchManley15} on magnetic heavy-fermion metals demonstrate that the low-energy dynamic spin susceptibility $\chi(\omega,\mathbf{Q})$,\index{dynamic spin susceptibility} measured with diffuse neutron scattering, provides direct information about the nesting vectors. Unlike ARPES,\index{angle-resolved photoemission}\index{photoelectron spectroscopy} in which momentum measured perpendicular to the cleavage plane depends on the photon energy, neutron scattering is capable of probing energy-momentum space without any restrictions. This advantage of INS combined with the bulk sensitivity are of particular importance for materials with highly 3-dimensional band structures.

In Ce$_{\text{1}-x}R_x$B$_\text{6}$ ($R$~=~La, Nd) solid solutions, the Ce sublattice is randomly diluted by nonmagnetic La ion or another magnetic Nd$^{3+}$ rare-earth ion, adding an extra dimension to the parameter space of the phase diagram, as shown in Fig.\,\ref{FigInosov:CeLaNdB6PhaseDiagram}. Suppression of both AFM and AFQ phases, as well as the appearance of a new ordered phase~IV\index{Ce$_{1-x}$La$_x$B$_6$!phase IV} upon La-doping of CeB$_6$ was already discussed in Sec.~\ref{Ino_SubSec:CeB6Introduction}. In the case of Nd doping, concentration as low as $\sim$10\% suppresses the AFQ phase at zero field. At the same time, a new phase~V emerges, in which the AFM propagation vector becomes slightly incommensurate. At $x\approx0.5$, the order finally changes to conventional AFM stacking of ferromagnetic layers with the ordering wave vector $\mathbf{q}_0=(00\frac{1}{2})$, like in the pure NdB$_6$~\cite{KobayashiYoshino03, YoshinoKobayashi04, MignotRobert09}.

It was demonstrated that both La and Nd doping of CeB$_6$ has a notable influence on the Fermi-surface geometry~\cite{ArkoCrabtree76, OnukiUmezawa89, GoodrichHarrison99, NeupaneAlidoust15}. The shape of the Fermi-surface, the effective mass of charge carriers, and the number of conduction electrons per unit cell are very similar for both NdB$_6$ and LaB$_6$ due to a strong localization of $4f$ electrons in NdB$_6$~\cite{OnukiUmezawa89, ArkoCrabtree76}. On the other hand, hybridization of Ce~$4f^1$ electrons with the conduction band qualitatively modifies the Fermi surface of CeB$_6$ as compared to LaB$_6$~\cite{NeupaneAlidoust15}. Therefore, both La and Nd doping of CeB$_6$ do not simply change the number or magnitude of localized $4f$ magnetic moments, but also induce an effective hole doping, decreasing the number of conduction electrons and modifying the Fermi surface geometry. In order to verify, to what extent the nesting instabilities\index{Fermi surface!instabilities} of the Fermi surface predetermine the order parameters of AFM and AFQ phases in these compounds, we can make use of the available single crystals of La and Nd-doped Ce$_{1-x}R_x$B$_6$ ($R$~=~La,\,Nd) solid solutions and demonstrate that the doping-induced changes in the electronic structure correlate with the changes of the ordered phases for compounds with various substitution levels. We already confirmed that diffuse neutron scattering may be used as a complementary method for probing the electronic structure, therefore in the next section we will discuss the experimentally observed redistribution of the diffuse magnetic spectral weight across the Brillouin zone upon La and Nd doping, which we associate with the changes in the Fermi-surface nesting properties related to the modified charge-carrier concentration.

\subsection{Momentum-space structure of the diffuse spin fluctuations}\label{Ino_SubSec:CeLaNdB6MomentumDependence}
\index{quasielastic magnetic scattering!momentum dependence|(}

According to the linear response theory for an electron gas, the imaginary part of the dynamic spin susceptibility $\chi(\mathbf{q},\omega)$\index{dynamic spin susceptibility} is proportional to the Lindhard function\index{Lindhard function}
\begin{equation}
\chi_\mathbf{q}=\sum_{\mathbf{k}}\frac{n_\text{F}(\epsilon_{\mathbf{k}+\mathbf{q}})-n_\text{F}(\epsilon_{\mathbf{k}})}{\epsilon_{\mathbf{k}}-\epsilon_{\mathbf{k}+\mathbf{q}}},
\end{equation}
where $\epsilon(\mathbf{k})$ is the dispersion relation for the conduction electrons, and $n(\epsilon)$ is the Fermi function. The Lindhard function contains information about Fermi-surface nesting properties, as its real part at $\omega\rightarrow0$ is peaked at the nesting vectors\index{nesting} and determines the propensity towards Fermi-surface instabilities\index{Fermi surface!instabilities} in charge- or spin-density-wave systems~\cite{ChanHeine73, Fawcett88, BorisenkoKordyuk08}. It also enters the expression for the oscillatory RKKY interaction\index{RKKY interaction|)} between localized Kondo spins in metals, which is mediated by the conduction electrons over long distances~\cite{RudermanKittel54, Kasuya56, Yafet87, KimLee96, Aristov97, LitvinovDugaev98}. Therefore, when localized magnetic impurities are added to a nonmagnetic metal, they tend to develop short-range dynamic correlations that are seen as quasielastic magnetic scattering (QEMS) in neutron spectroscopy\index{diffuse neutron scattering}\index{quasielastic magnetic scattering}\index{neutron scattering!quasielastic} or even lead to a long-range magnetic ordering of the impurity spins~\cite{BrownCaudron85, FretwellDugdale99}. The QEMS intensity can therefore develop maxima at the Fermi-surface nesting vectors in $\mathbf{Q}$ space even in dilute Kondo alloys that are far from any ordering instability.

\begin{figure}[t!]\vspace{-5pt}
\centerline{\includegraphics[width=0.75\textwidth]{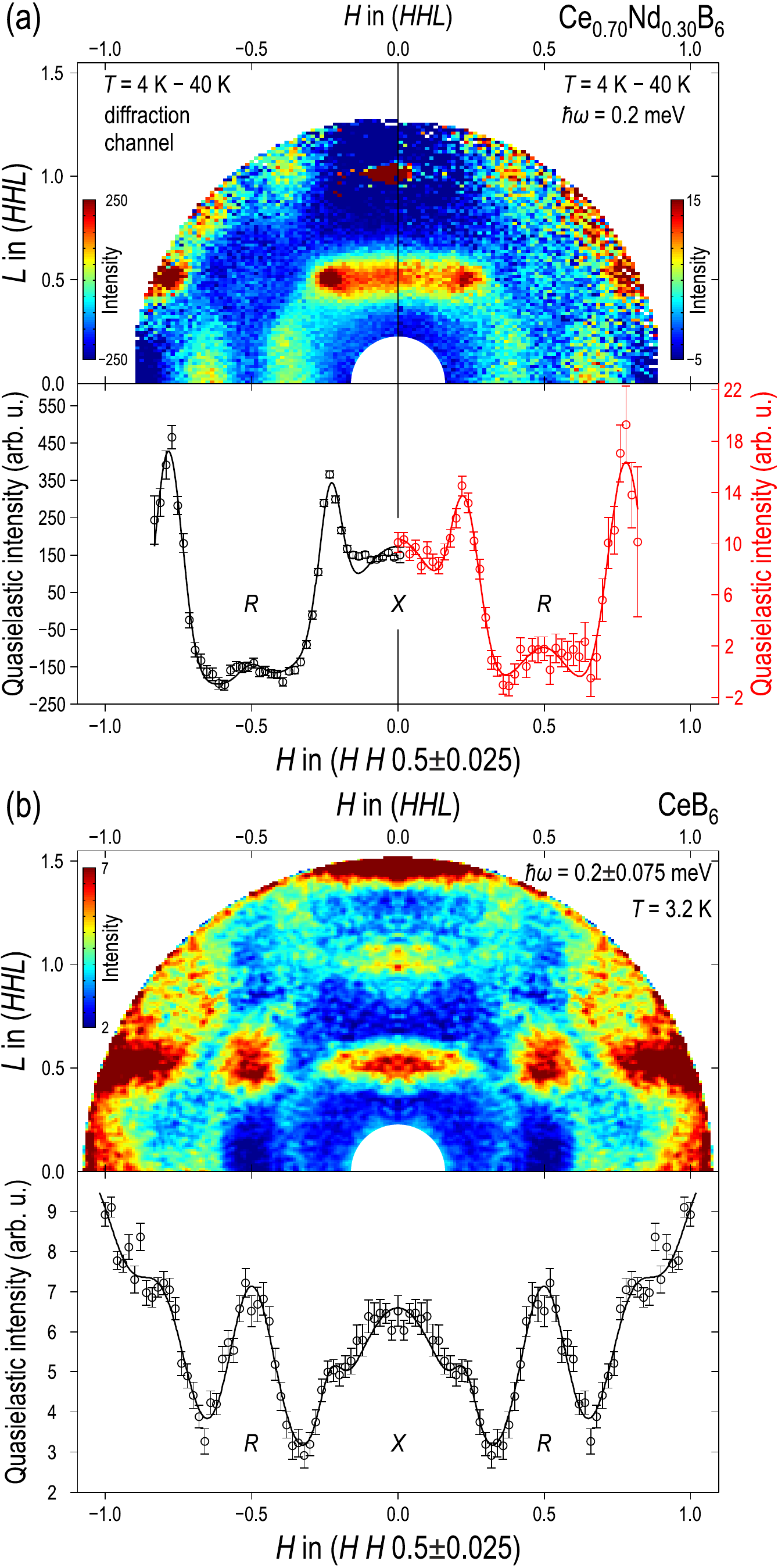}}
\caption{Summary of INS results, measured on Ce$_{1-x}$Nd$_x$B$_6$ with $x=0.3$ (top) and 0.3 (bottom).
(a)~Background-subtracted INS intensity measured on Ce$_\text{0.7}$Nd$_\text{0.3}$B$_\text{6}$ at an energy transfer $\Delta{}E=0.2$~meV and in the diffraction channel.
(b)~INS results for the parent compound CeB$_\text{6}$ measured at $T=3.2$~K. Reproduced from Nikitin \textit{et al.}~\cite{NikitinPortnichenko18}.\index{diffuse neutron scattering}\index{quasielastic magnetic scattering}\index{neutron scattering!quasielastic}
\vspace{-4em}}
\label{FigInosov:CeNdB6_QEMSmap}
\end{figure}

\begin{figure}[t!]
\centerline{\includegraphics[width=0.75\textwidth]{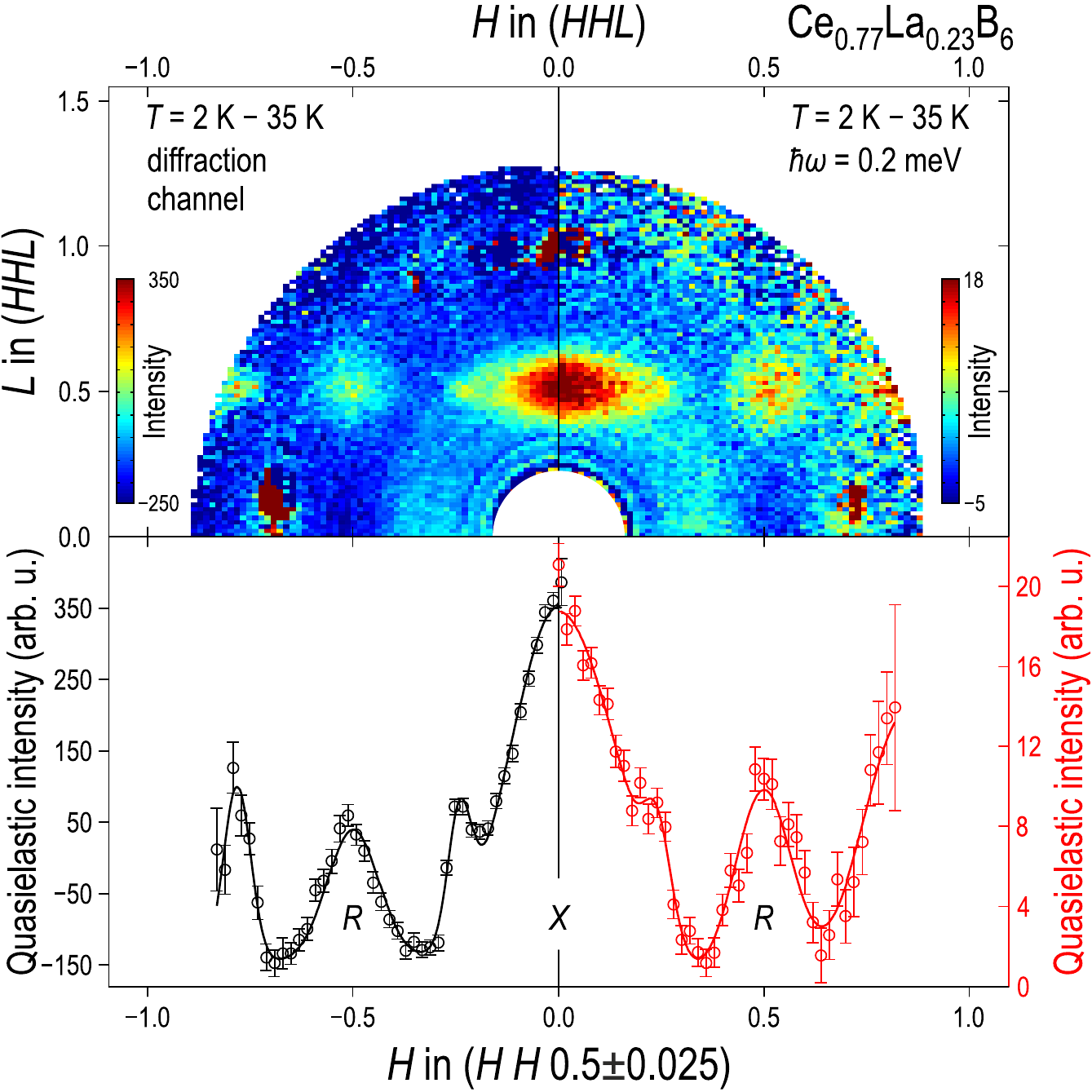}}
\caption{The background-subtracted data for the Ce$_{0.77}$La$_{0.23}$B$_6$ sample, measured and presented in the same way as the data in Fig.\,\ref{FigInosov:CeNdB6_QEMSmap}. Reproduced from Nikitin \textit{et al.}~\cite{NikitinPortnichenko18}.\index{diffuse neutron scattering}\index{quasielastic magnetic scattering}\index{neutron scattering!quasielastic}}
\label{FigInosov:Ce0p77La0p23B6_QEMSmap}
\end{figure}

\begin{figure}[t!]
\centerline{\includegraphics[width=0.75\textwidth]{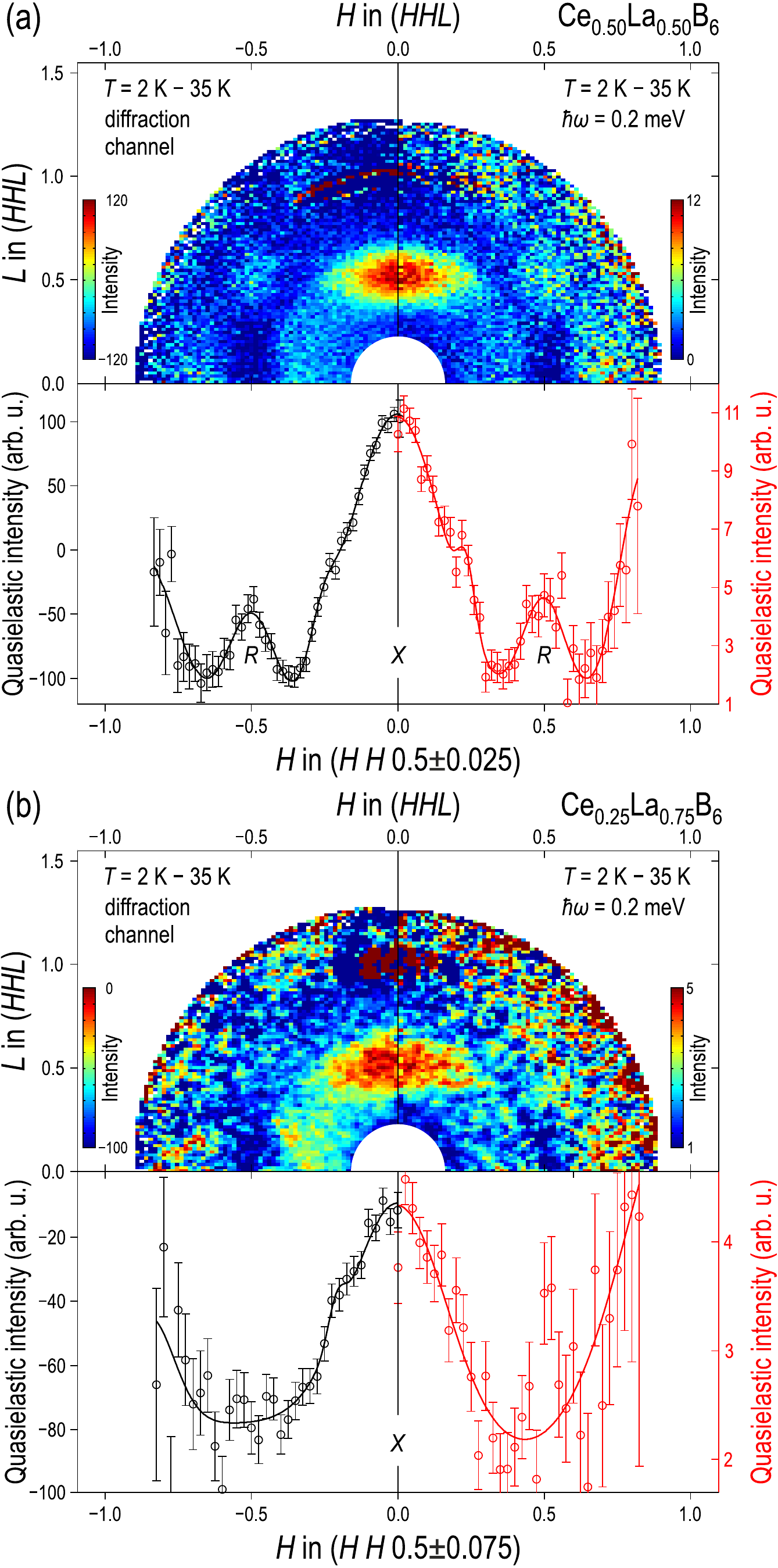}}
\caption{The background-subtracted data for the two samples of (a)~Ce$_{0.5}$La$_{0.5}$B$_6$ and (b)~Ce$_{0.25}$La$_{0.75}$B$_6$, measured and presented in the same way as the data in Fig.\,\ref{FigInosov:CeNdB6_QEMSmap}. Reproduced from Nikitin \textit{et al.}~\cite{NikitinPortnichenko18}.\index{diffuse neutron scattering}\index{quasielastic magnetic scattering}\index{neutron scattering!quasielastic}
\vspace{-1em}}
\label{FigInosov:CeLaB6_QEMSmap}
\end{figure}

In order to follow the evolution of the Fermi-surface nesting properties upon doping, observed as enhanced QEMS intensity at the above-mentioned points, we performed detailed mapping of the QEMS intensity distribution~\cite{NikitinPortnichenko18}, and the results of our INS measurements are summarized in Figs.\,\ref{FigInosov:CeNdB6_QEMSmap}--\ref{FigInosov:CeLaB6_QEMSmap}.

However, before we move further to the discussion of the results, it is necessary to briefly explain several technical details. Despite that fact that the quasielastic magnetic line has its maximum of intensity near zero energy transfer, in order to map out the $\mathbf{Q}$ dependence of QEMS intensity one usually performs measurements at nonzero energy transfer. This is a common procedure, as the quasielastic line is usually broader than the energy resolution \cite{Robinson00}, but at the position where the maximum of intensity is observed, a strong nonmagnetic background from the incoherent elastic line complicates the measurements. This approach has been successfully applied in many earlier studies of $f\!$-electron compounds \cite{Rossat-MignodRegnault88, RegnaultErkelens88a, SchroederAeppli98, StockertLoehneysen98, KadowakiSato04, StockSokolov11, SinghThamizhavel11, KimuraNakatsuji13}. Here, because of the specifics of the used spectrometer, in addition to the spectroscopic measurements at a preset value of energy transfer, an energy-integrated scattering over the full energy range (except for the preset value of the spectroscopic channel) is measured in parallel. This allows us to collect energy-integrated data in parallel to any spectroscopic measurement at no extra cost in acquisition time.

In our measurements, we chose an energy transfer of 0.2~meV for the spectroscopic channel, which allowed us to map out the QEMS intensity just above the incoherent elastic line. The second channel, which in this case is called diffraction, is not restricted to elastic scattering, but integrates over all neutron energies, thus the amplitude of the magnetic signal in this channel is strongly enhanced due to the broad width of the quasielastic Lorentzian peak as compared to the elastic line. To distinguish the diffuse magnetic signal from nonmagnetic background scattering on the sample and cryogenic environment, we mapped out the same area in momentum space at an elevated temperature and used the high-temperature datasets as background. Since the magnetic intensity in the diffraction channel is strongly dominated by inelastic scattering, whereas the background comes predominantly from the incoherent elastic line, after appropriate subtraction of the high-temperature background in each channel we could obtain clean momentum-space distributions of the magnetic intensity also for the diffraction channel.\footnote{Details of this procedure as well as the comparison of the signal-to-noise ratio in each channel can be found in Ref.~\cite{NikitinPortnichenko18}.}

Now we start with the discussion of the Nd-substituted compound. It is known that pure NdB$_6$\index{NdB$_6$} develops AFM order with the propagation vector $(\frac{1}{2}\,0\,0)$, which coincides with the $X$ point~\cite{MignotRobert09}. According to the phase diagram shown in Fig.~\ref{FigInosov:CeLaNdB6PhaseDiagram}, this phase persists down to $x\approx0.5$. One would therefore expect that at high Nd concentrations, critical paramagnon fluctuations\index{paramagnon fluctuations} of phase~VI are expected in Ce$_{1-x}$Nd$_x$B$_6$ around the $X$ point in momentum space above $T_{\rm N}$. Here we are interested in looking at an intermediate Nd concentration, shortly before reaching phase~VI, to see how the spectral-weight transfer to the $X$ point takes place. It is natural to expect a strong rise of intensity at the $X$ point with a simultaneous suppression of excitations at the $\Gamma$ and $R$ points\index{CeB$_6$!$R$-point resonant mode}\index{CeB$_6$!magnetic resonant mode} in Ce$_\text{0.7}$Nd$_\text{0.3}$B$_\text{6}$, which is indeed confirmed by the measurements in Fig.\,\ref{FigInosov:CeNdB6_QEMSmap}\,(a). The color map shows the $\mathbf{Q}$-dependence of QEMS intensity within the $(HHL)$ plane. The plot below it shows cuts along the $(HH\frac{1}{2})$ direction, obtained by integration within $\pm 0.025$ along the $L$ axis. The left and right parts of the figure show the data obtained in the diffraction and spectroscopic channels, respectively. For comparison, in Fig.\,\ref{FigInosov:CeNdB6_QEMSmap}\,(b) we show the reference measurement on pure CeB$_6$, obtained by integrating the $T=2.6$~K TOF data (same as in Figs.~\ref{FigInosov:CeB6BdepGamma_IN5} and \ref{FigInosov:CeB6PhotoTOM}) in the $[0.125,0.275]$~meV energy window (without subtraction of high-temperature background), symmetrized with respect to the vertical axis.

The QEMS map measured on the pure CeB$_6$ has several distinct local maxima in the following high-symmetry points within the Brillouin zone in pure CeB$_6$: the ferromagnetic $\Gamma$ point,\index{CeB$_6$!ferromagnetic resonance}\index{CeB$_6$!zone-center excitations} the AFQ propagation vector $R(\frac{1}{2}\,\frac{1}{2}\,\frac{1}{2})$ at the corner of the cubic Brillouin zone. Also, one can see a large elliptical hump around the $X$-point, where no static order is realized, that connects additional weaker peaks at the AFM wavevectors $\mathbf{q}_1=(\pm\!\frac{1}{4}\,\pm\!\frac{1}{4}\,\frac{1}{2})$, seen as a central maximum with two shoulders at the bottom of Fig.~\ref{FigInosov:CeNdB6_QEMSmap}\,(b). Upon Nd doping, one can see a strong reduction of magnetic intensity at the $R$ point,\index{CeB$_6$!$R$-point resonant mode}\index{CeB$_6$!magnetic resonant mode} which is consistent with the rapid suppression of the AFQ phase by Nd. Instead, we find a narrow diffuse peak, centered at the $X$ point, that connects the strong intensity maxima at the equivalent AFM wavevectors, $\mathbf{q}_1=(\pm\!\frac{1}{4}\,\pm\!\frac{1}{4}\,\frac{1}{2})$, and an elongated broad peak extending along $(\frac{1}{3}\frac{1}{3}L)$. The presence of extended peaks in momentum space with multiple local maxima of QEMS intensity is a signature of itinerant frustration\index{itinerant frustration}\index{frustration!itinerant} in this system, which can explain the proximity of multiple AFM phases in a small region of the phase diagram.

Since it is known that both La and Nd substitutions lead to an effective hole doping and should therefore cause similar changes of the Fermi surface~\cite{OnukiUmezawa89, ArkoCrabtree76}, one would expect similar changes in the QEMS intensity distribution on the La-doped side as in the case of Nd substitution. Indeed, Fig.~\ref{FigInosov:Ce0p77La0p23B6_QEMSmap} shows that 23\% La doping strongly suppresses the inelastic-scattering intensity at the zone center and at the AFM wavevectors $\mathbf{q}_1=(\frac{1}{4}\frac{1}{4}\frac{1}{2})$ and $\mathbf{q}_2=(\frac{1}{4}\frac{1}{4}0)$. Most of the spectral weight is now accumulated around the $X$ point. Figure\,\ref{FigInosov:CeLaB6_QEMSmap} shows that in samples with even higher La concentrations of 50\% and 75\%, the elliptical feature at the $X$ point dominates the QEMS intensity distribution. In spite of the overall decrease in magnetic intensity as expected for the nonmagnetic dilution with La, the relative accumulated spectral weight at the $X$ point (per mole Ce) goes up. The peak at the $R$ point\index{CeB$_6$!$R$-point resonant mode}\index{CeB$_6$!magnetic resonant mode} can be clearly seen up to a rather high La concentration of $x=0.5$, but gets fully suppressed in the most diluted Ce$_{0.25}$La$_{0.75}$B$_6$ sample.

We see that the substitution of Nd for Ce has a dual effect on the system. First, as already mentioned, it reduces the 4f\,--\,5d hybridization and shrinks the electron-like Fermi surfaces, that is equivalent to an effective hole doping. Second, it introduces large magnetic moments of Nd$^{3+}$ into the system, increasing its propensity towards magnetic ordering, which is an opposite effect to the nonmagnetic La$^{3+}$ dilution of the Ce$^{3+}$ moments. As follows from our results in Figs.\,\ref{FigInosov:CeNdB6_QEMSmap}--\ref{FigInosov:CeLaB6_QEMSmap}, the evolution of the Fermi-surface nesting properties in both systems is similar, leading to an enhanced QEMS intensity near the $X$ point at the expense of the suppressed peak at the $R$ point.\index{CeB$_6$!$R$-point resonant mode}\index{CeB$_6$!magnetic resonant mode} The corresponding fluctuations are observed above $T_{\rm N}$ as extended diffuse peaks with several local maxima at the corresponding wave vectors. However, dilution of the magnetic moments on the La-rich side of the phase diagram prevents the formation of any $(0\,0\,\frac{1}{2})$-type magnetic order despite the presence of nesting, in contrast to Ce$_{\text{1}-x}$Nd$_x$B$_\text{6}$, where three distinct AFM phases compete in the intermediate doping range until one of them (phase~VI) prevails on the Nd-rich side of the phase diagram. In contrast, Ce$_{\text{1}-x}$La$_x$B$_\text{6}$ tends to develop an elusive hidden-order phase~IV\index{Ce$_{1-x}$La$_x$B$_6$!phase IV} that vanishes at high La concentrations~\cite{JangPortnichenko17}.

These results demonstrate how diffuse neutron scattering can be used to probe the nesting vectors in complex $f\!$-electron systems directly, without reference to the single-particle band structure, and emphasize the role of Fermi-surface geometry in stabilizing magnetic order in rare-earth hexaborides.
\index{quasielastic magnetic scattering!momentum dependence|)}
\index{nesting|)}\index{Fermi surface!nesting properties|)}
\index{Ce$_{1-x}$La$_x$B$_6$!Fermi-surface nesting|)}\index{Ce$_{1-x}$Nd$_x$B$_6$!Fermi-surface nesting|)}

\subsection{Temperature dependence of the quasielastic magnetic scattering}
\index{quasielastic magnetic scattering!temperature dependence|(}

\index{Kondo temperature|(}
It is well known that in CeB$_6$, as well as in many other heavy-fermion metals, the half-width of the quasielastic line $\Gamma$ has the following phenomenological temperature dependence:
\begin{equation}\label{EqInosov:GammaSqRoot}
\Gamma(T)=k_\text{B}T_\text{K} + A\sqrt{T},
\end{equation}
where $T_\text{K}$ is the Kondo temperature, and $A$ is an empirical fitting parameter~\cite{LowenhauptFischer93, Robinson00}. This dependence corresponds to the conventional spin relaxation rate typical for the paramagnetic state of most heavy-fermion compounds, such as CeCu$_6$,\index{CeCu$_6$} CeAl$_3$,\index{CeAl$_3$} CePb$_3$,\index{CePb$_3$} CeRu$_2$Si$_2$,\index{CeRu$_2$Si$_2$} CeB$_6$, and many others. From the theory point of view, a $\sqrt{T}$ scaling of the relaxation rate with temperature is expected for a single Kondo impurity in the high-temperature limit \cite{CoxBickers86, BickersCox87}. Despite a long history of examples where this dependence is realized~\cite{Robinson00}, the physical meaning of the parameter $A$ is not yet clear. Also, there have been no systematic studies of the dependence of the neutron-deduced Kondo temperature $T_\text{K}$ and the parameter $A$ on doping.

The Kondo temperature $T_\text{K}$ arises as a parameter when the problem of conduction electrons scattering on localized magnetic impurities is considered, which is relevant to a number of alloys with magnetic impurities and metallic $f$-electron systems (Kondo lattices).\index{Kondo effect} It was found by de Haas~\textit{et~al.}~\cite{HaasBoer34} that with decreasing temperature, electrical resistivity increases logarithmically, while for a conventional metal one would expect a monotonic decrease. It took 30 years to find an appropriate explanation to this long-standing puzzle, until the work carried out in 1964 by Kondo, who showed that the coupling between the conduction electrons and the $f$ electrons can lead to a term proportional to $\ln(T)$ in the low-temperature resistivity~\cite{Kondo64, Hewson97}. The temperature at which this characteristic change in electrical resistivity takes place is referred to as the Kondo temperature. Below it, the impurity and the conduction-electron spins bind very strongly and form an overall nonmagnetic singlet state.

The parent compound CeB$_6$ shows a typical Kondo-type behavior, with a resistivity minimum that starts to increase upon cooling. Although CeB$_6$ represents a classical Kondo lattice, its temperature dependence of resistivity is well described with a model valid for the dilute Kondo state, with the corresponding $T_{\rm K}=5\kern1pt$--10~K~\cite{TakaseKojima80}. This result significantly differs from the Kondo temperature of $\sim$3~K, determined from the lowest-temperature quasielastic line width, measured with INS~\cite{HornSteglich81}. Measurements of the other extreme case, with just a few percent of Ce left after dilution with nonmagnetic La, represent the genuine impurity model. They yield much lower values of $T_\text{K}$ between 1 and 3~K~\cite{Winzer75, SamwerWinzer76, Felsch78}.

\begin{table}[t!]
\tbl{\hspace{-1ex}Comparison of the Kondo temperatures, determined with various methods for the two limiting cases: (i)~the isolated impurity model with very low Ce concentration and (ii)~a dense Kondo lattice.\index{Kondo temperature!neutron-deduced}\index{Ce$_{1-x}$La$_x$B$_6$!Kondo temperature}\smallskip}
{\begin{tabular}{@{~}c@{~~~~~~~~~~~~}c@{~~~~~~~~~~~~}c@{~~~~~~~~~~~~}c@{~}}
\toprule
$x$ in Ce$_{1-x}$La$_{x}$B$_{6}$~(at.\,\%)          & $T_{\text K}$~(K)                 & Method                        & Ref. \\
\midrule
        0                                           &5--10                          & resistivity                   & \cite{TakaseKojima80}   \\
        0                                           &3                              & INS$^\ast$                    & \cite{HornSteglich81}   \\
\midrule
$0 \leq x \leq75$                                   &2                                & resistivity                   & \cite{SatoSumiyama85}   \\
\midrule
97                                                  &1                                & resistivity                   & \cite{SatoSumiyama85}   \\
$97.10 \leq x \leq99.39$                            & $2.8$                             & magnetic susceptibility      & \cite{Felsch78}\\
$97.10 \leq x \leq99.39$                            & $1.05$                            & resistivity                   & \cite{SamwerWinzer76}   \\
$99.89 \leq x \leq99.93$                            & $1.40\pm0.05$                     & magnetic susceptibility      & \cite{Felsch78}\\
$98.80 \leq x \leq99.39$                            & $1.1\pm0.2$                       & resistivity                   & \cite{Winzer75}   \\
\bottomrule
\end{tabular}}
\begin{tabnote}
$^\ast$The INS value from Ref.~\cite{HornSteglich81} refers to the low-temperature ($T=5$\,K) residual half-width $\Gamma_0$ of the quasielastic line (so-called neutron-deduced Kondo temperature~\cite{LowenhauptFischer93}).\vspace{-2pt}
\end{tabnote}
\protect\label{TabInosov:KondoTemperature}
\end{table}

It is clear that in case of a dense arrangement of impurities, their interactions can no longer be neglected. This effect is observed as a doping dependence of the residual resistivity. According to Sato~\textit{et~al.}~\cite{SatoKunii84}, disregarding coherence effects might lead to an exaggerated value of the Kondo temperature, which might be the case in the work of Takase~\textit{et~al.}~\cite{TakaseKojima80}. In order to accurately estimate the Kondo temperature, one should fit the data at higher temperatures, at which coherence effects are less pronounced, and extrapolate them to \mbox{$T \rightarrow 0$}. By a lucky coincidence, the crystal field splitting in CeB$_6$ is large enough, therefore it will not affect the fit results at temperatures below the characteristic temperature of the $\Gamma_{8}$--$\Gamma_{7}$ splitting. Determined in this way, the Kondo temperature from resistivity measurements is claimed to be weakly dependent on La concentration up to a 75\% doping level, with the corresponding value of $T_\text{K}=2$\,K that agrees better with the neutron-deduced Kondo temperature~\cite{HornSteglich81}. Further increase of the La concentration gradually lowers the Kondo temperature to 1~K~\cite{SatoKunii84, SatoSumiyama85}. Kondo temperatures for various La concentrations determined using various techniques are summarized in Table~\ref{TabInosov:KondoTemperature}.

\begin{figure}[t!]\vspace{-2pt}
\centerline{\includegraphics[width=\textwidth]{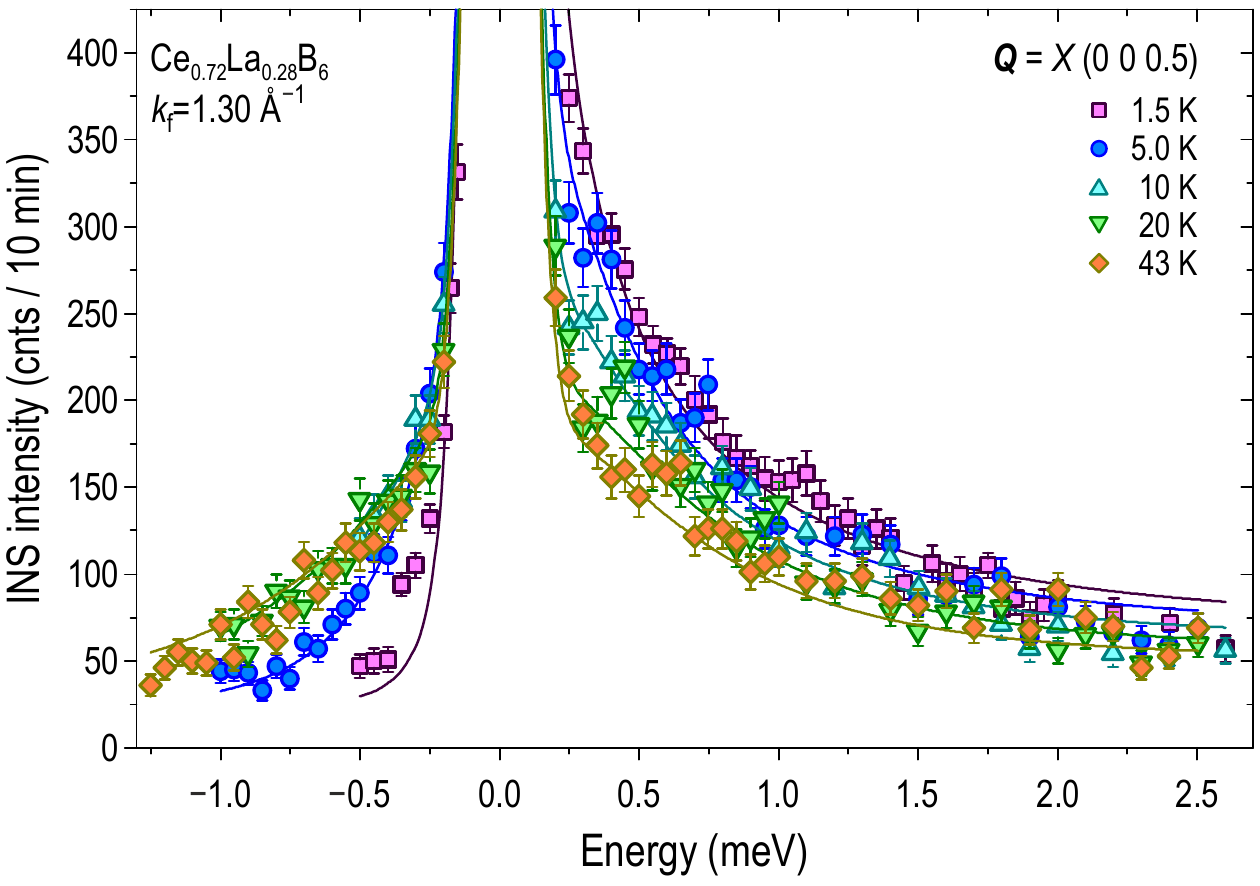}}
\caption{Temperature dependence of the low-energy magnetic scattering for the 28\% doped sample. The solid lines are fits with quasielastic Lorentzian profiles. Reproduced from Ref.~\cite{Portnichenko18}.\vspace{-2pt}}
\label{FigInosov:CeLa0p28B6TdepIN14}
\end{figure}

An alternative method to study the doping dependence of the Kondo energy scale is to measure the half-width of the quasielastic line $\Gamma(T)$ defined by Eq.~(\ref{EqInosov:Quasielastic}) as a function of temperature and extract the neutron-deduced Kondo temperature\index{Kondo temperature!neutron-deduced} $T_{\rm K}=\Gamma_0/k_{\rm B}$ from the residual quasielastic line width $\Gamma_0 = \lim_{T\rightarrow0}\Gamma(T)$~\cite{LowenhauptFischer93}. The availability of the La-doped series of Ce$_{\text{1}-x}$La$_x$B$_\text{6}$ single crystals gives us an opportunity to address these long-standing open questions and compare $T_\text{K}$ determined with neutron scattering with the results summarized in Table~\ref{TabInosov:KondoTemperature}.\enlargethispage{2pt}

\index{Kondo temperature!doping dependence|(}\index{Ce$_{1-x}$La$_x$B$_6$!Kondo temperature|(}
The evolution of the quasielastic magnetic scattering signal with temperature for the sample with the 28\% La doping measured at the $X$ point is shown in Fig.\,\ref{FigInosov:CeLa0p28B6TdepIN14}. Experimental results can be well described by a quasielastic Lorentzian profile of the form given by Eq.~(\ref{EqInosov:Quasielastic})~\cite{GoremychkinOsborn00}. Upon increasing the temperature, we observe a monotonic suppression of the signal, and the line width exhibits a gradual broadening. As will be discussed later, the temperature dependence of the quasielastic line width follows the conventional $\sqrt{T}$ law and is not very different from that for the parent compound.

\begin{figure}[t]
\centerline{\includegraphics[width=\textwidth]{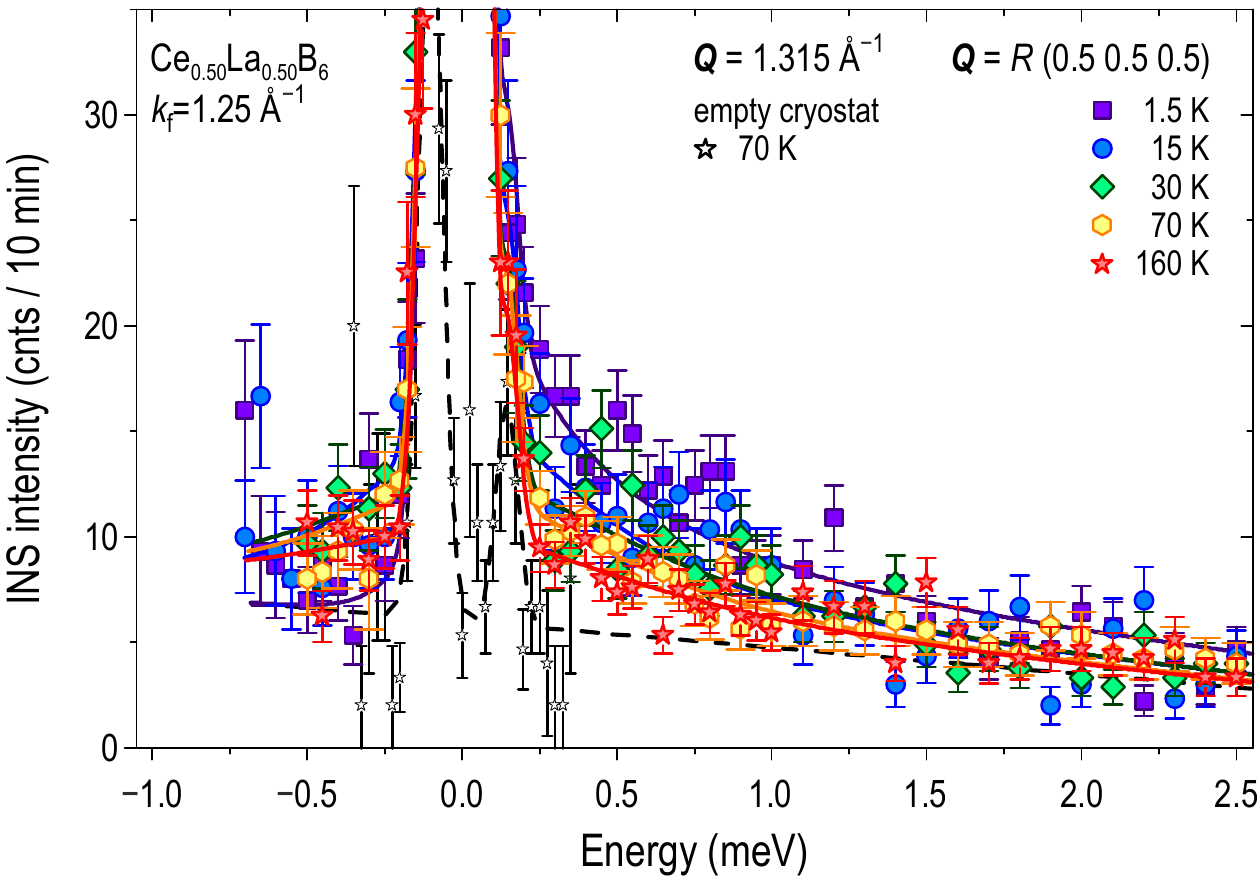}}
\caption{Temperature dependence of the low-energy magnetic scattering for the 50\% doped sample. Background signal from the empty cryostat, shown with black stars, has been considered in the global fit of the data. The solid lines are fits with quasielastic Lorentzian profiles. Reproduced from Ref.~\cite{Portnichenko18}.}
\label{FigInosov:CeLa0p50B6TdepFLEXX}
\end{figure}

The temperature dependence of the quasielastic signal for the 50\% La doping sample was measured at the $R$ point\footnote{According to Fig.\,\ref{FigInosov:CeLaB6_QEMSmap}(a), a certain fraction of the spectral weight is present in the $R$ point, and the residual quasielastic line width is within the error the same for both points~\cite{Portnichenko18}. Therefore, measurements of the line width at the $R$ point would only require longer counting time.} and is shown in Fig.\,\ref{FigInosov:CeLa0p50B6TdepFLEXX}. The inconsistency in measuring the $R$ point instead of the $X$ point, unlike in all other cases, was due to the technical limitations on the minimum $2\theta$ angle. In order to accurately extract the width of the quasielastic line, we had to take into account the background scattering originating from the sample environment. Compared with the parent compound, there are half as many Ce atoms, which is further aggravated by the suppression of spectral weight at the $R$ point reported above, and therefore all extrinsic effects\footnote{Scattering from the sample environment is discussed in detail in Ref.~\cite{Portnichenko18}.} have to be carefully considered. Except for the reduced signal strength, there is no difference with respect to the lower doping levels. In consistency with the previous sample, the intensity of the signal is strongly suppressed with doping and accompanied by a corresponding broadening of the line width at higher temperatures.

\begin{figure}[t]
\centerline{\includegraphics[width=\textwidth]{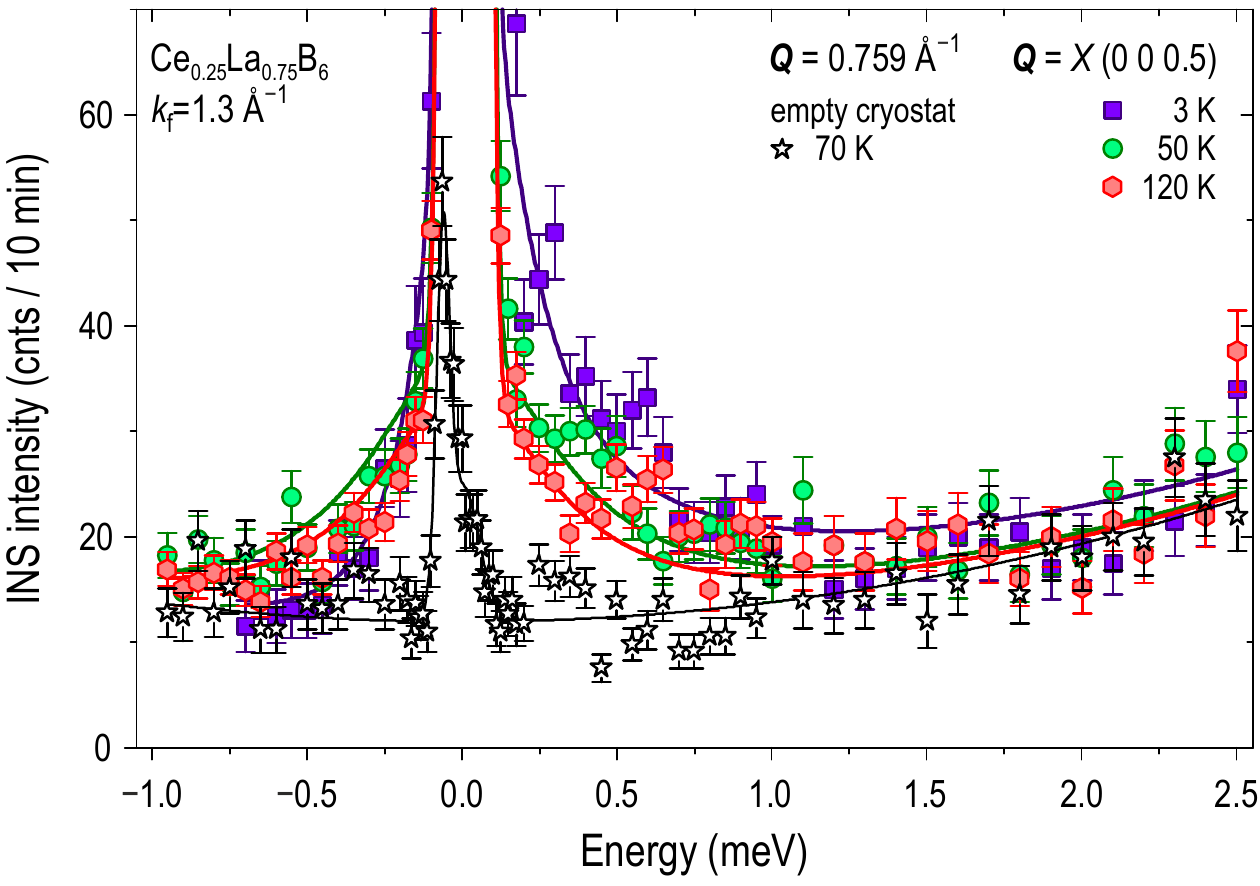}}
	\caption{Temperature dependence of the low-energy magnetic scattering for the 75\% doped sample. Background signal from the empty cryostat, shown with black stars, has been considered in the global fit of the data. The solid lines are fits with quasielastic Lorentzian profiles. Reproduced from Ref.~\cite{Portnichenko18}.}
\label{FigInosov:CeLa0p75B6TdepPANDA}
\end{figure}

Finally, measurements on the highly doped compound Ce$_{0.25}$La$_{0.75}$B$_{6}$ are shown in Fig.~\ref{FigInosov:CeLa0p75B6TdepPANDA}. We have already confirmed that elastic scattering from the sample environment may have a significant influence on the determined quasielastic line width. In order to accurately determine the Kondo temperature, every possible effort to suppress unnecessary scattering was made. In particular, the background from an empty cryostat was measured over the full energy range, and the measurements were carried out without He exchange gas. This allowed us to obtain clear evidence of the signal broadening with increasing temperature even in the most magnetically dilute sample with only 25\% of Ce ions, which should approximate the Kondo-impurity limit.

\begin{figure}[t]
\centerline{\includegraphics[width=0.9\textwidth]{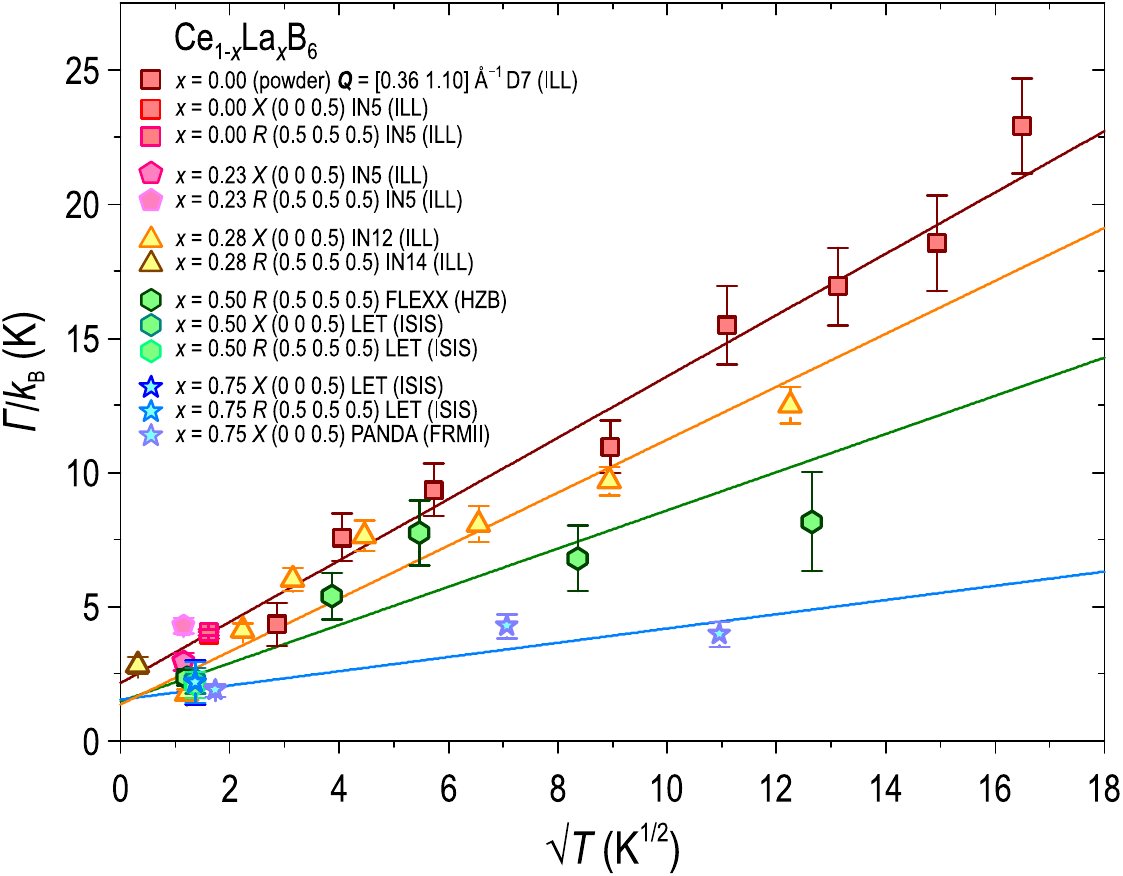}}
\caption{Summary of all the measurements on samples with different La-doping concentrations, adapted from Ref.~\cite{Portnichenko18}. The temperature dependence of the quasielastic line width, $\Gamma$, plotted vs. $\sqrt{T}$, is fitted with straight lines to the $T^{1/2}$ law given by Eq.~(\ref{EqInosov:GammaSqRoot}). Powder measurements for the parent compound, which correspond to the $X$ point within the momentum transfer range $\mathbf{Q}=[0.36~1.10]$\,\AA$^{-1}$, were plotted using the INS data from Horn \textit{et al.}~\cite{HornSteglich81}.\index{Ce$_{1-x}$La$_x$B$_6$!multipolar excitations}\index{multipolar excitations!in Ce$_{1-x}$La$_x$B$_6$}}
\label{FigInosov:SqrtGamma}
\end{figure}

The summary of all our measurements on samples with different La-doping concentrations on different instruments and at different wave vectors is presented in Fig.\,\ref{FigInosov:SqrtGamma}. All temperature dependencies were fitted to the empirical $T^{1/2}$ law given by Eq.~(\ref{EqInosov:GammaSqRoot}), and the corresponding Kondo temperatures were determined as $T_\text{K}=\Gamma_0/k_\text{B}$. As a reference measurement on the parent CeB$_6$ sample, we used the results published in Ref.~\cite{HornSteglich81}, which we extrapolated to $T \rightarrow 0$ to obtain a more accurate reference value of the neutron-deduced Kondo temperature. As one can see, all samples comply with the above-mentioned $T^{1/2}$ law. Upon 28\% La dilution, a slight decrease of the quasielastic line width at higher temperatures, compared with the parent compound, can be found. Results for the 50\% doped sample show a small deviation from $T^{1/2}$ law scaling, but we consider it an artificial effect.\footnote{At the time of our measurements, we did not yet fully appreciate the extremely important role of accurate background estimation, therefore as shown in Fig.~\ref{FigInosov:CeLa0p50B6TdepFLEXX}, the empty cryostat was measured only in the narrow energy range in the vicinity of the elastic line.} One can still fit the data with the same law and see that the reduction in the line width persists. Upon subsequent increase of La concentration up to 75\%, we still observe reliable evidence for the linear temperature dependence of the quasielastic line width on $\sqrt{T}$, but the low intensity of the signal only allowed us to measure a few temperature points for this particular doping level.

\begin{figure}[t]
\centerline{\includegraphics[width=0.9\textwidth]{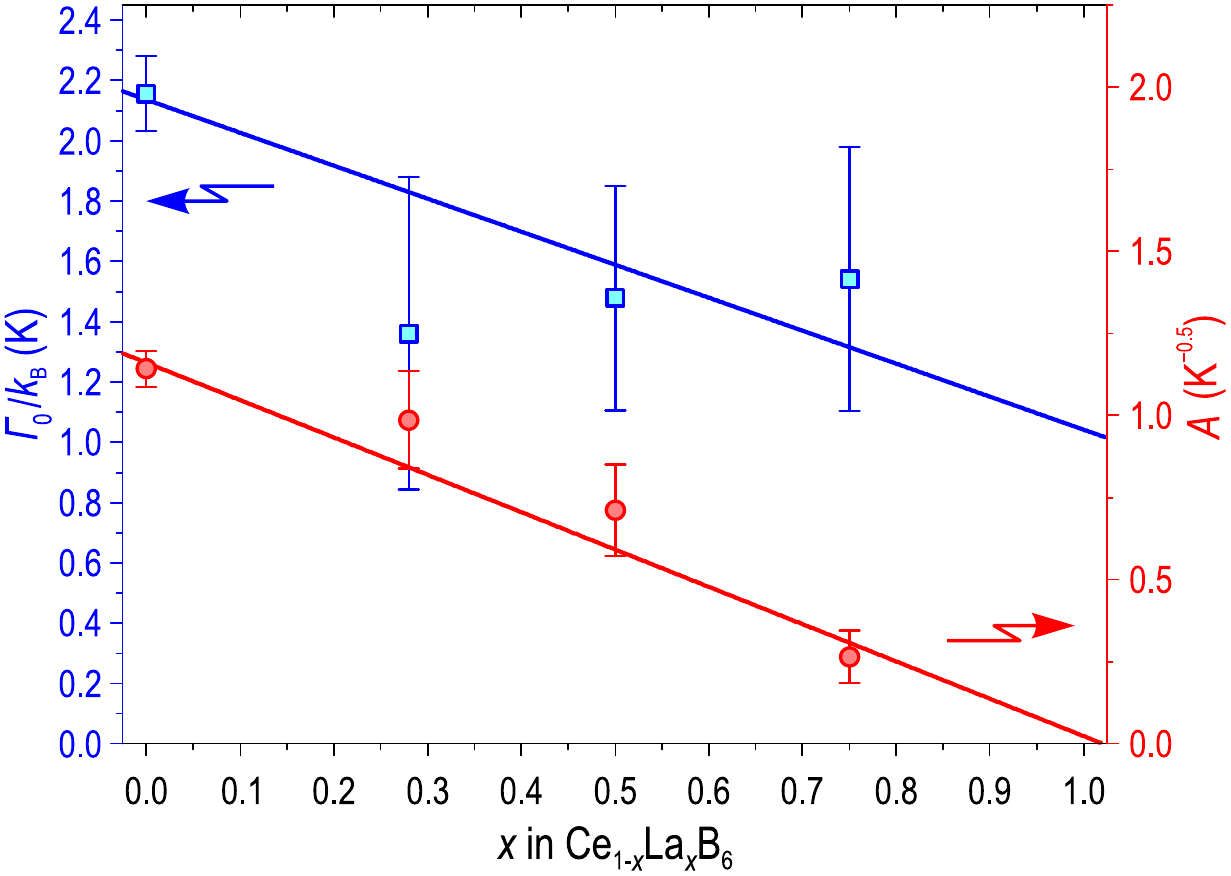}}
\caption{Dependence of the neutron-deduced Kondo temperature, $T_\text{K}=\Gamma_0/k_\text{B}$, and the $A$ parameter defined by Eq.~(\ref{EqInosov:GammaSqRoot}), on the La concentration in Ce$_{1-x}$La$_x$B$_6$, adapted from Ref.~\cite{Portnichenko18}.}
\label{FigInosov:KondoTemperature}
\end{figure}

By fitting the data in Fig.~\ref{FigInosov:SqrtGamma} to Eq.~(\ref{EqInosov:GammaSqRoot}), we obtain the doping dependence of the linear slope $A$ and the linear offset at $T \rightarrow 0$, which defines the neutron-deduced Kondo temperature~\cite{Portnichenko18}. These results are presented in Fig.~\ref{FigInosov:KondoTemperature}. In the Kondo-lattice limit of pure CeB$_{6}$, the Kondo temperature essentially coincides with the value obtained from the high-temperature resistivity data by Sato~\textit{et~al.}~\cite{SatoKunii84, SatoSumiyama85}. Even with the large uncertainties of the data points in La-doped samples, it is evident that the Kondo temperature tends to decrease when the Kondo lattice is diluted with nonmagnetic impurities, approaching the value of 1~K obtained by Sato \textit{et al.} from resistivity measurements on a highly dilute sample with only 3\% of Ce ions, which can be treated as isolated Kondo impurities~\cite{SatoSumiyama85}. We can therefore conclude that despite all the difficulties that accompanied our attempts to determine the quasielastic line width at high temperatures, we managed to get an unambiguous confirmation of the $T^{1/2}$ law scaling in a broad range of Ce concentrations. The doping dependence of the neutron-deduced Kondo temperature, extracted from the residual quasielastic line width, is in good agreement with earlier bulk measurements~\cite{SatoKunii84,SatoSumiyama85}, evidencing a slight reduction in $T_{\rm K}$ upon nonmagnetic dilution of the Kondo lattice.

Another important result from Fig.~\ref{FigInosov:SqrtGamma} is the doping dependence of parameter $A$, which has a pronounced linear dependence on the Ce concentration. To the best of our knowledge, its physical significance is not at all clear. While it reflects how quickly the relaxation rate grows with temperature, we are not aware of any microscopic theoretical model that would relate this parameter with the properties of the electronic structure, electron-electron interactions, or the quality of magnetic scattering centers. Our result suggests that parameter $A$ is proportional to the concentration of localized magnetic Ce moments in Ce$_{1-x}$La$_{x}$B$_{6}$ and extrapolates to zero in the infinitely dilute limit. It remains to be seen if this conclusion holds generally for other Kondo-lattice systems.
\index{quasielastic magnetic scattering!temperature dependence|)}
\index{Ce$_{1-x}$La$_x$B$_6$!Kondo temperature|)}\index{Kondo temperature!doping dependence|)}\index{Kondo temperature|)}
\index{diffuse neutron scattering|)}\index{quasielastic magnetic scattering|)}\index{neutron scattering!quasielastic|)}\index{paramagnon fluctuations|)}
\index{Ce$_{1-x}$La$_x$B$_6$!spin dynamics|)}\index{Ce$_{1-x}$Nd$_x$B$_6$!spin dynamics|)}
\index{CeB$_6$!magnetic excitations|)}\index{magnetic excitations!in CeB$_6$|)}

\vspace{-2pt}\subsection{Field-induced collective excitations in Ce$_{\text{1}-x}$La$_x$B$_\text{6}$}\label{Ino_SubSec:CeLaB6Collective}
\index{Ce$_{1-x}$La$_x$B$_6$!collective excitations|(}

As explained in Sec.\,\ref{Ino_SubSec:CollectiveExcitations}, the localized viewpoint on the dynamical magnetic properties of CeB$_6$ has been challenged by the new INS experiments demonstrating the appearance of a sharp resonant mode at $\mathbf{Q}_\text{AFQ}$ in the AFM phase, centered at an energy of $\hbar\omega_R=0.48$~meV~\cite{FriemelLi12}. It has been successfully explained by Akbari \textit{et al.}~\cite{AkbariThalmeier12} as a pole in the spin susceptibility of the itinerant heavy quasiparticles, calculated within the RPA formalism for the heavy-fermion ground state, signifying a close relationship to the sharp resonant modes\index{neutron resonant mode} observed in the superconducting state of some other heavy-fermion compounds,\index{heavy-fermion superconductors}\index{unconventional superconductors} such as CeCoIn$_5$~\cite{StockBroholm08, EreminZwicknagl08},\index{CeCoIn$_5$} CeCu$_2$Si$_2$~\cite{StockertArndt11},\index{CeCu$_2$Si$_2$} or the antiferromagnetic superconductor UPd$_2$Al$_3$~\cite{SatoAso01, BlackburnHiess06, ChangEremin07}.\index{UPd$_2$Al$_3$} Such sharp magnetic excitations of itinerant origin, which are usually well localized both in energy and momentum, are referred to as spin excitons~\cite{LiuZha95, AbanovChubukov99, EreminMorr05, ChubukovGorkov08}\index{spin exciton} to be distinguished from conventional magnons\index{magnon excitations} or crystal-field excitations\index{crystal-field excitations} in localized magnets.

Later Jang~\textit{et al.}~\cite{JangFriemel14} established that the $R$-point exciton in CeB$_6$\index{exciton modes!in CeB$_6$}\index{CeB$_6$!exciton modes} is continuously connected to a ferromagnetic collective mode, which is much more intense than the spin waves\index{CeB$_6$!spin waves}\index{spin waves!in CeB$_6$} emerging from the AFM wavevectors $\mathbf{q}_1$ and $\mathbf{q}_2$, putting CeB$_6$ close to a ferromagnetic instability. In Sec.\,\ref{Ino_SubSec:CeB6Dispersion} we already discussed the evolution of both excitations, observed at the $\Gamma$ and $R$ points, upon the application of an external magnetic field. Significant differences in behavior of the two resonances between the AFM phases~III\index{CeB$_6$!phase III} and III$^\prime$\index{CeB$_6$!phase III$'$} further suggest that they may have different origin.

\begin{figure}[t!]
\centerline{\includegraphics[width=0.9\textwidth]{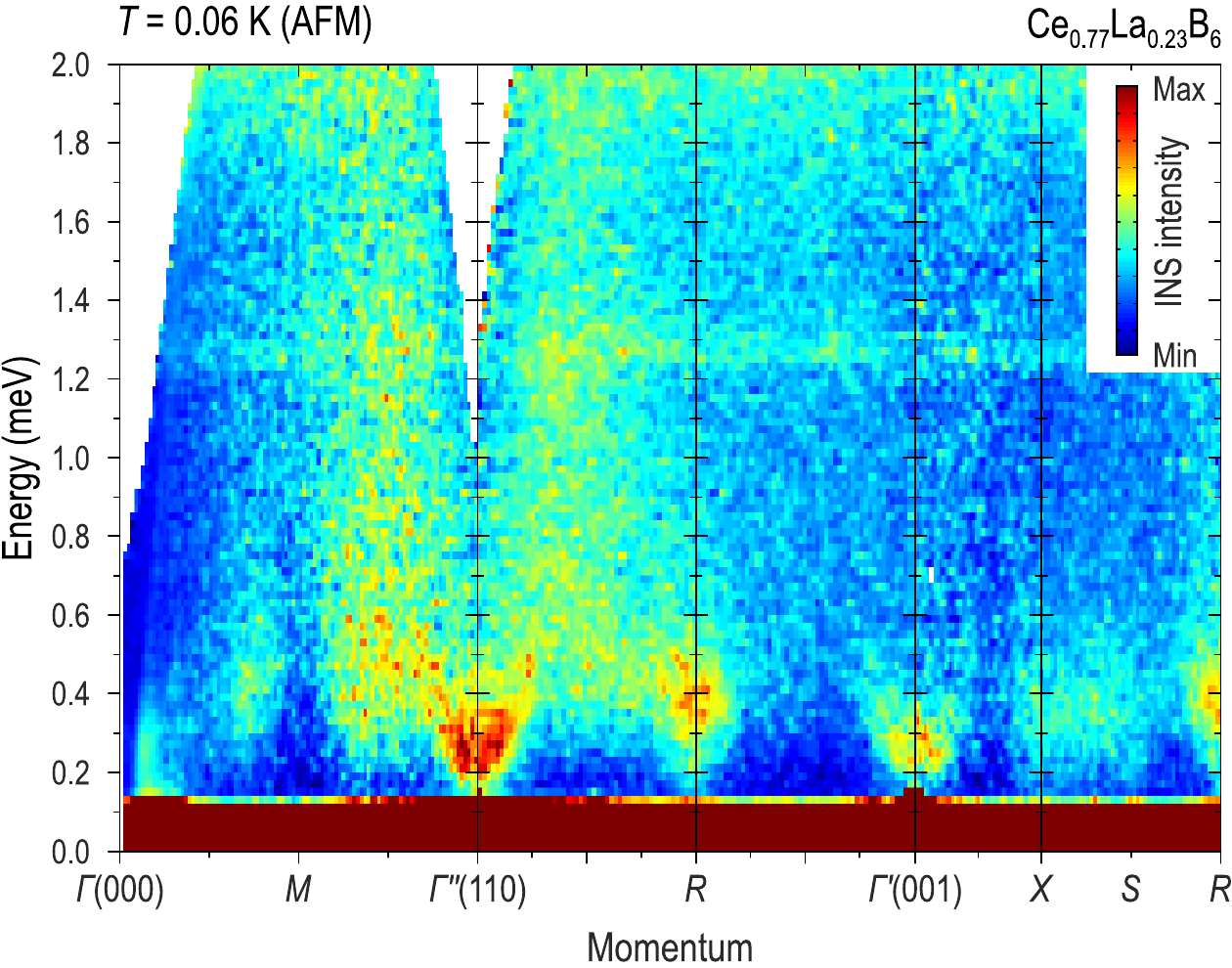}}
\caption{Color map of the raw INS intensity, measured by TOF spectroscopy on the Ce$_{0.77}$La$_{0.23}$B$_6$ single crystal at $T=60$~mK in zero magnetic field. The spectrum is plotted along a polygonal path in momentum space that contains all high-symmetry directions of the cubic Brillouin zone, adapted from Ref.~\cite{Portnichenko18}. The horizontal line around 1.25~meV is an experimental artefact.}
\label{FigInosov:CeLa0p23B6Map_IN5}
\end{figure}

According to the phase diagrams shown in Fig.~\ref{FigInosov:PhaseDiagrams}, in the absence of magnetic field CeB$_6$ develops two low-temperature ordered phases with the quadrupolar and dipolar order parameters. The AFM phase can be suppressed in a magnetic field of $B_\text{c}=1.05$~T.\index{CeB$_6$!antiferromagnetic order}\index{antiferromagnetic order!in CeB$_6$} For $B>B_\text{Q}=1.7$~T, the AFQ phase\index{CeB$_6$!antiferromagnetic order}\index{antiferromagnetic order!in CeB$_6$} is established and stabilized up to very high fields, and for $B_\text{c}<B<B_\text{Q}$, an intermediate magnetic phase III$^\prime$ persists.\index{CeB$_6$!phase III$'$} Substitution with nonmagnetic La in Ce$_{\text{1}-x}$La$_x$B$_\text{6}$ also leads to a suppression of the AFM phase with a critical doping level $x_\text{c}=0.3$. This offers an alternative way to investigate the nature of the resonant peak at the $R$ point by following its behavior with an increase in the La concentration.

It has already been shown that the resonance at the $R$ point within the AFM phase exhibits gradual broadening and shifts to lower energies upon warming, until it is transformed into a quasielastic like as soon as the AFM phase is suppressed~\cite{FriemelLi12}. It is natural to expect a similar behavior upon the suppression of the AFM order with La doping. The distribution of magnetic spectral weight in momentum space, on the other hand, is much less sensitive to the thermodynamic state of the sample and is mainly determined by the Fermi-surface geometry at the particular doping level. Notable changes in the quasielastic magnetic scattering have been observed within the paramagnetic phase even for the 50\% and 75\% La-doped samples, as discussed in Sec.\,\ref{Ino_SubSec:CeLaNdB6MomentumDependence}. However, here we will be mostly interested in the magnetic excitation spectrum at low doping levels, $x<x_\text{c}$, where the momentum-space redistribution of spectral weight can be neglected.

\begin{figure}[b!]
\centerline{\includegraphics[width=\textwidth]{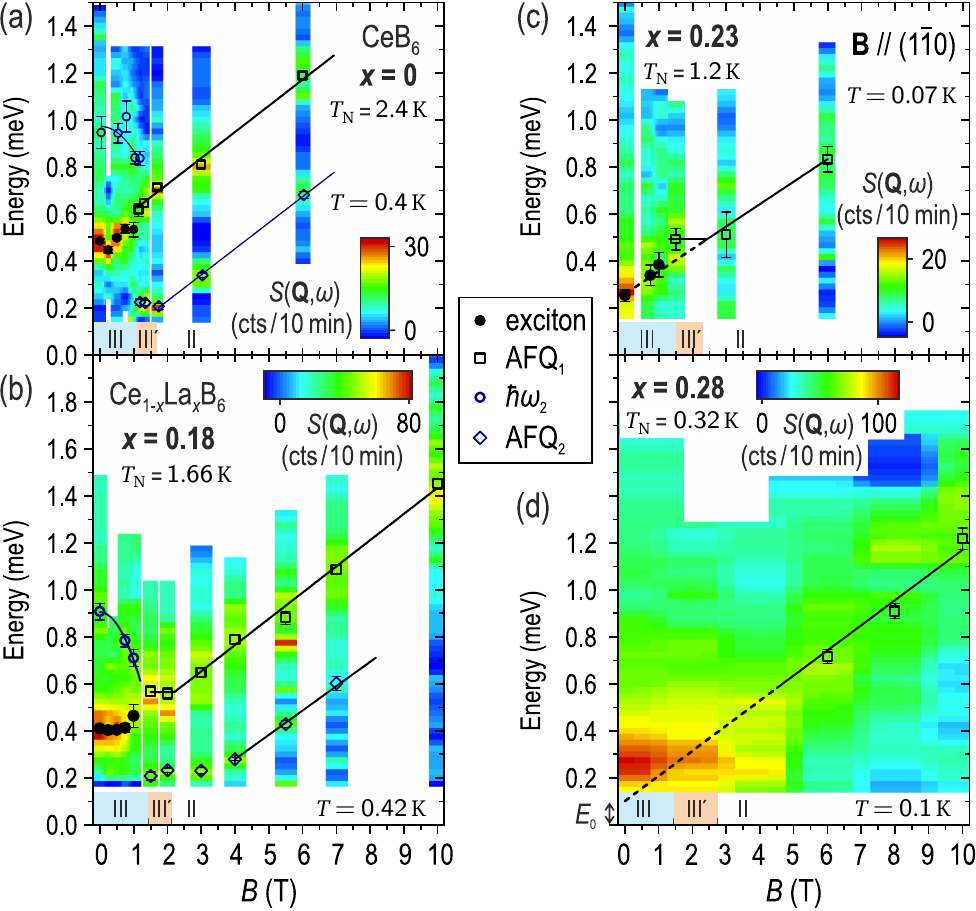}}
\caption{Color maps of the background-corrected INS intensity, $S(\mathbf{Q}_\text{AFQ},\omega)$, measured at the $R$ point for (a)~CeB$_6$ and (b)--(d)~Ce$_{\text{1}-x}$La$_x$B$_\text{6}$ with $x=0.18$, 0.23, and 0.28. The doping level is indicated in every panel together with the N\'eel temperature of the sample ($T_{\rm N}$) and the corresponding measurement temperature ($T$). The plotted intensity has been smoothed in order to reduce the statistical noise and enhance readability. The symbols denote peak positions derived from Lorentzian fits.\index{CeB$_6$!$R$-point resonant mode}\index{CeB$_6$!magnetic resonant mode}\index{Ce$_{1-x}$La$_x$B$_6$!$R$-point resonant mode}\index{exciton modes!in Ce$_{1-x}$La$_x$B$_6$}\index{Ce$_{1-x}$La$_x$B$_6$!exciton modes}}
\label{FigInosov:CeLaB6HdepRpoint}
\end{figure}

We start with presenting the spectrum of collective magnetic excitations in the absence of an external magnetic field. Figure~\ref{FigInosov:CeLa0p23B6Map_IN5} shows the low-temperature INS spectrum of Ce$_{0.77}$La$_{0.23}$B$_6$, measured by cold-neutron TOF neutron spectroscopy at $T=60$~mK (within phase~III), to be compared with the respective data in Fig.~\ref{FigInosov:CeB6INS} for the parent CeB$_6$ compound. The spectra are qualitatively similar, and the intensity maxima at the $R$ and $\Gamma$ points are still preserved. The energy of the exciton mode at the $R$ point,\index{exciton modes!in CeB$_6$}\index{CeB$_6$!exciton modes} $\hbar\omega_R$, is slightly reduced from 0.48~meV in CeB$_6$ to 0.38~meV in Ce$_{0.77}$La$_{0.23}$B$_6$, while the energy of the ferromagnetic resonance at the zone center, $\hbar\omega_\Gamma=0.25$~meV, remains practically unchanged. This result is consistent with the expectation that the spin gap\index{spin gap!in Ce$_{1-x}$La$_x$B$_6$}\index{Ce$_{1-x}$La$_x$B$_6$!spin gap} at the $R$ point (which is the propagation vector of the AFO phase) should close at the quantum-critical phase transition to phase~IV.\index{Ce$_{1-x}$La$_x$B$_6$!phase IV}\index{Ce$_{1-x}$La$_x$B$_6$!octupolar order}\index{magnetic octupoles} It also implies that the behavior of the two resonant modes is qualitatively different or even opposite to the one observed as a function of magnetic field, where the $\Gamma$-point mode softens to zero with the suppression of the AFM phase (Fig.~\ref{FigInosov:CeB6BdepGamma_IN5}), while the $R$-point mode stays at a constant energy until a second field-induced mode emerges below it (Fig.~\ref{FigInosov:CeB6BdepR_IN5}).

Friemel \textit{et al.}~\cite{FriemelJang15} compared the evolution of magnetic excitations at the $R$ point in Ce$_{\text{1}-x}$La$_x$B$_\text{6}$ samples with $x=0$, 0.18, 0.23, and 0.28 upon applying the magnetic field along the $[1\overline{1}0]$ crystal directions. These results are summarized in Fig.~\ref{FigInosov:CeLaB6HdepRpoint}, where the scattering function\footnote{In order to calculate $S(\mathbf{Q},\omega)$, the background intensity has been subtracted from the data. Details of this procedure can be found in Ref.~\cite{FriemelJang15}.} $S(\mathbf{Q}_\text{AFQ},\omega)$ at the wave vector $\mathbf{Q}_\text{AFQ}=(\frac{1}{2}\frac{1}{2}\frac{1}{2})$ is plotted vs. energy transfer and magnetic field. The field ranges of phases III,\index{CeB$_6$!phase III} III$^\prime$\index{CeB$_6$!phase III$'$}, and II\index{CeB$_6$!phase II} are indicated at the bottom of each panel. These excitations were fitted to a Lorentzian line shape given by Eq.\,(\ref{EqInosov:Quasielastic}), and the mode energy $\hbar\omega_0$ vs.~$B$ is overlayed in Fig.~\ref{FigInosov:CeLaB6HdepRpoint} as black data points. In zero field, the \mbox{$x=0$} ($T_{\rm N}=2.4$~K), \mbox{$x=0.18$} ($T_{\rm N}=1.66$~K), \mbox{$x=0.23$} ($T_{\rm N}=1.2$~K), and \mbox{$x=0.28$} ($T_{\rm N,\,onset} \approx 0.32$~K) doped samples exhibit the exciton\index{exciton modes!in Ce$_{1-x}$La$_x$B$_6$}\index{Ce$_{1-x}$La$_x$B$_6$!exciton modes} at $\hbar\omega_R=0.48$, 0.41, 0.25, and \raisebox{0.3pt}{\footnotesize$\lesssim$\,}0.1~meV, respectively.\footnote{The energy of the $R$-point exciton in the $x=0.23$ sample is somewhat lower than in Fig.~\ref{FigInosov:CeLa0p23B6Map_IN5} because of the higher measurement temperature. The TOF data in Fig.~\ref{FigInosov:CeLa0p23B6Map_IN5} were measured at 60~mK, whereas the triple-axis\index{triple-axis spectrometer} data in Fig.~\ref{FigInosov:CeLaB6HdepRpoint} were taken at 420~mK, which is only about 2.8 times lower than $T_{\rm N}$ of this particular sample.} In addition to the decrease in energy, the peak also broadens upon doping. Consequently, for $x=0.28$, only a quasielastic line can be observed at low fields. Another, much weaker and broader peak can be also seen near $\hbar\omega_2=0.94$~meV in the $x=0$ and $x=0.18$ samples.

In the AFM phase, spectra of the parent and 18\% La-substituted compounds show that the exciton energy\index{exciton modes!in Ce$_{1-x}$La$_x$B$_6$}\index{Ce$_{1-x}$La$_x$B$_6$!exciton modes} stays nearly constant vs. $\!B$, see Figs.~\ref{FigInosov:CeB6BdepR_IN5} and \ref{FigInosov:CeLaB6HdepRpoint}\,(a,\,b), while its amplitude shows a gradual suppression. This contrasts with the resonant mode in the SC state of CeCoIn$_5$, whose energy splits in magnetic field with the main part of the spectral weight carried by the lower Zeeman branch~\cite{StockBroholm12}. For neither of the modes do we observe any splitting in magnetic field, which agrees with the complete lifting of the degeneracy within the $\Gamma_8$ quartet ground state by the consecutive AFQ and AFM orderings. However, the energy of the high-energy mode $\hbar\omega_2$, shown with empty circles in Figs.~\ref{FigInosov:CeLaB6HdepRpoint}\,(a,\,b), diminishes and gets sharper with field with a varying slope between the $x=0$ and $x=0.18$ compounds and a rather concave order-parameter-like field dependence. These facts together with the vanishing of the mode above $T_\text{N}$ let us conclude that it might correspond to the onset of the particle-hole continuum\index{particle-hole continuum} at twice the AFM charge gap.\index{charge gap!in CeB$_6$} Its magnitude of $\hbar\omega_2=(0.94\pm 0.07)$~meV in zero field for CeB$_6$ agrees with the $\mathbf{Q}$-averaged gap size of $2\Delta_\text{AFM}\approx1.2$~meV determined by point-contact spectroscopy~\cite{PaulusVoss85}.

The integrated spectral weight of the exciton,\index{exciton modes!in Ce$_{1-x}$La$_x$B$_6$}\index{Ce$_{1-x}$La$_x$B$_6$!exciton modes} corresponding to the area of the peak, remains nearly constant with field below $T_\text{N}$ for $x=0$, 0.18, and 0.23. As the system enters the aforementioned phase III$^\prime$\index{CeB$_6$!phase III$'$} above $B_\text{c}$, the amplitude increases. The peak position in energy is changing abruptly [Fig.~\ref{FigInosov:CeLaB6HdepRpoint}\,(a,\,b)] or continuously [Fig.~\ref{FigInosov:CeLaB6HdepRpoint}\,(c)]. Upon eventually entering the AFQ phase, the excitation starts shifting to higher energies, as seen in the high-field spectra for $B>2$~T. Even for the heavily doped $x=0.28$ sample, a rather broad mode emerges for fields $B>6$~T. This mode (we will denote it here as AFQ$_{1}$) is dominating the spectrum in the AFQ phase for all samples. Its peak intensity changes rather continuously when crossing the III$^\prime$-II\index{CeB$_6$!phase II} phase boundary at $B_\text{Q}$ and remains nearly constant in the AFQ regime.

Moreover, upon entering phase III$^\prime$\index{CeB$_6$!phase III$'$} at $B_\text{c}$, we observe the appearance of the second low-energy mode, which can be seen for the $x=0$ and $x=0.18$ compounds at approximately 0.2~meV in Fig.~\ref{FigInosov:CeLaB6HdepRpoint}\,(a,\,b). This excitation, denoted here as AFQ$_2$, is very sharp and evolves monotonically and continuously into phase~II,\index{CeB$_6$!phase II} its energy increasing in parallel to that of the AFQ$_1$ mode. The discussion about the nature of this mode in pure CeB$_6$ can be found in Sec.~\ref{Ino_Sec:BdepCeB6}; here we only note that it also persists in La-doped samples, demonstrating that the AFM and AFQ phases\index{CeB$_6$!phase II}\index{CeB$_6$!phase III}\index{CeB$_6$!phase III$'$} have clearly distinct spin dynamics. The linear monotonic increase of both the AFQ$_1$ and AFQ$_2$ mode energies with magnetic field in phase~II\index{CeB$_6$!phase II} is characterized by a common slope $g=(0.11\pm0.004)\,\text{meV/T}=(1.90\pm0.07)\mu_\text{B}$, which is doping independent. This can be qualitatively explained by a transition between two Zeeman-split energy levels, consistent with the purely localized description of the spin dynamics in a mean-field model of ordered multipoles in magnetic field~\cite{ThalmeierShiina98, PortnichenkoAkbari20}, see also Sec.~\ref{Ino_Sec:BdepCeB6}. The localized model would also naturally explain the increasing line width $\Gamma$ of the AFQ$_1$ mode with La doping, as the La-substitution randomly alters the environment of the Ce$^{3+}$ ion, composed of six nearest neighbors.

\begin{figure}[t]
\centerline{\includegraphics[width=\textwidth]{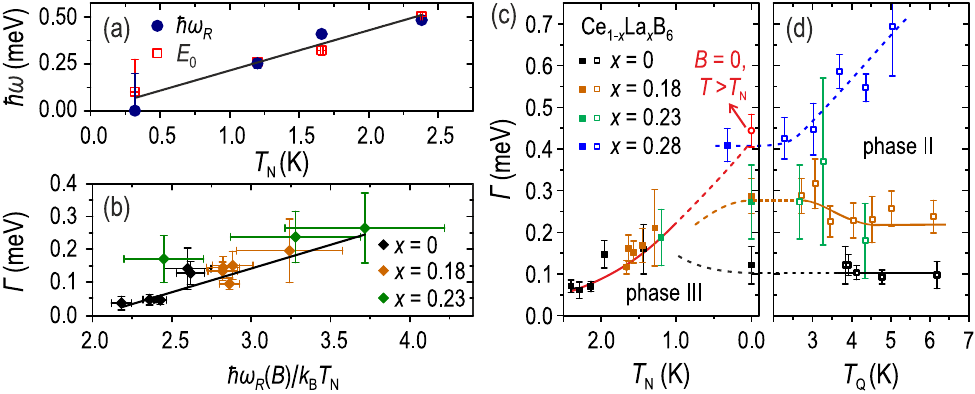}}
\caption{(a)~Zero-field exciton energy, $\hbar\omega_R$, and $B\rightarrow0$ extrapolation of the AFQ$_1$ mode, $E_0$, as a function of $T_{\rm N}$. (b)~Half width at half maximum of the exciton, $\Gamma$, plotted vs. $\hbar\omega_R/k_{\rm B}T_{\rm N}$. (c)~The same vs. $\!T_\text{N}(B)$ in the AFM phase for all doping levels. Note the inverted direction of the horizontal axis. (d)~${\Gamma}$ vs. $\!T_\text{Q}$ for the AFQ$_1$ mode in the AFQ phase for all doping levels. The field-dependent transition temperatures $T_\text{N}(B)$ and $T_\text{Q}(B)$ were determined from measurements of the specific heat or from interpolation of the published phase diagrams ($x=0$, $x=0.2$, $x=0.25$)~\cite{EffantinRossat-Mignod85, KobayashiSera00, KobayashiYoshino03, SuzukiNakamura05}. $T_\text{N}(B)$ for the $x=0.18$ sample was estimated from the AFM charge gap $\hbar\omega_2$.\index{charge gap!in CeB$_6$} All lines are guides to the eyes.}
\label{FigInosov:CeLaB6Rpoint_param}
\end{figure}

It still remains to be clarified how the exciton\index{exciton modes!in Ce$_{1-x}$La$_x$B$_6$}\index{Ce$_{1-x}$La$_x$B$_6$!exciton modes} and the AFQ$_1$ mode are related. One possible scenario~\cite{AkbariThalmeier12} describes the exciton as a collective mode below the onset of the particle-hole continuum at $2\Delta_\text{AFM}$.\index{particle-hole continuum} An alternative approach would understand the exciton as a multipolar excitation,\index{spin exciton}\index{multipolar excitations} which is overdamped by the coupling to the conduction electrons in the AFQ state $T>T_\text{N}$, but emerges as a sharp peak in the AFM state where the damping\index{spin exciton!damping} is removed by the opening of a partial charge gap~\cite{FriemelLi12, JangFriemel14}.\index{charge gap!in CeB$_6$} On the one hand, it would be an oversimplification to identify the exciton with the AFQ$_1$ mode, according to the second scenario, since the field dependence of the energy and the amplitude is completely different for both excitations. On the other hand, the zero-field extrapolation of the AFQ$_1$ mode energy $E_0$ almost coincides with the exciton energy $\hbar\omega_R$, both following the suppression of the magnetic energy scale, $k_\text{B}T_\text{N}$, as shown in Fig.~\ref{FigInosov:CeLaB6Rpoint_param}\,(a).

\index{lattice disorder!in Ce$_{1-x}$La$_x$B$_6$|(}\index{Ce$_{1-x}$La$_x$B$_6$!disorder effects|(}
Another piece of information is given by the doping and field dependencies of the exciton line width, $\Gamma$.\index{spin exciton!line width}\index{spin exciton!doping dependence} Figure~\ref{FigInosov:CeLaB6Rpoint_param}\,(b) shows that it increases with the ratio of the exciton energy to the AFM ordering temperature, $\hbar\omega_R/ k_{\rm B}T_{\rm N}$, which can be considered as a rough measure of the relative distance between the exciton and the onset of the particle-hole continuum\index{particle-hole continuum} under the assumption that the charge gap $\Delta_\text{AFM}$\index{charge gap!in CeB$_6$}\index{charge gap!in Ce$_{1-x}$La$_x$B$_6$} is proportional to $T_{\rm N}$. The points for all samples in which the exciton has been observed appear to fall on the same line, indicating that proximity to the continuum dominates the mode damping.\index{spin exciton!damping} A similar picture is given in Fig.~\ref{FigInosov:CeLaB6Rpoint_param}\,(c), where the line width $\Gamma$ is plotted directly vs. $T_\text{N}$, whose dependence on the magnetic field has been taken into account. The universality of these dependencies among all the measured samples suggests that the suppression of the AFM order and the associated closing of the partial charge gap lead to a broadening of the exciton, rather than the chemical disorder from the La substitution. This ultimately leads to a quasielastic line shape in the limit of the absent phase~III\index{CeB$_6$!phase III} in zero field, reached either by temperature for $T>T_\text{N}$ (point indicated by an arrow) or by doping (for $x=0.28$), resulting in identical line widths for both cases within the experimental uncertainty. In contrast, the line width of the AFQ$_1$ mode in phase~II\index{CeB$_6$!phase II} is independent of the respective AFQ energy scale, $k_{\rm B}T_\text{Q}$, as shown in Fig.~\ref{FigInosov:CeLaB6Rpoint_param}\,(d). The line widths for $x=0.18$ and $x=0.23$ are comparable, which can be explained with the similar disorder effect because of chemical substitution. Were the AFQ$_1$ mode and the exciton of the same origin, we would expect a more similar response to disorder for both. Therefore, the exciton must be derived from itinerant HF quasiparticles that are not as sensitive to the randomized local molecular field of the Ce$^{3+}$ ion as the localized AFQ$_1$ mode. The contrasting field dependencies for the energies for the exciton and the AFQ$_1$ mode in Fig.~\ref{FigInosov:CeLaB6HdepRpoint} further substantiate this conclusion.\index{lattice disorder!in Ce$_{1-x}$La$_x$B$_6$|)}\index{Ce$_{1-x}$La$_x$B$_6$!disorder effects|)}
\index{Ce$_{1-x}$La$_x$B$_6$!collective excitations|)}\index{Ce$_{1-x}$La$_x$B$_6$!magnetic excitations|)}\index{magnetic excitations!in Ce$_{1-x}$La$_x$B$_6$|)}\index{CeB$_6$!La doped|)}
\index{CeB$_6$|)}\index{neutron scattering|)}\index{inelastic neutron scattering|)}

\section*{Acknowledgments}
\addcontentsline{toc}{section}{Acknowledgments}

The authors of this chapter thank Peter Thalmeier, Alireza Akbari, Gerd Friemel, Hoyoung Jang, Yuan Li, Stanislav Nikitin, Bernhard Keimer, Vladimir Hinkov, George Jackeli, Andreas Koitzsch, Nikolay Sluchanko, Sergey Demishev, Alexey Semeno, Takeshi Matsumura, Takemi Yamada, Silke B\"uhler-Paschen, Vladislav Kataev, Oliver Stockert, Dongjin Jang, and Manuel Brando for many stimulating discussions and fruitful collaborations. Most of the results presented here would be impossible without the high-quality single crystal provided by Natalya Shitsevalova, Anatoliy Dukhnenko, and Volodymyr Filipov at the I.~M. Frantsevich Institute for Problems of Material Sciences of NAS, Kyiv, Ukraine, and the assistance of the instrument scientists\index{instrument scientists} at various neutron facilities: Astrid Schneidewind, Petr \v{C}erm\'ak and Igor Radelytskyi at the J\"{u}lich Centre for Neutron Science, JCNS-MLZ, Germany; Alexandre Ivanov, Jacques Ollivier, Arno Hiess, Paul Steffens and Martin B\"{o}hm at the Institute Laue-Langevin (ILL), Grenoble, France; Jean-Michel Mignot, Yvan Sidis, Sylvain Petit and Philippe Bourges from Laboratoire L\'{e}on Brillouin (LLB), Saclay, France; Robert Bewley and Tatiana Guidi at the ISIS Neutron and Muon Source, Rutherford Appleton Laboratory, United Kingdom; Andrey Podlesnyak and Tao Hong at the Spallation Neutron Source, Oak Ridge National Laboratory, USA; Jose Rodriguez-Rivera, Nicholas Butch and Yiming Qiu at the National Institute of Standards and Technology (NIST), Maryland, USA; Zita H\"{u}sges, Zhilun Lu, Jianhui Xu, Michael Tovar, Karel Prokes, Ilya Glavatskyy, Diana Lucia Quintero-Castro and Konrad Siemensmeyer at the Helmholtz-Zentrum Berlin, Germany; as well as Leonid Lev and Vladimir Strocov at the ADRESS-Beamline of the Swiss Light Source (SLS), Paul Scherrer Institute, Switzerland.

We acknowledge financial support by the German Research Foundation (DFG) under individual research grants \mbox{IN\,209/3-2} and \mbox{IN\,209/4-1}, as well as the W\"urzburg-Dresden Cluster of Excellence on Complexity and Topology in Quantum Matter\,---\,\textit{ct.qmat} (EXC~2147, project-id 39085490) and project C03 of the Collaborative Research Center SFB~1143 in Dresden (project-id 247310070).\vspace{-2pt}


\putbib[Chapter_Inosov]

\end{bibunit} 

%


\begin{thebibliography}{100}
\newcommand{\enquote}[1]{``#1''}
\providecommand{\url}[1]{\texttt{#1}}
\providecommand{\urlprefix}{URL }
\providecommand{\eprint}[2][]{\url{#2}}

\bibitem{Kusunose08}
Kusunose, H.; \enquote{Description of multipole in $f$-electron systems};
  \emph{J.~Phys. Soc. Jpn.} \textbf{77}, 064710 (2008).

\bibitem{KuramotoKusunose09}
Kuramoto, Y., Kusunose, H., and Kiss, A.; \enquote{Multipole orders and
  fluctuations in strongly correlated electron systems}; \emph{J.~Phys. Soc.
  Jpn.} \textbf{78}, 072001 (2009).

\bibitem{StackelbergNeumann32}
{von Stackelberg}, M.~V. and Neumann, F.; \enquote{The crystal structure of
  borides on the consistency of \textit{Me}B$_{6}$.}; \emph{Z.~Physik. Chem.}
  \textbf{B19}, 314--320 (1932).

\bibitem{ErkelensRegnault87}
Erkelens, W. A.~C., Regnault, L.~P., Burlet, P., Rossat-Mignod, J., Kunii, S.,
  and Kasuya, T.; \enquote{Neutron scattering study of the antiferroquadrupolar
  ordering in {C}e{B}$_{6}$ and {C}e$_{0.75}${L}a$_{0.25}${B}$_{6}$};
  \emph{J.~Magn. Magn. Mater.} \textbf{63--64}, 61--63 (1987).

\bibitem{NagaoIgarashi01}
Nagao, T. and i.~Igarashi, J.; \enquote{Resonant x-ray scattering from the
  quadrupolar ordering phase of CeB$_{6}$}; \emph{J.~Phys. Soc. Jpn.}
  \textbf{70}, 2892 (2001).

\bibitem{NagaoIgarashi06}
Nagao, T. and Igarashi, J.-i.; \enquote{Electric quadrupole contribution to
  resonant x-ray scattering: Application to multipole ordering phases in
  ${\mathrm{Ce}}_{1\ensuremath{-}x}{\mathrm{La}}_{x}{\mathrm{B}}_{6}$};
  \emph{Phys. Rev.~B} \textbf{74}, 104404 (2006).

\bibitem{MatsumuraYonmura09}
Matsumura, T., Yonmura, T., Kunimori, K., Sera, M., and Iga, F.;
  \enquote{Magnetic field induced 4f octupole in CeB$_{6}$ probed by resonant
  x-ray diffraction}; \emph{Phys. Rev. Lett.} \textbf{103}, 017203 (2009).

\bibitem{NagaoIgarashi10}
Nagao, T. and Igarashi, J.-i.; \enquote{Spectral analysis of resonant x-ray
  scattering in ${\text{CeB}}_{6}$ under an external magnetic field};
  \emph{Phys. Rev.~B} \textbf{82}, 024402 (2010).

\bibitem{NakamuraGoto96}
Nakamura, S., Goto, T., and Kunii, S.; \enquote{Ultrasonic investigation of
  quadrupolar response in Konod system Ce$_{x}$La$_{1-x}$B$_{6}$};
  \emph{Physica~B: Condens. Matter} \textbf{219--220}, 89 (1996).

\bibitem{YanagisawaMombetsu18}
Yanagisawa, T., Mombetsu, S., Hidaka, H., Amitsuka, H., Cong, P.~T., Yasin, S.,
  Zherlitsyn, S., Wosnitza, J., Huang, K., Kanchanavatee, N., Janoschek, M.,
  Maple, M.~B., and Aoki, D.; \enquote{Search for multipolar instability in
  URu$_{2}$Si$_{2}$ studied by ultrasonic measurements under pulsed magnetic
  field}; \emph{Phys. Rev.~B} \textbf{97}, 155137 (2018).

\bibitem{ShenLiu19}
Shen, Z., Liu, C., Qin, Z., Shen, S., Li, Y.-D., Bewley, R., Schneidewind, A.,
  Chen, G., and Zhao, J.; \enquote{Intertwined dipolar and multipolar order in
  the triangular-lattice magnet TmMgGaO$_{4}$}; \emph{Nat. Commun.}
  \textbf{10}, 4530 (2019).

\bibitem{PortnichenkoNikitin19}
Portnichenko, P.~Y., Nikitin, S.~E., Prokofiev, A., Paschen, S., Mignot, J.-M.,
  Ollivier, J., Podlesnyak, A., Meng, S., Lu, Z., and Inosov, D.~S.;
  \enquote{Evolution of the propagation vector of antiferroquadrupolar phases
  in ${\mathrm{Ce}}_{3}{\mathrm{Pd}}_{20}{\mathrm{Si}}_{6}$ under magnetic
  field}; \emph{Phys. Rev.~B} \textbf{99}, 214431 (2019).

\bibitem{Bouvet93}
Bouvet, A.; \emph{{\rm ``\'{E}tude par diffusion in\'elastique des neutrons des
  propri\'et\'es magn\'etiques des borures de terre rare: {C}e{B}$_{6}$,
  {P}r{B}$_{6}$ et {Y}b{B}$_{12}$''}}; Ph.D. thesis; L'Universit\'e Joseph
  Fourrier (1993).

\bibitem{ShiinaShiba03}
Shiina, R., Shiba, H., Thalmeier, P., Takahashi, A., and Sakai, O.;
  \enquote{Dynamics of multipoles and neutron scattering spectra in quadrupolar
  ordering phase of {C}e{B}$_{6}$}; \emph{J.~Phys. Soc. Jpn.} \textbf{72},
  1216--1225 (2003).

\bibitem{ThalmeierShiina98}
Thalmeier, P., Shiina, R., Shiba, H., and Sakai, O.; \enquote{Theory of
  multipolar excitations in {C}e{B}$_{6}$}; \emph{J.~Phys. Soc. Jpn.}
  \textbf{67}, 2363--2371 (1998).

\bibitem{ThalmeierShiina03}
Thalmeier, P., Shiina, R., Shiba, H., Takahashi, A., and Sakai, O.;
  \enquote{Temperature and field dependence of multipolar excitations in
  {C}e{B}$_{6}$}; \emph{J.~Phys. Soc. Jpn.} \textbf{72}, 3219--3225 (2003).

\bibitem{ThalmeierShiina04}
Thalmeier, P., Shiina, R., Shiba, H., Takahashi, A., and Sakai, O.;
  \enquote{Multipolar excitations in the antiferroquadrupolar phase of
  {C}e{B}$_{6}$}; \emph{Physica~B: Condens. Matter} \textbf{350}, E35--E38
  (2004).

\bibitem{TakaseKojima80}
Takase, A., Kojima, K., Komatsubara, T., and Kasuya, T.; \enquote{Electrical
  resistivity and magnetoresistance of {C}e{B}$_{6}$}; \emph{Solid State
  Commun.} \textbf{36}, 461--464 (1980).

\bibitem{HiroiSera97}
Hiroi, M., Sera, M., Kobayashi, N., and Kunii, S.; \enquote{Competition between
  the antiferro-quadrupolar and antiferro-exchange interactions in
  {C}e$_x${L}a$_{1\mathrm{-}x}${B}$_6$}; \emph{Phys. Rev.~B} \textbf{55},
  8339--8346 (1997).

\bibitem{HiroiKobayashi98}
Hiroi, M., Kobayashi, S.-I., Sera, M., Kobayashi, N., and Kunii, S.;
  \enquote{Drastic change of the magnetic phase diagram of {C}e$_{x}${L}a$_{1-
  x}${B}$_{6}$ between $x = 0.75$ and $0.5$}; \emph{J.~Phys. Soc. Jpn.}
  \textbf{67}, 53--56 (1998).

\bibitem{KobayashiSera00}
Kobayashi, S., Sera, M., Hiroi, M., Kobayashi, N., and Kunii, S.;
  \enquote{Transport properties in phase~IV of
  {C}e$_{x}${L}a$_{1-x}${B}$_{6}$}; \emph{J.~Phys. Soc. Jpn.} \textbf{69},
  926--936 (2000).

\bibitem{SuzukiNakamura05}
Suzuki, O., Nakamura, S., Akatsu, M., Nemoto, Y., Goto, T., and Kunii, S.;
  \enquote{Elastic properties and magnetic phase diagrams of dense {K}ondo
  compound {Ce}$_{0.75}${L}a$_{0.25}${B}$_{6}$.}; \emph{J.~Phys. Soc. Jpn.}
  \textbf{74}, 735--741 (2005).

\bibitem{FriemelJang15}
Friemel, G., Jang, H., Schneidewind, A., Ivanov, A., Dukhnenko, A.~V.,
  Shitsevalova, N.~Y., Filipov, V.~B., Keimer, B., and Inosov, D.~S.;
  \enquote{Magnetic field and doping dependence of low-energy spin fluctuations
  in the antiferroquadrupolar compound {C}e$_{1-x}${L}a$_{x}${B}$_{6}$};
  \emph{Phys. Rev.~B} \textbf{92}, 014410 (2015).

\bibitem{JangPortnichenko17}
Jang, D., Portnichenko, P.~Y., Cameron, A.~S., Friemel, G., Dukhnenko, A.~V.,
  Shitsevalova, N.~Y., Filipov, V.~B., Schneidewind, A., Ivanov, A., Inosov,
  D.~S., and Brando, M.; \enquote{Large positive correlation between the
  effective electron mass and the multipolar fluctuation in the heavy-fermion
  metal {C}e$_{1-x}${L}a$_{x}${B}$_6$}; \emph{npj Quantum Mater.} \textbf{2}
  (2017); 62.

\bibitem{TayamaSakakibara97}
Tayama, T., Sakakibara, T., Tenya, K., Amitsuka, H., and Kunii, S.;
  \enquote{Magnetic phase diagram of {C}e$_{x}${L}a$_{1- x}${B}$_{6}$ studied
  by static magnetization measurement at very low temperatures}; \emph{J.~Phys.
  Soc. Jpn.} \textbf{66}, 2268 (1997).

\bibitem{FurunoSato85}
Furuno, T., Sato, N., Kunii, S., Kasuya, T., and Sasaki, W.; \enquote{Specific
  heat measurements of {C}e$_{1-x}${L}a$_{x}${B}$_{6}$}; \emph{J.~Phys. Soc.
  Jpn.} \textbf{54}, 1899--1905 (1985).

\bibitem{KobayashiYoshino03}
Kobayashi, S., Yoshino, Y., Tsuji, S., Tou, H., Sera, M., and Iga, F.;
  \enquote{Appearance of the phase {IV} in {C}e$_{x}${L}a$_{1-x}${B}$_{6}$ at
  $x \approx 0.8$}; \emph{J.~Phys. Soc. Jpn.} \textbf{72}, 2947--2954 (2003).

\bibitem{Friemel14}
Friemel, G.; \emph{{\rm ``Itinerant spin dynamics in iron-based superconductors
  and cerium-based heavy-fermion antiferromagnets''}}; Ph.D. thesis; Fakult\"at
  Mathematik und Physik, Universit\"at Stuttgart (2014).

\bibitem{PortnichenkoDemishev16}
Portnichenko, P.~Y., Demishev, S.~V., Semeno, A.~V., Ohta, H., Cameron, A.~S.,
  Surmach, M.~A., Jang, H., Friemel, G., Dukhnenko, A.~V., Shitsevalova, N.~Y.,
  Filipov, V.~B., Schneidewind, A., Ollivier, J., Podlesnyak, A., and Inosov,
  D.~S.; \enquote{Magnetic field dependence of the neutron spin resonance in
  {C}e{B}$_{6}$}; \emph{Phys. Rev.~B} \textbf{94}, 035114 (2016).

\bibitem{MatthiasGeballe68}
Matthias, B.~T., Geballe, T.~H., Andres, K., Corenzwit, E., Hull, G.~W., and
  Maita, J.~P.; \enquote{Superconductivity and antiferromagnetism in boron-rich
  lattices}; \emph{Science} \textbf{159}, 530 (1968).

\bibitem{KunimoriTanida10}
Kunimori, K., Tanida, H., Matsumura, T., Sera, M., and Iga, F.; \enquote{Stable
  existence of phase {IV} inside phase {II} under pressure in
  {C}e$_{0.8}${L}a$_{0.2}${B}$_{6}$}; \emph{J.~Phys. Soc. Jpn.} \textbf{79},
  073703 (2010).

\bibitem{ZirngieblHillebrands84}
Zirngiebl, E., Hillebrands, B., Blumenr\"{o}der, S., G\"{u}ntherodt, G.,
  Loewenhaupt, M., Carpenter, J.~M., Winzer, K., and Fisk, Z.;
  \enquote{Crystal-field excitations in {C}e{B}$_{6}$ studied by {R}aman and
  neutron spectroscopy}; \emph{Phys. Rev.~B} \textbf{30}, 4052--4054 (1984).

\bibitem{EffantinRossat-Mignod85}
Effantin, J.~M., Rossat-Mignod, J., Burlet, P., Bartholin, H., Kunii, S., and
  Kasuya, T.; \enquote{Magnetic phase diagram of {C}e{B}$_{6}$}; \emph{J.~Magn.
  Magn. Mater.} \textbf{47--48}, 145--148 (1985).

\bibitem{MurakamiKawada98}
Murakami, Y., Kawada, H., Kawata, H., Tanaka, M., Arima, T., Moritomo, Y., and
  Tokura, Y.; \enquote{Direct observation of charge and orbital ordering in
  {L}a$_{0.5}${S}r$_{1.5}${M}n{O}$_{4}$}; \emph{Phys. Rev. Lett.} \textbf{80},
  1932--1935 (1998).

\bibitem{MannixTanaka05}
Mannix, D., Tanaka, Y., Carbone, D., Bernhoeft, N., and Kunii, S.;
  \enquote{Order parameter segregation in {Ce}$_{0.7}${L}a$_{0.3}${B}$_{6}$:
  $4f$ octopole and $5d$ dipole magnetic order}; \emph{Phys. Rev. Lett.}
  \textbf{95}, 117206 (2005).

\bibitem{KuwaharaIwasa07}
Kuwahara, K., Iwasa, K., Kohgi, M., Aso, N., Sera, M., and Iga, F.;
  \enquote{Detection of neutron scattering from phase {IV} of
  {C}e$_{0.7}${L}a$_{0.3}${B}$_{6}$: {A} confirmation of the octupole order};
  \emph{J.~Phys. Soc. Jpn.} \textbf{76}, 093702 (2007).

\bibitem{LoveseyFernandez-Rodriguez07}
Lovesey, S.~W., Fern\'andez-Rodr\'{\i}guez, J., Blanco, J.~A., and Tanaka, Y.;
  \enquote{Ce multipoles in phase IV of
  ${\mathrm{Ce}}_{0.7}{\mathrm{La}}_{0.3}{\mathrm{B}}_{6}$ inferred from
  resonant x-ray Bragg diffraction}; \emph{Phys. Rev.~B} \textbf{75}, 054401
  (2007).

\bibitem{KuwaharaIwasa09}
Kuwahara, K., Iwasa, K., Kohgi, M., Aso, N., Sera, M., Iga, F., Matsuura, M.,
  and Hirota, K.; \enquote{Magnetic octupole order in
  Ce$_{0.7}$La$_{0.3}$B$_6$: A polarized neutron diffraction study};
  \emph{Physica~B: Condens. Matter} \textbf{404}, 2527--2528 (2009).

\bibitem{SobczakSienko79}
Sobczak, R.~J. and Sienko, M.; \enquote{Superconductivity in the hexaborides};
  \emph{J.~Less Common Metals} \textbf{67}, 167--171 (1979).

\bibitem{SchellWinter82}
Schell, G., Winter, H., Rietschel, H., and Gompf, F.; \enquote{Electronic
  structure and superconductivity in metal hexaborides}; \emph{Phys. Rev.~B}
  \textbf{25}, 1589 (1982).

\bibitem{CameronFriemel16}
Cameron, A.~S., Friemel, G., and Inosov, D.~S.; \enquote{Multipolar phases and
  magnetically hidden order: review of the heavy-fermion compound
  {C}e$_{1-x}${L}a$_{x}${B}$_{6}$}; \emph{Rep. Prog. Phys.} \textbf{79}, 066502
  (2016).

\bibitem{EffantinBurlet82}
Effantin, J.~M., Burlet, P., Rossat-Mignod, J., Kunii, S., and Kasuya, T.;
  \emph{{\rm ``A neutron scattering investigation of the magnetic phase diagram
  of CeB$_{6}$''}}; in Wachter, P., and Boppart, H. (eds.), ``Valence
  Instabilities'', \textit{Proc. Int. Z\"urich Conf. Valence Instabilities},
  p.~559,  (North-Holland, Amsterdam,~1982).

\bibitem{Effantin85}
Effantin, J.-M.; \emph{{\rm ``\'{E}tude par diffusion des neutrons des
  composes''}}; Ph.D. thesis; L'Universit\'e Scientifique et Medical de
  Grenoble (1985).

\bibitem{ZaharkoFischer03}
Zaharko, O., Fischer, P., Schenck, A., Kunii, S., Brown, P.-J., Tasset, F., and
  Hansen, T.; \enquote{Zero-field magnetic structure in {C}e{B}$_{6}$
  reinvestigated by neutron diffraction and muon spin relaxation}; \emph{Phys.
  Rev.~B} \textbf{68}, 214401 (2003).

\bibitem{PadernoPokrzywnicki67}
Paderno, Y.~B., Pokrzywnicki, S., and Stali\'{n}ski, B.; \enquote{Magnetic
  properties of some rare-earth hexaborides}; \emph{phys. stat. sol. (b)}
  \textbf{24}, K73--K76 (1967).

\bibitem{HornSteglich81}
Horn, S., Steglich, F., Loewenhaupt, M., Scheuer, H., Felsch, W., and Winzer,
  K.; \enquote{The magnetic behavior of {C}e{B}$_{6}$: Comparison between
  elastic and inelastic neutron scattering, initial susceptibility and
  high-field magnetization}; \emph{Z.~Physik B: Condens. Matter} \textbf{42},
  125--134 (1981).

\bibitem{BurletRossat-Mignod82}
Burlet, P., Rossat-Mignod, J., Effantin, J.~M., Kasuya, T., Kunii, S., and
  Komatsubara, T.; \enquote{{M}agnetic ordering in cerium hexaboride
  {C}e{B}$_{6}$}; \emph{J.~Appl. Phys.} \textbf{53}, 2149--2151 (1982).

\bibitem{KunimoriKotani11}
Kunimori, K., Kotani, M., Funaki, H., Tanida, H., Sera, M., Matsumura, T., and
  Iga, F.; \enquote{Existence region of phase {III}' in {C}e{B}$_{6}$};
  \emph{J.~Phys. Soc. Jpn.} \textbf{80}, SA056 (2011).

\bibitem{FeyerhermAmato94}
Feyerherm, R., Amato, A., Gygax, F., Schenck, A., \={O}nuki, Y., and Sato, N.;
  \enquote{Muon spin rotation ($\mu${SR}) studies of magnetic ordering of
  {C}e{B}$_{6}$}; \emph{Physica~B: Condens. Matter} \textbf{194--196}, 357--358
  (1994).

\bibitem{FeyerhermAmato95}
Feyerherm, R., Amato, A., Gygax, F., Schenck, A., \={O}nuki, Y., and Sato, N.;
  \enquote{Problems of the magnetic structure of {C}e{B}$_{6}$}; \emph{J.~Magn.
  Magn. Mater.} \textbf{140--144}, 1175--1176, Part 2 (1995); international
  Conference on Magnetism.

\bibitem{FujitaSuzuki80}
Fujita, T., Suzuki, M., Komatsubara, T., Kunii, S., Kasuya, T., and Ohtsuka,
  T.; \enquote{Anomalous specific heat of {C}e{B}$_{6}$}; \emph{Solid State
  Commun.} \textbf{35}, 569--572 (1980).

\bibitem{KomatsubaraSato83}
Komatsubara, T., Sato, N., Kunii, S., Oguro, I., Furukawa, Y., Onuki, Y., and
  Kasuya, T.; \enquote{{D}ense {K}ondo behavior in {C}e{B}$_{6}$ and its
  alloys}; \emph{J.~Magn. Magn. Mater.} \textbf{31--34, part 1}, 368--372
  (1983).

\bibitem{TakigawaYasuoka83}
Takigawa, M., Yasuoka, H., Tanaka, T., and Ishizawa, Y.; \enquote{{NMR} study
  on the spin structure of {C}e{B}$_{6}$}; \emph{J.~Phys. Soc. Jpn.}
  \textbf{52}, 728--731 (1983).

\bibitem{Lovesey15}
Lovesey, S.~W.; \enquote{Theory of neutron scattering by electrons in magnetic
  materials}; \emph{Phys. Scr.} \textbf{90}, 108011 (2015).

\bibitem{Ohkawa85}
Ohkawa, J.~F.; \enquote{Orbital antiferromagnetism in {C}e{B}$_{6}$};
  \emph{J.~Phys. Soc. Jpn.} \textbf{54}, 3909--3914 (1985).

\bibitem{ShiinaShiba97}
Shiina, R., Shiba, H., and Thalmeier, P.; \enquote{Magnetic-field effects on
  quadrupolar ordering in a ${\Gamma}_{8}$-quartet system {C}e{B}$_{6}$};
  \emph{J.~Phys. Soc. Jpn.} \textbf{66}, 1741--1755 (1997).

\bibitem{HanzawaKasuya84}
Hanzawa, K. and Kasuya, T.; \enquote{Antiferro-quadrupolar ordering in
  {C}e{B}$_{6}$}; \emph{J.~Phys. Soc. Jpn.} \textbf{53}, 1809--1818 (1984).

\bibitem{NakaoMagishi01}
Nakao, H., Magishi, K.-i., Wakabayashi, Y., Murakami, Y., Koyama, K., Hirota,
  K., Endoh, Y., and Kunii, S.; \enquote{Antiferro-quadrupole ordering of
  {C}e{B}$_{6}$ studied by resonant x-ray scattering}; \emph{J.~Phys. Soc.
  Jpn.} \textbf{70}, 1857--1860 (2001).

\bibitem{YakhouPlakhty01}
Yakhou, F., Plakhty, V., Suzuki, H., Gavrilov, S., Burlet, P., Paolasini, L.,
  Vettier, C., and Kunii, S.; \enquote{$k = 2\pi/a
  [\frac{1}{2}\frac{1}{2}\frac{1}{2}]$ zero-field ordering in the intermediate
  phase of {C}e{B}$_{6}$ observed by {X}-ray scattering: what orders?};
  \emph{Phys. Lett.~A} \textbf{285}, 191--196 (2001).

\bibitem{SeraIchikawa01}
Sera, M., Ichikawa, H., Yokoo, T., Akimitsu, J., Nishi, M., Kakurai, K., and
  Kunii, S.; \enquote{Anomalous temperature dependence of the
  magnetic-field-induced antiferromagnetic moment in the antiferroquadrupolar
  ordered state of {C}e{B}$_{6}$}; \emph{Phys. Rev. Lett.} \textbf{86},
  1578--1581 (2001).

\bibitem{Rossat-Mignod87}
Rossat-Mignod, J.; \emph{{\rm ``Magnetic Structures''}}; chap.~19 in Sk\"old,
  K. and Price, D. L. (eds.), {\rm ``Neutron Scattering''}, \textit{Methods of
  Experimental Physics}, vol.~23C, pp.~69--157,  (Academic Press,
  Amsterdam,~1987).

\bibitem{SeraKobayashi99}
Sera, M. and Kobayashi, S.; \enquote{Magnetic properties of the 4-sublattice
  model for the antiferro ({AF}) quadrupolar order dominated by the {AF}
  octupolar and {AF} exchange interactions\,---\,a simple model for
  {C}e{B}$_{6}$}; \emph{J.~Phys. Soc. Jpn.} \textbf{68}, 1664--1678 (1999).

\bibitem{FriemelLi12}
Friemel, G., Li, Y., Dukhnenko, A., Shitsevalova, N., Sluchanko, N., Ivanov,
  A., Filipov, V., Keimer, B., and Inosov, D.; \enquote{Resonant magnetic
  exciton mode in the heavy-fermion antiferromagnet {C}e{B}$_{6}$}; \emph{Nat.
  Commun.} \textbf{3}, 830 (2012).

\bibitem{PlakhtyRegnault05}
Plakhty, V.~P., Regnault, L.~P., Goltsev, A.~V., Gavrilov, S.~V., Yakhou, F.,
  Flouquet, J., Vettier, C., and Kunii, S.; \enquote{Itinerant magnetism in the
  Kondo crystal {C}e{B}$_{6}$ as indicated by polarized neutron scattering.};
  \emph{Phys. Rev.~B} \textbf{71}, 100407 (2005).

\bibitem{SluchankoBogach07}
Sluchanko, N., Bogach, A., Glushkov, V., Demishev, S., Ivanov, V., Ignatov, M.,
  Kuznetsov, A., Samarin, N., Semeno, A., and Shitsevalova, N.;
  \enquote{Enhancement of band magnetism and features of the magnetically
  ordered state in the {C}e{B}$_{6}$ compound with strong electron
  correlations}; \emph{J.~Exp. Theor. Phys.} \textbf{104}, 120--138 (2007).

\bibitem{SatoKunii84}
Sato, N., Kunii, S., Oguro, I., Komatsubara, T., and Kasuya, T.;
  \enquote{Magnetic properties of single crystals of
  {C}e$_{x}${L}a$_{1-x}${B}$_{6}$}; \emph{J.~Phys. Soc. Jpn.} \textbf{53},
  3967--3979 (1984).

\bibitem{LoewenhauptCarpenter85}
Loewenhaupt, M., Carpenter, J., and Loong, C.-K.; \enquote{Magnetic excitations
  in {C}e{B}$_{6}$}; \emph{J.~Magn. Magn. Mater.} \textbf{52}, 245--249 (1985).

\bibitem{Ohkawa83}
Ohkawa, J.~F.; \enquote{Ordered states in periodic {A}nderson Hamiltonian with
  orbital degeneracy and with large {C}oulomb correlation}; \emph{J.~Phys. Soc.
  Jpn.} \textbf{52}, 3897--3906 (1983).

\bibitem{UiminKuramoto96}
Uimin, G., Kuramoto, Y., and Fukushima, N.; \enquote{Mode coupling effects on
  the quadrupolar ordering in {C}e{B}$_{6}$}; \emph{Solid State Commun.}
  \textbf{97}, 595--598 (1996).

\bibitem{ShiinaSakai98}
Shiina, R., Sakai, O., Shiba, H., and Thalmeier, P.; \enquote{Interplay of
  field-induced multipoles in {C}e{B}$_{6}$.}; \emph{J.~Phys. Soc. Jpn.}
  \textbf{67}, 941--949 (1998).

\bibitem{SakaiShiina97}
Sakai, O., Shiina, R., Shiba, H., and Thalmeier, P.; \enquote{A new
  interpretation of {NMR} in quadrupolar ordering phase of
  {C}e{B}$_{6}$\,---\,consistency with neutron scattering}; \emph{J.~Phys. Soc.
  Jpn.} \textbf{66}, 3005--3007 (1997).

\bibitem{KawakamiKunii81}
Kawakami, M., Kunii, S., Mizuno, K., Sugita, M., Kasuya, T., and Kume, K.;
  \enquote{The $^{11}${B} nuclear magnetic resonance in {C}e{B}$_{6}$ single
  crystal}; \emph{J.~Phys. Soc. Jpn.} \textbf{50}, 432--437 (1981).

\bibitem{JangFriemel14}
Jang, H., Friemel, G., Ollivier, J., Dukhnenko, A.~V., Shitsevalova, N.~Y.,
  Filipov, V.~B., Keimer, B., and Inosov, D.~S.; \enquote{Intense low-energy
  ferromagnetic fluctuations in the antiferromagnetic heavy-fermion metal
  {C}e{B}$_{6}$}; \emph{Nat. Mater.} \textbf{13}, 682--687 (2014).

\bibitem{FongKeimer95}
Fong, H.~F., Keimer, B., Anderson, P.~W., Reznik, D., Do\ifmmode~\breve{g}\else
  \u{g}\fi{}an, F., and Aksay, I.~A.; \enquote{Phonon and magnetic neutron
  scattering at 41~me{V} in {YBa}$_{2}${Cu}$_{3}${O}$_{7}$}; \emph{Phys. Rev.
  Lett.} \textbf{75}, 316--319 (1995).

\bibitem{InosovPark10}
Inosov, D.~S., Park, J.~T., Bourges, P., Sun, D.~L., Sidis, Y., Schneidewind,
  A., Hradil, K., Haug, D., Lin, C.~T., Keimer, B., and Hinkov, V.;
  \enquote{Normal-state spin dynamics and temperature-dependent spin-resonance
  energy in optimally doped {B}a{F}e$_{1.85}${C}o$_{0.15}${A}s$_{2}$.};
  \emph{Nat. Phys.} \textbf{6}, 178--181 (2010).

\bibitem{StockBroholm08}
Stock, C., Broholm, C., Hudis, J., Kang, H.~J., and Petrovic, C.; \enquote{Spin
  resonance in the $d$-wave superconductor {C}e{C}o{I}n$_{5}$.}; \emph{Phys.
  Rev. Lett.} \textbf{100}, 087001 (2008).

\bibitem{StockertArndt11}
Stockert, O., Arndt, J., Faulhaber, E., Geibel, C., Jeevan, H.~S., Kirchner,
  S., Loewenhaupt, M., Schmalzl, K., Schmidt, W., Si, Q., and Steglich, F.;
  \enquote{Magnetically driven superconductivity in {C}e{C}u$_{2}${S}i$_{2}$.};
  \emph{Nat. Phys.} \textbf{7}, 119--124 (2011).

\bibitem{LiuZha95}
Liu, D.~Z., Zha, Y., and Levin, K.; \enquote{Theory of neutron scattering in
  the normal and superconducting states of
  $\mathrm{Y}{\mathrm{Ba}}_{2}{\mathrm{Cu}}_{3}{\mathrm{O}}_{6+x}$};
  \emph{Phys. Rev. Lett.} \textbf{75}, 4130--4133 (1995).

\bibitem{AbanovChubukov99}
Abanov, A. and Chubukov, A.~V.; \enquote{A relation between the resonance
  neutron peak and ARPES data in cuprates}; \emph{Phys. Rev. Lett.}
  \textbf{83}, 1652--1655 (1999).

\bibitem{EreminMorr05}
Eremin, I., Morr, D.~K., Chubukov, A.~V., Bennemann, K.~H., and Norman, M.~R.;
  \enquote{Novel neutron resonance mode in
  ${d}_{{x}^{2}\ensuremath{-}{y}^{2}}$-wave superconductors}; \emph{Phys. Rev.
  Lett.} \textbf{94}, 147001 (2005).

\bibitem{ChubukovGorkov08}
Chubukov, A.~V. and Gor'kov, L.~P.; \enquote{Spin resonance in
  three-dimensional superconductors: the case of {CeCoIn}$_{5}$}; \emph{Phys.
  Rev. Lett.} \textbf{101}, 147004 (2008).

\bibitem{PaulusVoss85}
Paulus, E. and Voss, G.; \enquote{Point contact spectra of cerium compounds};
  \emph{J.~Magn. Magn. Mater.} \textbf{47--48}, 539--541 (1985).

\bibitem{AlekseevMignot95}
Alekseev, P.~A., Mignot, J.~M., Rossat-Mignod, J., Lazukov, V.~N., Sadikov,
  I.~P., Konovalova, E.~S., and Paderno, Y.~B.; \enquote{{M}agnetic excitation
  spectrum of mixed-valence {S}m{B}$_{6}$ studied by neutron scattering on a
  single crystal}; \emph{J.~Phys.: Condens. Matter} \textbf{7}, 289 (1995).

\bibitem{FuhrmanLeiner15}
Fuhrman, W.~T., Leiner, J., Nikoli\ifmmode~\acute{c}\else \'{c}\fi{}, P.,
  Granroth, E., G.\, Stone, B., M.\, Lumsden, D., M.\, DeBeer-Schmitt, L.,
  Alekseev, A., P.\, Mignot, J.-M., Koohpayeh, M., S.\, Cottingham, P., Phelan,
  W.~A., Schoop, L., McQueen, M., T.\, and Broholm, C.; \enquote{Interaction
  driven subgap spin exciton in the {K}ondo insulator {S}m{B}$_{6}$};
  \emph{Phys. Rev. Lett.} \textbf{114}, 036401 (2015).

\bibitem{AkbariThalmeier12}
Akbari, A. and Thalmeier, P.; \enquote{Spin exciton formation inside the hidden
  order phase of ${\mathrm{CeB}}_{6}$}; \emph{Phys. Rev. Lett.} \textbf{108},
  146403 (2012).

\bibitem{SchaufusKataev09}
Schaufu\ss, U., Kataev, V., Zvyagin, A.~A., B\"uchner, B., Sichelschmidt, J.,
  Wykhoff, J., Krellner, C., Geibel, C., and Steglich, F.; \enquote{Evolution
  of the {K}ondo state of {Y}b{R}h$_2${S}i$_2$ probed by high-field {ESR}};
  \emph{Phys. Rev. Lett.} \textbf{102}, 076405 (2009).

\bibitem{DemishevSemeno05}
Demishev, S.~V., Semeno, A.~V., Paderno, Y.~B., Shitsevalova, N.~Y., and
  Sluchanko, N.~E.; \enquote{Experimental evidence for magnetic resonance in
  the antiferro-quadrupole phase}; \emph{phys. stat. sol. (b)} \textbf{242},
  R27--R29 (2005).

\bibitem{DemishevSemeno09}
Demishev, S.~V., Semeno, A.~V., Bogach, A.~V., Samarin, N.~A., Ishchenko,
  T.~V., Filipov, V.~B., Shitsevalova, N.~Y., and Sluchanko, N.~E.;
  \enquote{Magnetic spin resonance in {C}e{B}$_{6}$}; \emph{Phys. Rev.~B}
  \textbf{80}, 245106 (2009).

\bibitem{YamadaHanzawa19}
Yamada, T. and Hanzawa, K.; \enquote{Derivation of {RKKY} interaction between
  multipole moments in CeB$_6$ by the effective {W}annier model based on the
  band-structure calculation}; \emph{J.~Phys. Soc. Jpn.} \textbf{88}, 084703
  (2019).

\bibitem{HanzawaYamada19}
Hanzawa, K. and Yamada, T.; \enquote{Origin of anisotropic RKKY interactions in
  CeB$_6$}; \emph{J.~Phys. Soc. Jpn.} \textbf{88}, 124710 (2019).

\bibitem{RegnaultErkelens88}
Regnault, L.~P., Erkelens, W. A.~C., Rossat-Mignod, J., Vettier, C., Kunii, S.,
  and Kasuya, T.; \enquote{Inelastic neutron scattering study of the rare earth
  hexaboride {C}e{B}$_{6}$.}; \emph{J.~Magn. Magn. Mater.} \textbf{76--77},
  413--414 (1988).

\bibitem{MoriyaTakimoto95}
Moriya, T. and Takimoto, T.; \enquote{Anomalous properties around magnetic
  instability in heavy-electron systems}; \emph{J.~Phys. Soc. Jpn.}
  \textbf{64}, 960--969 (1995).

\bibitem{StockertFaulhaber04}
Stockert, O., Faulhaber, E., Zwicknagl, G., St\"u\ss{}er, N., Jeevan, H.~S.,
  Deppe, M., Borth, R., K\"uchler, R., Loewenhaupt, M., Geibel, C., and
  Steglich, F.; \enquote{Nature of the $A$ Phase in
  ${\mathrm{CeCu}}_{2}{\mathrm{Si}}_{2}$}; \emph{Phys. Rev. Lett.} \textbf{92},
  136401 (2004).

\bibitem{WiebeJanik07}
Wiebe, C.~R., Janik, J.~A., MacDougall, G.~J., Luke, G.~M., Garrett, J.~D.,
  Zhou, H.~D., Jo, Y.-J., Balicas, L., Qiu, Y., Copley, J. R.~D., Yamani, Z.,
  and Buyers, W. J.~L.; \enquote{Gapped itinerant spin excitations account for
  missing entropy in the hidden-order state of {U}{R}u$_2${S}i$_2$}; \emph{Nat.
  Phys.} \textbf{3}, 96 (2007).

\bibitem{KoitzschHeming16}
Koitzsch, A., Heming, N., Knupfer, M., B\"uchner, B., Portnichenko, P.~Y.,
  Dukhnenko, A.~V., Shitsevalova, N.~Y., Filipov, V.~B., Lev, L.~L., Strocov,
  V.~N., Ollivier, J., and Inosov, D.~S.; \enquote{Nesting-driven multipolar
  order in {C}e{B}$_6$ from photoemission tomography}; \emph{Nat. Commun.}
  \textbf{7}, 10876 (2016).

\bibitem{OnukiUmezawa89}
\={O}nuki, Y., Umezawa, A., Kwok, W.~K., Crabtree, G.~W., Nishihara, M.,
  Yamazaki, T., Omi, T., and Komatsubara, T.; \enquote{High-field
  magnetoresistance and de {H}aas\,--\,van {A}lphen effect in antiferromagnetic
  {P}r{B}$_{6}$ and {N}d{B}$_{6}$}; \emph{Phys. Rev.~B} \textbf{40},
  11195--11207 (1989).

\bibitem{JossRuitenbeek87}
Joss, W., van Ruitenbeek, J.~M., Crabtree, G.~W., Tholence, J.~L., van Deursen,
  A. P.~J., and Fisk, Z.; \enquote{Observation of the magnetic field dependence
  of the cyclotron mass in the Kondo lattice {C}e{B}$_{6}$}; \emph{Phys. Rev.
  Lett.} \textbf{59}, 1609--1612 (1987).

\bibitem{DeursenPols85}
{van Deursen}, A. P.~J., Pols, R.~E., de~Vroomen, A.~R., and Fisk, Z.;
  \enquote{Fermi surfaces and effective masses in {L}n{B}$_{6}$, {L}n: {L}a,
  {C}e and {P}r}; \emph{J.~Less Common Metals} \textbf{111}, 331--334 (1985).

\bibitem{EndoNakamura06}
Endo, M., Nakamura, S., Isshiki, T., Kimura, N., Nojima, T., Aoki, H., Harima,
  H., and Kunii, S.; \enquote{Evolution of {F}ermi {S}urface {P}roperties in
  {C}e$_{x}${L}a$_{1-x}${B}$_{6}$ and {P}r$_{x}${L}a$_{1-x}${B}$_{6}$};
  \emph{J.~Phys. Soc. Jpn.} \textbf{75}, 114704 (2006).

\bibitem{GoodrichHarrison99}
Goodrich, R.~G., Harrison, N., Teklu, A., Young, D., and Fisk, Z.;
  \enquote{Development of the high-field heavy-fermion ground state in
  {C}e$_{1-x}${L}a$_{x}${B}$_{6}$ intermetallics}; \emph{Phys. Rev. Lett.}
  \textbf{82}, 3669--3672 (1999).

\bibitem{IshizawaTanaka77}
Ishizawa, Y., Tanaka, T., Bannai, E., and Kawai, S.; \enquote{de
  {H}aas\,--\,van {A}lphen effect and {F}ermi surface of {L}a{B}$_{6}$};
  \emph{J.~Phys. Soc. Jpn.} \textbf{42}, 112--118 (1977).

\bibitem{HarrisonMeeson93}
Harrison, N., Meeson, P., Probst, P.~A., and Springford, M.;
  \enquote{Quasiparticle and thermodynamic mass in the heavy-fermion system
  {C}e{B}$_{6}$}; \emph{J.~Phys.: Condens. Matter} \textbf{5}, 7435 (1993).

\bibitem{HarrisonHall98}
Harrison, N., Hall, D.~W., Goodrich, R.~G., Vuillemin, J.~J., and Fisk, Z.;
  \enquote{Quantum interference in the spin-polarized heavy-fermion compound
  {C}e{B}$_{6}$: Evidence for topological deformation of the {F}ermi surface in
  strong magnetic fields}; \emph{Phys. Rev. Lett.} \textbf{81}, 870--873
  (1998).

\bibitem{TekluGoodrich00}
Teklu, A.~A., Goodrich, R.~G., Harrison, N., Hall, D., Fisk, Z., and Young, D.;
  \enquote{Fermi surface properties of low-concentration {C}e$_{x}${L}a$_{1-
  x}${B}$_{6}$: de {H}aas\,--\,van {A}lphen.}; \emph{Phys. Rev.~B} \textbf{62},
  12875--12881 (2000).

\bibitem{SoumaIida04}
Souma, S., Iida, Y., Sato, T., Takahashi, T., and Kunii, S.; \enquote{Band
  structure and Fermi surface of {C}e{B}$_{6}$ studied by angle-resolved
  photoemission spectroscopy}; \emph{Physica~B: Condens. Matter} \textbf{351},
  283--285 (2004).

\bibitem{NeupaneAlidoust15}
Neupane, M., Alidoust, N., Belopolski, I., Bian, G., Xu, S.-Y., Kim, D.-J.,
  Shibayev, P.~P., Sanchez, D.~S., Zheng, H., Chang, T.-R., Jeng, H.-T.,
  Riseborough, P.~S., Lin, H., Bansil, A., Durakiewicz, T., Fisk, Z., and
  Hasan, M.~Z.; \enquote{Fermi surface topology and hot spot distribution in
  the Kondo lattice system CeB$_{6}$}; \emph{Phys. Rev.~B} \textbf{92}, 104420
  (2015).

\bibitem{PatilAdhikary10}
Patil, S., Adhikary, G., Balakrishnan, G., and Maiti, K.; \enquote{Influence of
  4$f$ electronic states on the surface states of rare-earth hexaborides};
  \emph{Appl. Phys. Lett.} \textbf{96}, 092106 (2010).

\bibitem{Trenary12}
Trenary, M.; \enquote{Surface science studies of metal hexaborides}; \emph{Sci.
  Technol. Adv. Mater.} \textbf{13}, 023002 (2012).

\bibitem{ZhangButch13}
Zhang, X., Butch, N.~P., Syers, P., Ziemak, S., Greene, R.~L., and Paglione,
  J.; \enquote{Hybridization, inter-ion correlation, and surface states in the
  Kondo insulator {S}m{B}$_6$}; \emph{Phys. Rev. X} \textbf{3}, 011011 (2013).

\bibitem{HemingTreske14}
Heming, N., Treske, U., Knupfer, M., B\"uchner, B., Inosov, D.~S.,
  Shitsevalova, N.~Y., Filipov, V.~B., Krause, S., and Koitzsch, A.;
  \enquote{Surface properties of {S}m{B}$_6$ from x-ray photoelectron
  spectroscopy}; \emph{Phys. Rev.~B} \textbf{90}, 195128 (2014).

\bibitem{NikitinPortnichenko18}
Nikitin, S.~E., Portnichenko, P.~Y., Dukhnenko, A.~V., Shitsevalova, N.~Y.,
  Filipov, V.~B., Qiu, Y., Rodriguez-Rivera, J.~A., Ollivier, J., and Inosov,
  D.~S.; \enquote{Doping-induced redistribution of magnetic spectral weight in
  the substituted hexaborides
  ${\mathrm{Ce}}_{1\ensuremath{-}x}{\mathrm{La}}_{x}{\mathrm{B}}_{6}$ and
  ${\mathrm{Ce}}_{1\ensuremath{-}x}{\mathrm{Nd}}_{x}{\mathrm{B}}_{6}$};
  \emph{Phys. Rev.~B} \textbf{97}, 075116 (2018).

\bibitem{BalcarLovesey89}
Balcar, E. and Lovesey, S.~W.; \emph{{\rm ``Theory of magnetic neutron and
  photon scattering''}}; vol.~2,  (Clarendon Press, Oxford,~1989); chap. 2.

\bibitem{JensenMackintosh91}
Jensen, J. and Mackintosh, A.~R.; \emph{{\rm ``Rare earth magnetism''}};
  International Series of Monographs on Physics, vol. 81,  (Clarendon Press,
  Oxford,~1991); chap. 4.

\bibitem{ShiinaSakai07}
Shiina, R., Sakai, O., and Shiba, H.; \enquote{Magnetic form factor of elastic
  neutron scattering expected for octupolar phases in
  {C}e$_{1-x}${L}a$_{x}${B}$_{6}$ and {N}p{O}$_{2}$}; \emph{J.~Phys. Soc. Jpn.}
  \textbf{76}, 094702 (2007).

\bibitem{SantiniCarretta09}
Santini, P., Carretta, S., Amoretti, G., Caciuffo, R., Magnani, N., and Lander,
  G.~H.; \enquote{Multipolar interactions in $f$-electron systems: The paradigm
  of actinide dioxides}; \emph{Rev. Mod. Phys.} \textbf{81}, 807--863 (2009).

\bibitem{Shiina12}
Shiina, R.; \enquote{Anisotropic Form Factors of Neutron Scattering by Magnetic
  Octupole in CeB$_{6}$}; \emph{J.~Phys.: Condens. Matter} \textbf{391}, 012064
  (2012).

\bibitem{PortnichenkoAkbari20}
Portnichenko, P.~Y., Akbari, A., Nikitin, S.~E., Cameron, A.~S., Dukhnenko,
  A.~V., Filipov, V.~B., Shitsevalova, N.~Y., \v{C}erm\'{a}k, P., Radelytskyi,
  I., Schneidewind, A., Ollivier, J., Podlesnyak, A., Huesges, Z., Xu, J.,
  Ivanov, A., Sidis, Y., Petit, S., Mignot, J.-M., Thalmeier, P., and Inosov,
  D.~S.; \enquote{Field-angle-resolved magnetic excitations as a probe of
  hidden-order symmetry in CeB$_{6}$}; \emph{Phys. Rev.~X} \textbf{10}, 021010
  (2020).

\bibitem{GoremychkinOsborn00}
Goremychkin, E.~A., Osborn, R., Rainford, B.~D., and Murani, A.~P.;
  \enquote{Evidence for anisotropic Kondo behavior in
  ${\mathrm{Ce}}_{0.8}{\mathrm{La}}_{0.2}{\mathrm{Al}}_{3}$}; \emph{Phys. Rev.
  Lett.} \textbf{84}, 2211--2214 (2000).

\bibitem{DemishevSemeno06}
Demishev, S.~V., Semeno, A.~V., Bogach, A.~V., Paderno, Y.~B., Shitsevalova,
  N.~Y., and Sluchanko, N.~E.; \enquote{Magnetic resonance in cerium hexaboride
  caused by quadrupolar ordering}; \emph{J.~Magn. Magn. Mater.} \textbf{300},
  e534--e537 (2006).

\bibitem{DemishevSemeno08}
Demishev, S.~V., Semeno, A.~V., Ohta, H., Okubo, S., Paderno, Y.~B.,
  Shitsevalova, N.~Y., and Sluchanko, N.~E.; \enquote{High-frequency study of
  the orbital ordering resonance in the strongly correlated heavy fermion metal
  CeB$_{6}$}; \emph{Appl. Magn. Reson.} \textbf{35}, 319--326 (2008).

\bibitem{Schlottmann12}
Schlottmann, P.; \enquote{Electron spin resonance in
  antiferro-quadrupolar-ordered CeB${}_{6}$}; \emph{Phys. Rev.~B} \textbf{86},
  075135 (2012).

\bibitem{Schlottmann13}
{Schlottmann}, P.; \enquote{{Electron spin resonance in CeB$_{6}$}};
  \emph{J.~Appl. Phys.} \textbf{113}, 17E109 (2013).

\bibitem{Schlottmann18}
Schlottmann, P.; \enquote{Theory of electron spin resonance in
  ferromagnetically correlated heavy fermion compounds};
  \emph{Magnetochemistry} \textbf{4}, 27 (2018).

\bibitem{LeQuintero-Castro13}
Le, M., Quintero-Castro, D., Toft-Petersen, R., Groitl, F., Skoulatos, M.,
  Rule, K., and Habicht, K.; \enquote{Gains from the upgrade of the cold
  neutron triple-axis spectrometer FLEXX at the BER-II reactor}; \emph{Nucl.
  Instrum. Methods Phys. Res. Sect.~A} \textbf{729}, 220--226 (2013).

\bibitem{SemenoGilmanov16}
Semeno, A.~V., Gilmanov, M.~I., Bogach, A.~V., Krasnorussky, V.~N., Samarin,
  A.~N., Samarin, N.~A., Sluchanko, N.~E., Shitsevalova, N.~Y., Filipov, V.~B.,
  Glushkov, V.~V., and Demishev, S.~V.; \enquote{Magnetic resonance anisotropy
  in CeB$_{6}$: an entangled state of the art}; \emph{Sci. Rep.} \textbf{6},
  39196 (2016).

\bibitem{SemenoGilmanov17}
Semeno, A., Gilmanov, M., Sluchanko, N., Krasnorussky, V., Shitsevalova, N.,
  Filipov, V., Flachbart, K., and Demishev, S.; \enquote{Angular dependences of
  ESR parameters in antiferroquadrupolar phase of CeB$_{6}$}; \emph{Acta
  Physica Polonica A} \textbf{131}, 1060--1062 (2017).

\bibitem{Gilmanov19}
Gilmanov, M.~I.; \emph{{\rm ``Electron spin resonance in rare-earth hexaborides
  $R$B$_{6}$ ($R$~=~Gd, Ce, Sm)'' (in Russian)}}; Ph.D. thesis; A. M. Prokhorov
  General Physics Institute of RAS, Moscow (2019).

\bibitem{SeraSato87}
Sera, M., Sato, N., and Kasuya, T.; \enquote{Magnetoelastic studies on
  Ce$_{1-x}$La$_{x}$B$_{6}$ single crystals}; \emph{J.~Magn. Magn. Mater.}
  \textbf{63--64}, 64--66 (1987).

\bibitem{YoshinoKobayashi04}
Yoshino, Y., Kobayashi, S., Tsuji, S., Tou, H., Sera, M., Iga, F., Zenitani,
  Y., and Akimitsu, J.; \enquote{{N}d ion doping effects on the multipolar
  interactions in {C}e{B}$_{6}$}; \emph{J.~Phys. Soc. Jpn.} \textbf{73}, 29--32
  (2004).

\bibitem{SchenckGygax07}
Schenck, A., Gygax, F.~N., and Solt, G.; \enquote{Complex phase diagram of
  {C}e$_{0.75}${L}a$_{0.25}${B}$_{6}$ studied by muon spin rotation and
  relaxation in zero and nonzero external fields}; \emph{Phys. Rev.~B}
  \textbf{75}, 024428 (2007).

\bibitem{MignotRobert09}
Mignot, J.-M., Robert, J., Andr\'e, G., Sera, M., and Iga, F.; \enquote{Effect
  of Nd substitution on the magnetic order in
  ${\text{Ce}}_{x}{\text{Nd}}_{1\ensuremath{-}x}{\text{B}}_{6}$ solid
  solutions}; \emph{Phys. Rev.~B} \textbf{79}, 224426 (2009).

\bibitem{ArkoCrabtree75}
Arko, A.~J., Crabtree, G., Ketterson, J.~B., Mueller, F.~M., Walch, P.~F.,
  Windmiller, L.~R., Fisk, Z., Hoyt, R.~F., Mota, A.~C., Viswanathan, R.,
  Ellis, D.~E., Freeman, A.~J., and Rath, J.; \enquote{Large electron-phonon
  interaction but low-temperature superconductivity in LaB$_{6}$}; \emph{Int.
  J. Quant. Chem.} \textbf{9}, 569--578 (1975).

\bibitem{BatkoBatkova95}
Bat'ko, I., Bat'kov\'a, M., Flachbart, K., Filippov, V., Paderno, Y.,
  Shicevalova, N., and Wagner, T.; \enquote{Electrical resistivity and
  superconductivity of {LaB}$_{6}$ and {LuB}$_{12}$}; \emph{J.~Alloy. Compd.}
  \textbf{217}, L1--L3 (1995).

\bibitem{InosovEvtushinsky09}
Inosov, D.~S., Evtushinsky, D.~V., Koitzsch, A., Zabolotnyy, V.~B., Borisenko,
  S.~V., Kordyuk, A.~A., Frontzek, M., Loewenhaupt, M., L\"oser, W., Mazilu,
  I., Bitterlich, H., Behr, G., Hoffmann, J.-U., Follath, R., and B\"uchner,
  B.; \enquote{Electronic structure and nesting-driven enhancement of the RKKY
  interaction at the magnetic ordering propagation vector in
  ${\mathrm{Gd}}_{2}{\mathrm{PdSi}}_{3}$ and
  ${\mathrm{Tb}}_{2}{\mathrm{PdSi}}_{3}$}; \emph{Phys. Rev. Lett.}
  \textbf{102}, 046401 (2009).

\bibitem{ButchManley15}
Butch, N.~P., Manley, M.~E., Jeffries, J.~R., Janoschek, M., Huang, K., Maple,
  M.~B., Said, A.~H., Leu, B.~M., and Lynn, J.~W.; \enquote{Symmetry and
  correlations underlying hidden order in URu$_{2}$Si$_{2}$}; \emph{Phys.
  Rev.~B} \textbf{91}, 035128 (2015).

\bibitem{ArkoCrabtree76}
Arko, A.~J., Crabtree, G., Karim, D., Mueller, F.~M., Windmiller, L.~R.,
  Ketterson, J.~B., and Fisk, Z.; \enquote{de Haas\,--\,van Alphen effect and
  the Fermi surface of La${\mathrm{B}}_{6}$}; \emph{Phys. Rev.~B} \textbf{13},
  5240--5247 (1976).

\bibitem{ChanHeine73}
Chan, S.~K. and Heine, V.; \enquote{Spin density wave and soft phonon mode from
  nesting Fermi surfaces}; \emph{J.~Phys. F: Metal Phys.} \textbf{3}, 795
  (1973).

\bibitem{Fawcett88}
Fawcett, E.; \enquote{Spin-density-wave antiferromagnetism in chromium};
  \emph{Rev. Mod. Phys.} \textbf{60}, 209--283 (1988).

\bibitem{BorisenkoKordyuk08}
Borisenko, S.~V., Kordyuk, A.~A., Yaresko, A.~N., Zabolotnyy, V.~B., Inosov,
  D.~S., Schuster, R., B\"uchner, B., Weber, R., Follath, R., Patthey, L., and
  Berger, H.; \enquote{Pseudogap and charge density waves in two dimensions};
  \emph{Phys. Rev. Lett.} \textbf{100}, 196402 (2008).

\bibitem{RudermanKittel54}
Ruderman, M.~A. and Kittel, C.; \enquote{Indirect exchange coupling of nuclear
  magnetic moments by conduction electrons}; \emph{Phys. Rev.} \textbf{96},
  99--102 (1954).

\bibitem{Kasuya56}
Kasuya, T.; \enquote{{A theory of metallic ferro- and antiferromagnetism on
  Zener's model}}; \emph{Prog. Theor. Phys.} \textbf{16}, 45--57 (1956).

\bibitem{Yafet87}
Yafet, Y.; \enquote{Ruderman-Kittel-Kasuya-Yosida range function of a
  one-dimensional free-electron gas}; \emph{Phys. Rev.~B} \textbf{36},
  3948--3949 (1987).

\bibitem{KimLee96}
Kim, J.~G., Lee, E.~K., and Lee, S.; \enquote{One-dimensional free-electron
  spin susceptibility at finite temperature}; \emph{Phys. Rev.~B} \textbf{54},
  6077--6080 (1996).

\bibitem{Aristov97}
Aristov, D.~N.; \enquote{Indirect RKKY interaction in any dimensionality};
  \emph{Phys. Rev.~B} \textbf{55}, 8064--8066 (1997).

\bibitem{LitvinovDugaev98}
Litvinov, V.~I. and Dugaev, V.~K.; \enquote{RKKY interaction in one- and
  two-dimensional electron gases}; \emph{Phys. Rev.~B} \textbf{58}, 3584--3585
  (1998).

\bibitem{BrownCaudron85}
Brown, P., Caudron, R., Fert, A., Givord, D., and Pureur, P.;
  \enquote{Helimagnetic structure in diluted Y-Gd alloys}; \emph{J.~Physique
  Lett.} \textbf{46}, 1139--1141 (1985).

\bibitem{FretwellDugdale99}
Fretwell, H.~M., Dugdale, S.~B., Alam, M.~A., Hedley, D. C.~R.,
  Rodriguez-Gonzalez, A., and Palmer, S.~B.; \enquote{Fermi surface as the
  driving mechanism for helical antiferromagnetic ordering in Gd-Y Alloys};
  \emph{Phys. Rev. Lett.} \textbf{82}, 3867--3870 (1999).

\bibitem{Robinson00}
Robinson, R.~A.; \emph{{\rm ``Magnetism in heavy fermion systems''}},  (World
  Scientific, Singapore2000).

\bibitem{Rossat-MignodRegnault88}
Rossat-Mignod, J., Regnault, L., Jacoud, J., Vettier, C., Lejay, P., Flouquet,
  J., Walker, E., Jaccard, D., and Amato, A.; \enquote{Inelastic neutron
  scattering study of cerium heavy fermion compounds}; \emph{J.~Magn. Magn.
  Mater.} \textbf{76--77}, 376--384 (1988).

\bibitem{RegnaultErkelens88a}
Regnault, L.~P., Erkelens, W. A.~C., Rossat-Mignod, J., Lejay, P., and
  Flouquet, J.; \enquote{Neutron scattering study of the heavy-fermion compound
  ${\mathrm{CeRu}}_{2}$${\mathrm{Si}}_{2}$}; \emph{Phys. Rev.~B} \textbf{38},
  4481--4487 (1988).

\bibitem{SchroederAeppli98}
Schr\"oder, A., Aeppli, G., Bucher, E., Ramazashvili, R., and Coleman, P.;
  \enquote{Scaling of magnetic fluctuations near a quantum phase transition};
  \emph{Phys. Rev. Lett.} \textbf{80}, 5623--5626 (1998).

\bibitem{StockertLoehneysen98}
Stockert, O., L\"ohneysen, H.~v., Rosch, A., Pyka, N., and Loewenhaupt, M.;
  \enquote{Two-dimensional fluctuations at the quantum-critical point of
  {C}e{C}u$_{6-x}${A}u$_{x}$}; \emph{Phys. Rev. Lett.} \textbf{80}, 5627--5630
  (1998).

\bibitem{KadowakiSato04}
Kadowaki, H., Sato, M., and Kawarazaki, S.; \enquote{Spin fluctuation in heavy
  fermion
  ${\mathrm{C}\mathrm{e}\mathrm{R}\mathrm{u}}_{2}{\mathrm{S}\mathrm{i}}_{2}$};
  \emph{Phys. Rev. Lett.} \textbf{92}, 097204 (2004).

\bibitem{StockSokolov11}
Stock, C., Sokolov, D.~A., Bourges, P., Tobash, P.~H., Gofryk, K., Ronning, F.,
  Bauer, E.~D., Rule, K.~C., and Huxley, A.~D.; \enquote{Anisotropic critical
  magnetic fluctuations in the ferromagnetic superconductor UCoGe}; \emph{Phys.
  Rev. Lett.} \textbf{107}, 187202 (2011).

\bibitem{SinghThamizhavel11}
Singh, D.~K., Thamizhavel, A., Lynn, J.~W., Dhar, S., Rodriguez-Rivera, J., and
  Herman, T.; \enquote{Field-induced quantum fluctuations in the heavy fermion
  superconductor {C}e{C}u$_{2}${G}e$_{2}$}; \emph{Sci. Rep.} \textbf{1}, 117
  (2011).

\bibitem{KimuraNakatsuji13}
Kimura, K., Nakatsuji, S., Wen, J.-J., Broholm, C., Stone, M.~B., Nishibori,
  E., and Sawa, H.; \enquote{Quantum fluctuations in spin-ice-like
  Pr$_{2}$Zr$_{2}$O$_{7}$}; \emph{Nat. Commun.} \textbf{4}, 1934-- (2013).

\bibitem{LowenhauptFischer93}
Lowenhaupt, M. and Fischer, K.~H.; \emph{{ \rm `Valence fluctuations and
  heavy-fermion $4f$ systems''}}; chap.~105 in Gschneidner Jr., K. A., and
  Eyring, L. (eds.), \textit{Handbook on the Physics and Chemistry of Rare
  Earths}, vol. 16, pp. 1--105,  (North-Holland, Amsterdam,~1993).

\bibitem{CoxBickers86}
Cox, D.~L., Bickers, N.~E., and Wilkins, J.~W.; \enquote{Calculated properties
  of valence fluctuators}; \emph{J.~Magn. Magn. Mater.} \textbf{54--57},
  333--337 (1986).

\bibitem{BickersCox87}
Bickers, N.~E., Cox, D.~L., and Wilkins, J.~W.; \enquote{Self-consistent
  large-$N$ expansion for normal-state properties of dilute magnetic alloys};
  \emph{Phys. Rev.~B} \textbf{36}, 2036--2079 (1987).

\bibitem{HaasBoer34}
de~Haas, W., de~Boer, J., and van D\"{e}n~Berg, G.; \enquote{The electrical
  resistance of gold, copper and lead at low temperatures}; \emph{Physica}
  \textbf{1}, 1115--1124 (1934).

\bibitem{Kondo64}
Kondo, J.; \enquote{Resistance minimum in dilute magnetic alloys}; \emph{Prog.
  Theor. Phys.} \textbf{32}, 37 (1964).

\bibitem{Hewson97}
Hewson, A.~C.; \emph{{\rm ``The Kondo problem to heavy fermions''}}; Cambridge
  Studies in Magnetism,  (Cambridge University Press1997); \!\!, ISBN
  0-521-36382-9.

\bibitem{Winzer75}
Winzer, K.; \enquote{Giant {Kondo} resistivity in ({L}a,\,{C}e){B}$_{6}$};
  \emph{Solid State Commun.} \textbf{16}, 521--524 (1975).

\bibitem{SamwerWinzer76}
Samwer, K. and Winzer, K.; \enquote{Magnetoresistivity of the Kondo system
  (La,\,Ce)B$_{6}$}; \emph{Z.~Physik B: Condens. Matter} \textbf{25}, 269--274
  (1976).

\bibitem{Felsch78}
Felsch, W.; \enquote{{K}ondo effect and impurity-impurity interaction in
  (La,\,Ce)B$_6$ alloys}; \emph{Z.~Physik B: Condens. Matter} \textbf{29},
  211--222 (1978).

\bibitem{SatoSumiyama85}
Sato, N., Sumiyama, A., Kunii, S., Nagano, H., and Kasuya, T.;
  \enquote{Interaction between Kondo states and the Hall effect of dense
  {K}ondo system {C}e$_{x}${L}a$_{1-x}${B}$_{6}$}; \emph{J.~Phys. Soc. Jpn.}
  \textbf{54}, 1923--1932 (1985).

\bibitem{Portnichenko18}
Portnichenko, P.~Y.; \emph{{\rm ``Magnetic dynamics in heavy-fermion systems
  with multipolar ordering studied by neutron scattering''}}; Ph.D. thesis;
  Fakult\"{a}t Physik der Technischen Universit\"{a}t Dresden (2018).

\bibitem{EreminZwicknagl08}
Eremin, I., Zwicknagl, G., Thalmeier, P., and Fulde, P.; \enquote{Feedback spin
  resonance in superconducting ${\mathrm{CeCu}}_{2}{\mathrm{Si}}_{2}$ and
  ${\mathrm{CeCoIn}}_{5}$}; \emph{Phys. Rev. Lett.} \textbf{101}, 187001
  (2008).

\bibitem{SatoAso01}
Sato, N.~K., Aso, N., Miyake, K., Shiina, R., Thalmeier, P., Varelogiannis, G.,
  Geibel, C., Steglich, F., Fulde, P., and Komatsubara, T.; \enquote{Strong
  coupling between local moments and superconducting `heavy' electrons in
  {UPd$_{2}$Al$_{3}$}}; \emph{Nature (London)} \textbf{410}, 340--343 (2001).

\bibitem{BlackburnHiess06}
Blackburn, E., Hiess, A., Bernhoeft, N., and Lander, G.~H.; \enquote{Inelastic
  neutron scattering from ${\mathrm{UPd}}_{2}{\mathrm{Al}}_{3}$ under high
  magnetic fields}; \emph{Phys. Rev. B} \textbf{74}, 024406 (2006).

\bibitem{ChangEremin07}
Chang, J., Eremin, I., Thalmeier, P., and Fulde, P.; \enquote{Theory of
  magnetic excitons in the heavy-fermion superconductor
  ${\mathrm{UPd}}_{2}{\mathrm{Al}}_{3}$}; \emph{Phys. Rev. B} \textbf{75},
  024503 (2007).

\bibitem{StockBroholm12}
Stock, C., Broholm, C., Demmel, F., Van~Duijn, J., Taylor, J.~W., Kang, H.~J.,
  Hu, R., and Petrovic, C.; \enquote{From incommensurate correlations to
  mesoscopic spin resonance in ${\mathrm{YbRh}}_{2}{\mathrm{Si}}_{2}$};
  \emph{Phys. Rev. Lett.} \textbf{109}, 127201 (2012).

\end{thebibliography}



\setcounter{chapter}{0}

\end{document}